\documentclass[12pt]{article} 
\usepackage{amsthm} 
\usepackage{amsmath,amssymb}         
\usepackage{stmaryrd}     
\usepackage{float}

\usepackage{xspace}  
\usepackage{color} 
\usepackage{epsfig}    
\graphicspath{{.}{./figures/}}   

\usepackage{macros/tikzfig} 
\usepackage{keycommand} 

\usepackage[ruled,vlined]{algorithm2e}

\newkeycommand{\boxmap}[style=small box][1]{\,\tikz{\node[style=\commandkey{style}] (x) {$#1$};\draw(0,-1.25)--(x)--(0,1.25);}\,}
\newkeycommand{\dboxmap}[style=dbox][1]{\,\tikz{\node[style=\commandkey{style}] (x) {$#1$};\draw[boldedge] (0,-1.25)--(x)--(0,1.25);}\,}

\newkeycommand{\wirelabel}[style=white label]{\,\tikz{\node[style=\commandkey{style}]{};}\,\xspace}

\newkeycommand{\pointmap}[style=point][1]{\,\tikz{\node[style=\commandkey{style}] (x) at (0,-0.3) {$#1$};\node [style=none] (3) at (0, 0.9) {};\node [style=none] (3) at (0, -0.9) {};\draw (x)--(0,0.7);}\,}

\newkeycommand{\point}[style=point,estyle=][1]{\,\tikz{\node[style=\commandkey{style}] (x) at (0,-0.05) {$#1$};\draw [\commandkey{estyle}] (x)--(0,1.05);}\,}

\newkeycommand{\copointmap}[style=copoint][1]{\,\tikz{\node[style=\commandkey{style}] (x) at (0,0.3) {$#1$};\draw (x)--(0,-0.7);}\,}

\newkeycommand{\dpoint}[style=point][1]{\,\tikz{\node[style=\commandkey{style},doubled] (x) at (0,-0.3) {$#1$};\draw[boldedge] (x)--(0,0.7);}\,}

\newkeycommand{\kpoint}[style=kpoint,estyle=][1]{\,\tikz{\node[style=\commandkey{style}] (x) at (0,-0.05) {$#1$};\draw [\commandkey{estyle}] (x)--(0,1.05);}\,}

\newkeycommand{\kpointgray}[style=gray kpoint,estyle=][1]{\,\tikz{\node[style=\commandkey{style}] (x) at (0,-0.05) {$#1$};\draw [\commandkey{estyle}] (x)--(0,1.05);}\,}

\newkeycommand{\typedkpoint}[style=kpoint,edgestyle=][2]{%
\,\begin{tikzpicture}
  \begin{pgfonlayer}{nodelayer}
    \node [style=none] (0) at (0, 1) {};
    \node [style=\commandkey{style}] (1) at (0, -0.75) {$#2$};
    \node [style=right label] (2) at (0.25, 0.5) {$#1$};
  \end{pgfonlayer}
  \begin{pgfonlayer}{edgelayer}
    \draw [\commandkey{edgestyle}] (1) to (0.center);
  \end{pgfonlayer}
\end{tikzpicture}\,}

\newkeycommand{\typedkpointdag}[style=kpoint adjoint,edgestyle=][2]{%
\,\begin{tikzpicture}
  \begin{pgfonlayer}{nodelayer}
    \node [style=none] (0) at (0, -1) {};
    \node [style=\commandkey{style}] (1) at (0, 0.75) {$#2$};
    \node [style=right label] (2) at (0.25, -0.5) {$#1$};
  \end{pgfonlayer}
  \begin{pgfonlayer}{edgelayer}
    \draw [\commandkey{edgestyle}] (0.center) to (1);
  \end{pgfonlayer}
\end{tikzpicture}\,}

\newkeycommand{\bistate}[style=kpoint][1]{\,\begin{tikzpicture}
  \begin{pgfonlayer}{nodelayer}
    \node [style=\commandkey{style}, minimum width=1 cm, inner sep=2pt] (0) at (0, -0.25) {$#1$};
    \node [style=none] (1) at (-0.75, 0) {};
    \node [style=none] (2) at (0.75, 0) {};
    \node [style=none] (3) at (-0.75, 0.88) {};
    \node [style=none] (4) at (0.75, 0.88) {};
  \end{pgfonlayer}
  \begin{pgfonlayer}{edgelayer}
    \draw (1.center) to (3.center);
    \draw (2.center) to (4.center);
  \end{pgfonlayer}
\end{tikzpicture}\,}

\newkeycommand{\bistatebig}[style=kpoint][1]{\,\begin{tikzpicture}
  \begin{pgfonlayer}{nodelayer}
    \node [style=\commandkey{style}, minimum width=, minimum width=1.26 cm, inner sep=2pt] (0) at (0, -0.25) {$#1$};
    \node [style=none] (1) at (-1, 0) {};
    \node [style=none] (2) at (1, 0) {};
    \node [style=none] (3) at (-1, 0.88) {};
    \node [style=none] (4) at (1, 0.88) {};
  \end{pgfonlayer}
  \begin{pgfonlayer}{edgelayer}
    \draw (1.center) to (3.center);
    \draw (2.center) to (4.center);
  \end{pgfonlayer}
\end{tikzpicture}\,}

\newkeycommand{\dbistate}[style=dkpoint][1]{\,\begin{tikzpicture}
  \begin{pgfonlayer}{nodelayer}
    \node [style=\commandkey{style}, minimum width=1 cm, inner sep=2pt] (0) at (0, -0.25) {$#1$};
    \node [style=none] (1) at (-0.75, 0) {};
    \node [style=none] (2) at (0.75, 0) {};
    \node [style=none] (3) at (-0.75, 1.0) {};
    \node [style=none] (4) at (0.75, 1.0) {};
  \end{pgfonlayer}
  \begin{pgfonlayer}{edgelayer}
    \draw [boldedge] (1.center) to (3.center);
    \draw [boldedge] (2.center) to (4.center);
  \end{pgfonlayer}
\end{tikzpicture}\,}

\newkeycommand{\bistatebraket}[style1=kpoint,style2=kpointadj][2]{\,\begin{tikzpicture}
  \begin{pgfonlayer}{nodelayer}
    \node [style=\commandkey{style1}, minimum width=1 cm, inner sep=2pt] (0) at (0, -0.75) {$#1$};
    \node [style=none] (1) at (-0.75, -0.5) {};
    \node [style=none] (2) at (0.75, -0.5) {};
    \node [style=none] (3) at (-0.75, 0.5) {};
    \node [style=none] (4) at (0.75, 0.5) {};
    \node [style=\commandkey{style2}, minimum width=1 cm, inner sep=2pt] (5) at (0, 0.75) {$#2$};
  \end{pgfonlayer}
  \begin{pgfonlayer}{edgelayer}
    \draw (1.center) to (3.center);
    \draw (2.center) to (4.center);
  \end{pgfonlayer}
\end{tikzpicture}\,}

\newkeycommand{\weight}[style=dkpoint][1]{\,\begin{tikzpicture}
  \begin{pgfonlayer}{nodelayer}
    \node [style=\commandkey{style}] (0) at (0, -0.5) {$#1$};
    \node [style=upground] (1) at (0, 0.75) {};
  \end{pgfonlayer}
  \begin{pgfonlayer}{edgelayer}
    \draw [style=boldedge] (0) to (1);
  \end{pgfonlayer}
\end{tikzpicture}\,}

\newkeycommand{\dkpoint}[style=dkpoint][1]{\,\tikz{\node[style=\commandkey{style}] (x) at (0,-0.1) {$#1$};\draw[boldedge] (x)--(0,1);}\,}
\newkeycommand{\dkpointadj}[style=dkpointadj][1]{\,\tikz{\node[style=\commandkey{style}] (x) at (0,0.1) {$#1$};\draw[boldedge] (x)--(0,-1);}\,}
\newkeycommand{\dkpointtrans}[style=dkpointtrans][1]{\,\tikz{\node[style=\commandkey{style}] (x) at (0,0.1) {$#1$};\draw[boldedge] (x)--(0,-1);}\,}
\newkeycommand{\dkpointconj}[style=dkpointconj][1]{\,\tikz{\node[style=\commandkey{style}] (x) at (0,-0.1) {$#1$};\draw[boldedge] (x)--(0,1);}\,}

\newkeycommand{\blackkpoint}[style=black kpoint][1]{\,\tikz{\node[style=\commandkey{style}] (x) at (0,-0.1) {$#1$};\draw (x)--(0,1);}\,}
\newkeycommand{\blackkpointadj}[style=black kpointadj][1]{\,\tikz{\node[style=\commandkey{style}] (x) at (0,0.1) {$#1$};\draw (x)--(0,-1);}\,}
\newkeycommand{\blackdkpoint}[style=black dkpoint][1]{\,\tikz{\node[style=\commandkey{style}] (x) at (0,-0.1) {$#1$};\draw[boldedge] (x)--(0,1);}\,}
\newkeycommand{\blackdkpointadj}[style=black dkpointadj][1]{\,\tikz{\node[style=\commandkey{style}] (x) at (0,0.1) {$#1$};\draw[boldedge] (x)--(0,-1);}\,}

\newcommand{\trace}{\,\begin{tikzpicture}
  \begin{pgfonlayer}{nodelayer}
    \node [style=none] (0) at (0, -0.60) {};
    \node [style=upground] (1) at (0, 0.40) {};
  \end{pgfonlayer}
  \begin{pgfonlayer}{edgelayer}
    \draw [style=none] (0.center) to (1);
  \end{pgfonlayer}
\end{tikzpicture}\,}

\newcommand\discard\trace

\newkeycommand{\pointketbra}[style1=copoint,style2=point,estyle=][2]{\,%
\begin{tikzpicture}
  \begin{pgfonlayer}{nodelayer}
    \node [style=none] (0) at (0, -1.75) {};
    \node [style=\commandkey{style1}] (1) at (0, -0.75) {$#1$};
    \node [style=\commandkey{style2}] (2) at (0, 0.75) {$#2$};
    \node [style=none] (3) at (0, 1.75) {};
  \end{pgfonlayer}
  \begin{pgfonlayer}{edgelayer}
    \draw [\commandkey{estyle}] (0.center) to (1);
    \draw [\commandkey{estyle}] (2) to (3.center);
  \end{pgfonlayer}
\end{tikzpicture}\,}

\newkeycommand{\twocopointketbra}[style1=copoint,style2=copoint,style3=point][3]{\,%
\begin{tikzpicture}
  \begin{pgfonlayer}{nodelayer}
    \node [style=none] (0) at (-0.75, -1.75) {};
    \node [style=none] (1) at (0.75, -1.75) {};
    \node [style=\commandkey{style1}] (2) at (-0.75, -0.75) {$#1$};
    \node [style=\commandkey{style2}] (3) at (0.75, -0.75) {$#2$};
    \node [style=\commandkey{style3}] (4) at (0, 0.75) {$#3$};
    \node [style=none] (5) at (0, 1.75) {};
  \end{pgfonlayer}
  \begin{pgfonlayer}{edgelayer}
    \draw (0.center) to (2);
    \draw (1.center) to (3);
    \draw (4) to (5.center);
  \end{pgfonlayer}
\end{tikzpicture}\,}

\newkeycommand{\twopointketbra}[style1=copoint,style2=point,style3=point][3]{\,%
\begin{tikzpicture}
  \begin{pgfonlayer}{nodelayer}
    \node [style=none] (0) at (0, -1.75) {};
    \node [style=none] (1) at (-0.75, 1.75) {};
    \node [style=\commandkey{style1}] (2) at (0, -0.75) {$#1$};
    \node [style=\commandkey{style2}] (3) at (-0.75, 0.75) {$#2$};
    \node [style=\commandkey{style3}] (4) at (0.75, 0.75) {$#3$};
    \node [style=none] (5) at (0.75, 1.75) {};
  \end{pgfonlayer}
  \begin{pgfonlayer}{edgelayer}
    \draw (0.center) to (2);
    \draw (1.center) to (3);
    \draw (4) to (5.center);
  \end{pgfonlayer}
\end{tikzpicture}\,}

\newkeycommand{\pointbraket}[style1=point,style2=copoint,estyle=][2]{\,%
\begin{tikzpicture}
  \begin{pgfonlayer}{nodelayer}
    \node [style=\commandkey{style1}] (0) at (0, -0.625) {$#1$};
    \node [style=\commandkey{style2}] (1) at (0, 0.625) {$#2$};
  \end{pgfonlayer}
  \begin{pgfonlayer}{edgelayer}
    \draw [\commandkey{estyle}] (0) to (1);
  \end{pgfonlayer}
\end{tikzpicture}\,}

\newkeycommand{\kpointketbra}[style1=kpointdag,style2=kpoint,estyle=][2]{\,%
\begin{tikzpicture}
  \begin{pgfonlayer}{nodelayer}
    \node [style=none] (0) at (0, -1.9) {};
    \node [style=\commandkey{style1}] (1) at (0, -0.95) {$#1$};
    \node [style=\commandkey{style2}] (2) at (0, 0.95) {$#2$};
    \node [style=none] (3) at (0, 1.9) {};
  \end{pgfonlayer}
  \begin{pgfonlayer}{edgelayer}
    \draw [\commandkey{estyle}] (0.center) to (1);
    \draw [\commandkey{estyle}] (2) to (3.center);
  \end{pgfonlayer}
\end{tikzpicture}\,}

\newkeycommand{\kpointmap}[style1=kpoint,style2=map][2]{\,%
\begin{tikzpicture}
  \begin{pgfonlayer}{nodelayer}
    \node [style=\commandkey{style1}] (0) at (0, -1.1) {$#1$};
    \node [style=\commandkey{style2}] (1) at (0, 0.5) {$#2$};
    \node [style=none] (2) at (0, 1.75) {};
  \end{pgfonlayer}
  \begin{pgfonlayer}{edgelayer}
    \draw (0) to (1);
    \draw (1) to (2.center);
  \end{pgfonlayer}
\end{tikzpicture}%
\,}

\newkeycommand{\kpointmaptrans}[style1=kpointconj,style2=mapadj][2]{\,%
\begin{tikzpicture}
  \begin{pgfonlayer}{nodelayer}
    \node [style=\commandkey{style1}] (0) at (0, -1.1) {$#1$};
    \node [style=\commandkey{style2}] (1) at (0, 0.5) {$#2$};
    \node [style=none] (2) at (0, 1.75) {};
  \end{pgfonlayer}
  \begin{pgfonlayer}{edgelayer}
    \draw (0) to (1);
    \draw (1) to (2.center);
  \end{pgfonlayer}
\end{tikzpicture}%
\,}

\newkeycommand{\kpointmapadj}[style1=kpointadj,style2=map][2]{\,%
\begin{tikzpicture}
  \begin{pgfonlayer}{nodelayer}
    \node [style=\commandkey{style1}] (0) at (0, 1.1) {$#1$};
    \node [style=\commandkey{style2}] (1) at (0, -0.5) {$#2$};
    \node [style=none] (2) at (0, -1.75) {};
  \end{pgfonlayer}
  \begin{pgfonlayer}{edgelayer}
    \draw (0) to (1);
    \draw (1) to (2.center);
  \end{pgfonlayer}
\end{tikzpicture}%
\,}

\newkeycommand{\dkpointmap}[style1=dkpoint,style2=dmap][2]{\,%
\begin{tikzpicture}
  \begin{pgfonlayer}{nodelayer}
    \node [style=\commandkey{style1}] (0) at (0, -1.2) {$#1$};
    \node [style=\commandkey{style2}] (1) at (0, 0.4) {$#2$};
    \node [style=none] (2) at (0, 1.7) {};
  \end{pgfonlayer}
  \begin{pgfonlayer}{edgelayer}
    \draw[style=boldedge] (0) to (1);
    \draw[style=boldedge] (1) to (2.center);
  \end{pgfonlayer}
\end{tikzpicture}%
\,}

\newkeycommand{\dkpointmapx}[style1=dkpoint,style2=dmap][2]{\,%
\begin{tikzpicture}
  \begin{pgfonlayer}{nodelayer}
    \node [style=\commandkey{style1}] (0) at (0, -1.4) {$#1$};
    \node [style=\commandkey{style2}] (1) at (0, 0.6) {$#2$};
    \node [style=none] (2) at (0, 1.9) {};
  \end{pgfonlayer}
  \begin{pgfonlayer}{edgelayer}
    \draw[style=boldedge] (0) to (1);
    \draw[style=boldedge] (1) to (2.center);
  \end{pgfonlayer}
\end{tikzpicture}%
\,}

\newkeycommand{\boxcopointmapdag}[style1=map,style2=copoint][2]{\,%
\begin{tikzpicture}
  \begin{pgfonlayer}{nodelayer}
    \node [style=none] (0) at (0, -1.75) {};
    \node [style=\commandkey{style1}] (1) at (0, -0.5) {$#1$};
    \node [style=\commandkey{style2}] (2) at (0, 1) {$#2$};
  \end{pgfonlayer}
  \begin{pgfonlayer}{edgelayer}
    \draw (0.center) to (1);
    \draw (1) to (2);
  \end{pgfonlayer}
\end{tikzpicture}%
\,}

\newkeycommand{\kpointdagmap}[style1=map,style2=kpointdag][2]{\,%
\begin{tikzpicture}
  \begin{pgfonlayer}{nodelayer}
    \node [style=none] (0) at (0, -1.75) {};
    \node [style=\commandkey{style1}] (1) at (0, -0.5) {$#1$};
    \node [style=\commandkey{style2}] (2) at (0, 1.1) {$#2$};
  \end{pgfonlayer}
  \begin{pgfonlayer}{edgelayer}
    \draw (0.center) to (1);
    \draw (1) to (2);
  \end{pgfonlayer}
\end{tikzpicture}%
\,}

\newkeycommand{\sandwichmap}[style1=point,style2=map,style3=copoint][3]{\,%
\begin{tikzpicture}
  \begin{pgfonlayer}{nodelayer}
    \node [style=\commandkey{style1}] (0) at (0, -1.50) {$#1$};
    \node [style=\commandkey{style2}] (1) at (0, 0) {$#2$};
    \node [style=\commandkey{style3}] (2) at (0, 1.50) {$#3$};
  \end{pgfonlayer}
  \begin{pgfonlayer}{edgelayer}
    \draw (0) to (1);
    \draw (1) to (2);
  \end{pgfonlayer}
\end{tikzpicture}%
}

\newkeycommand{\kpointsandwichmap}[style1=kpoint,style2=map,style3=kpointdag][3]{\,%
\begin{tikzpicture}
  \begin{pgfonlayer}{nodelayer}
    \node [style=\commandkey{style1}] (0) at (0, -1.5) {$#1$};
    \node [style=\commandkey{style2}] (1) at (0, 0) {$#2$};
    \node [style=\commandkey{style3}] (2) at (0, 1.5) {$#3$};
  \end{pgfonlayer}
  \begin{pgfonlayer}{edgelayer}
    \draw (0) to (1);
    \draw (1) to (2);
  \end{pgfonlayer}
\end{tikzpicture}%
}

\newkeycommand{\sandwichmapconj}[style1=point,style2=mapconj,style3=copoint][3]{\,% 
\begin{tikzpicture}
  \begin{pgfonlayer}{nodelayer}
    \node [style=\commandkey{style1}] (0) at (0, -1.50) {$#1$};
    \node [style=\commandkey{style2}] (1) at (0, 0) {$#2$};
    \node [style=\commandkey{style3}] (2) at (0, 1.50) {$#3$};
  \end{pgfonlayer}
  \begin{pgfonlayer}{edgelayer}
    \draw (0) to (1);
    \draw (1) to (2);
  \end{pgfonlayer}
\end{tikzpicture}%
}

\newkeycommand{\sandwichtwo}[style1=point,style2=map,style3=map,style4=copoint][4]{\,%
\begin{tikzpicture}
  \begin{pgfonlayer}{nodelayer}
    \node [style=\commandkey{style2}] (0) at (0, -1) {$#2$};
    \node [style=\commandkey{style3}] (1) at (0, 1) {$#3$};
    \node [style=\commandkey{style1}] (2) at (0, -2.6) {$#1$};
    \node [style=\commandkey{style4}] (3) at (0, 2.6) {$#4$};
  \end{pgfonlayer}
  \begin{pgfonlayer}{edgelayer}
    \draw (2) to (0);
    \draw (0) to (1);
    \draw (1) to (3);
  \end{pgfonlayer}
\end{tikzpicture}\,}

\newkeycommand{\longbraket}[style1=point,style2=copoint][2]{\,% 
\begin{tikzpicture}
  \begin{pgfonlayer}{nodelayer}
    \node [style=\commandkey{style1}] (0) at (0, -2.6) {$#1$};
    \node [style=\commandkey{style2}] (1) at (0, 2.6) {$#2$};
  \end{pgfonlayer}
  \begin{pgfonlayer}{edgelayer}
    \draw (0) to (1);
  \end{pgfonlayer}
\end{tikzpicture}\,}

\newkeycommand{\onb}[style=point]{\ensuremath{\{\,\tikz{\node[style=\commandkey{style}] (x) at (0,-0.3) {$j$};\draw (x)--(0,0.7);}\,\}}\xspace}

\newkeycommand{\phase}[style=white phase dot][1]{\,\begin{tikzpicture}
    \begin{pgfonlayer}{nodelayer}
        \node [style=none] (0) at (0, 1) {};
        \node [style=\commandkey{style}] (2) at (0, -0) {$#1$}; 
        \node [style=none] (3) at (0, -1) {};
    \end{pgfonlayer}
    \begin{pgfonlayer}{edgelayer}
        \draw (2) to (0.center);
        \draw (3.center) to (2);
    \end{pgfonlayer}
\end{tikzpicture}\,}

\newkeycommand{\dphase}[style=white phase ddot][1]{\,\begin{tikzpicture}
    \begin{pgfonlayer}{nodelayer}
        \node [style=none] (0) at (0, 1.25) {};
        \node [style=\commandkey{style}] (2) at (0, -0) {$#1$};
        \node [style=none] (3) at (0, -1.25) {};
    \end{pgfonlayer}
    \begin{pgfonlayer}{edgelayer}
        \draw [boldedge] (2) to (0.center);
        \draw [boldedge] (3.center) to (2);
    \end{pgfonlayer}
\end{tikzpicture}\,}

\newkeycommand{\dphasepoint}[style=white phase ddot][1]{\,\begin{tikzpicture}[yshift=-3mm]
  \begin{pgfonlayer}{nodelayer}
    \node [style=none] (0) at (0, 1) {};
    \node [style=\commandkey{style}] (1) at (0, 0) {$#1$};
  \end{pgfonlayer}
  \begin{pgfonlayer}{edgelayer}
    \draw [style=boldedge] (0.center) to (1);
  \end{pgfonlayer}
\end{tikzpicture}\,}

\newkeycommand{\dphasepointgray}[style=gray phase ddot][1]{\,\begin{tikzpicture}[yshift=-3mm]
  \begin{pgfonlayer}{nodelayer}
    \node [style=none] (0) at (0, 1) {};
    \node [style=\commandkey{style}] (1) at (0, 0) {$#1$};
  \end{pgfonlayer}
  \begin{pgfonlayer}{edgelayer}
    \draw [style=boldedge] (0.center) to (1);
  \end{pgfonlayer}
\end{tikzpicture}\,}

\newkeycommand{\dphasecopoint}[style=white phase ddot][1]{\,\begin{tikzpicture}[yshift=3mm,yscale=-1]
  \begin{pgfonlayer}{nodelayer}
    \node [style=none] (0) at (0, 1) {};
    \node [style=\commandkey{style}] (1) at (0, 0) {$#1$};
  \end{pgfonlayer}
  \begin{pgfonlayer}{edgelayer}
    \draw [style=boldedge] (0.center) to (1);
  \end{pgfonlayer}
\end{tikzpicture}\,}

\newkeycommand{\dphasepointsm}[style=white phase ddot][1]{\,\begin{tikzpicture}[yshift=-3mm]
  \begin{pgfonlayer}{nodelayer}
    \node [style=none] (0) at (0, 1) {};
    \node [style=\commandkey{style}] (1) at (0, 0) {\footnotesize\!$#1$\!};
  \end{pgfonlayer}
  \begin{pgfonlayer}{edgelayer}
    \draw [style=boldedge] (0.center) to (1);
  \end{pgfonlayer}
\end{tikzpicture}\,}

\newkeycommand{\phasepoint}[style=white phase dot][1]{\,\begin{tikzpicture}[yshift=-3mm]
  \begin{pgfonlayer}{nodelayer}
    \node [style=none] (0) at (0, 1) {};
    \node [style=\commandkey{style}] (1) at (0, 0) {$#1$};
  \end{pgfonlayer}
  \begin{pgfonlayer}{edgelayer}
    \draw (0.center) to (1);
  \end{pgfonlayer}
\end{tikzpicture}\,}

\newkeycommand{\phasecopoint}[style=white phase dot][1]{\,\begin{tikzpicture}[yshift=3mm]
  \begin{pgfonlayer}{nodelayer}
    \node [style=none] (0) at (0, -1) {};
    \node [style=\commandkey{style}] (1) at (0, 0) {$#1$};
  \end{pgfonlayer}
  \begin{pgfonlayer}{edgelayer}
    \draw (0.center) to (1);
  \end{pgfonlayer}
\end{tikzpicture}\,}

\newkeycommand{\kpointbraket}[style1=kpoint,style2=kpoint adjoint,estyle=][2]{\,%
\begin{tikzpicture}
  \begin{pgfonlayer}{nodelayer}
    \node [style=\commandkey{style1}] (0) at (0, -0.735) {$#1$};
    \node [style=\commandkey{style2}] (1) at (0, 0.735) {$#2$};
  \end{pgfonlayer}
  \begin{pgfonlayer}{edgelayer}
    \draw [\commandkey{estyle}] (0) to (1);
  \end{pgfonlayer}
\end{tikzpicture}\,}

\newkeycommand{\kkpointbraket}[style1=kpoint,style2=kpoint adjoint,estyle=][2]{\,%
\begin{tikzpicture}
  \begin{pgfonlayer}{nodelayer}
    \node [style=\commandkey{style1}] (0) at (0, -0.685) {$#1$};
    \node [style=\commandkey{style2}] (1) at (0, 0.685) {$#2$};
  \end{pgfonlayer}
  \begin{pgfonlayer}{edgelayer}
    \draw [\commandkey{estyle}] (0) to (1);
  \end{pgfonlayer}
\end{tikzpicture}\,}

\newkeycommand{\kpointbraketx}[style1=kpoint,style2=kpoint adjoint,estyle=][2]{\,%
\begin{tikzpicture}
  \begin{pgfonlayer}{nodelayer}
    \node [style=\commandkey{style1}] (0) at (0, -0.885) {$#1$};
    \node [style=\commandkey{style2}] (1) at (0, 0.885) {$#2$};
  \end{pgfonlayer}
  \begin{pgfonlayer}{edgelayer}
    \draw [\commandkey{estyle}] (0) to (1);
  \end{pgfonlayer}
\end{tikzpicture}\,}

\tikzstyle{cdiag}=[matrix of math nodes, row sep=3em, column sep=3em, text height=1.5ex, text depth=0.25ex,inner sep=0.5em]
\tikzstyle{arrow above}=[transform canvas={yshift=0.5ex}]
\tikzstyle{arrow below}=[transform canvas={yshift=-0.5ex}]

\usepackage{tikz}
\usetikzlibrary{decorations.markings}
\usetikzlibrary{shapes.geometric}

\pgfdeclarelayer{edgelayer}
\pgfdeclarelayer{nodelayer}
\pgfsetlayers{edgelayer,nodelayer,main}

\tikzstyle{none}=[inner sep=0pt]
\definecolor{hexcolor0xff0000}{rgb}{1.000,0.000,0.000}
\definecolor{hexcolor0x000000}{rgb}{0.000,0.000,0.000}
\definecolor{hexcolor0x00ff00}{rgb}{0.000,1.000,0.000}
\definecolor{hexcolor0x000000}{rgb}{0.000,0.000,0.000}
\definecolor{hexcolor0xffff00}{rgb}{1.000,1.000,0.000}
\definecolor{hexcolor0xffffff}{rgb}{1.000,1.000,1.000}

\tikzstyle{rn}=[circle,fill=hexcolor0xff0000,draw=hexcolor0x000000,line width=0.8 pt]
\tikzstyle{gn}=[circle,fill=hexcolor0x00ff00,draw=hexcolor0x000000,line width=0.8 pt]
\tikzstyle{yn}=[circle,fill=hexcolor0xffff00,draw=hexcolor0x000000,line width=0.8 pt]
\tikzstyle{wn}=[circle,fill=hexcolor0xffffff,draw=hexcolor0x000000,line width=0.8 pt]
\tikzstyle{wnthick}=[circle,fill=hexcolor0xffffff,draw=hexcolor0x000000,line width=2.500]

\tikzstyle{simple}=[-,draw=hexcolor0x000000,line width=2.000]
\tikzstyle{arrow}=[-,draw=hexcolor0x000000,postaction={decorate},decoration={markings,mark=at position .5 with {\arrow{>}}},line width=2.000]
\tikzstyle{tick}=[-,draw=hexcolor0x000000,postaction={decorate},decoration={markings,mark=at position .5 with {\draw (0,-0.1) -- (0,0.1);}},line width=2.000]
\tikzstyle{halfthickness}=[-,draw=hexcolor0x000000,line width=0.500]
\tikzstyle{thick}=[-,draw=hexcolor0x000000,line width=2.500]
\tikzstyle{thicker}=[-,draw=hexcolor0x000000,line width=4.000]

\tikzstyle{env}=[copoint,regular polygon rotate=0,minimum width=0.2cm, fill=black]

\tikzstyle{probs}=[shape=semicircle,fill=white,draw=black,shape border rotate=180,minimum width=1.2cm]

\tikzstyle{every picture}=[baseline=-0.25em,scale=0.5]
\tikzstyle{dotpic}=[] % for backwards-compatibility
\tikzstyle{diredges}=[every to/.style={diredge}]
\tikzstyle{math matrix}=[matrix of math nodes,left delimiter=(,right delimiter=),inner sep=2pt,column sep=1em,row sep=0.5em,nodes={inner sep=0pt},text height=1.5ex, text depth=0.25ex]

\tikzstyle{inline text}=[text height=1.5ex, text depth=0.25ex,yshift=0.5mm]
\tikzstyle{label}=[font=\footnotesize,text height=1.5ex, text depth=0.25ex,yshift=0.5mm]
\tikzstyle{left label}=[label,anchor=east,xshift=1.5mm]
\tikzstyle{right label}=[label,anchor=west,xshift=-1.5mm]

\tikzstyle{braceedge}=[decorate,decoration={brace,amplitude=2mm,raise=-1mm}]
\tikzstyle{small braceedge}=[decorate,decoration={brace,amplitude=1mm,raise=-1mm}]

\tikzstyle{doubled}=[line width=1.6pt] % set the line width for all doubled (quantum) maps/wires
\tikzstyle{boldedge}=[doubled,shorten <=-0.17mm,shorten >=-0.17mm]
\tikzstyle{boldedgegray}=[doubled,gray,shorten <=-0.17mm,shorten >=-0.17mm]

\tikzstyle{semidoubled}=[line width=1.4pt] % set the line width for all doubled (quantum) maps/wires
\tikzstyle{semiboldedgegray}=[semidoubled,gray,shorten <=-0.17mm,shorten >=-0.17mm]

\tikzstyle{boldedgedashed}=[very thick,dashed,shorten <=-0.17mm,shorten >=-0.17mm]
\tikzstyle{vboldedgedashed}=[doubled,dashed,shorten <=-0.17mm,shorten >=-0.17mm]
\tikzstyle{left hook arrow}=[left hook-latex]
\tikzstyle{right hook arrow}=[right hook-latex]
\tikzstyle{sembracket}=[line width=0.5pt,shorten <=-0.07mm,shorten >=-0.07mm]

\tikzstyle{causal edge}=[->,thick,gray]
\tikzstyle{causal nondir}=[thick,gray]
\tikzstyle{timeline}=[thick,gray, dashed]

\tikzstyle{cedge}=[<->,thick,gray!70!white]

\tikzstyle{empty diagram}=[draw=gray!40!white,dashed,shape=rectangle,minimum width=1cm,minimum height=1cm]
\tikzstyle{empty diagram small}=[draw=gray!50!white,dashed,shape=rectangle,minimum width=0.6cm,minimum height=0.5cm]

\tikzstyle{dot}=[inner sep=0mm,minimum width=2mm,minimum height=2mm,draw,shape=circle]
\tikzstyle{ddot}=[inner sep=0mm, doubled, minimum width=2.5mm,minimum height=2.5mm,draw,shape=circle]

\tikzstyle{black dot}=[dot,fill=black]
\tikzstyle{white dot}=[dot,fill=white,,text depth=-0.2mm]
\tikzstyle{green dot}=[white dot] % for backwards-compatibility
\tikzstyle{gray dot}=[dot,fill=gray!40!white,,text depth=-0.2mm]
\tikzstyle{red dot}=[gray dot] % for backwards-compatibility

\tikzstyle{black ddot}=[ddot,fill=black]
\tikzstyle{white ddot}=[ddot,fill=white]
\tikzstyle{gray ddot}=[ddot,fill=gray!40!white]

\tikzstyle{gray edge}=[gray!40!white]

\tikzstyle{small dot}=[inner sep=0.5mm,minimum width=0pt,minimum height=0pt,draw,shape=circle]

\tikzstyle{small black dot}=[small dot,fill=black]
\tikzstyle{small white dot}=[small dot,fill=white]
\tikzstyle{small gray dot}=[small dot,fill=gray!40!white]

\tikzstyle{causal dot}=[inner sep=0.4mm,minimum width=0pt,minimum height=0pt,draw=white,shape=circle,fill=gray!40!white]

\tikzstyle{phase dimensions}=[minimum size=5mm,font=\footnotesize,rectangle,rounded corners=2.5mm,inner sep=0.2mm,outer sep=-2mm]
\tikzstyle{dphase dimensions}=[minimum size=5mm,font=\footnotesize,rectangle,rounded corners=2.5mm,inner sep=0.2mm,outer sep=-2mm]
\tikzstyle{white phase dot}=[dot,fill=white,phase dimensions]
\tikzstyle{white phase ddot}=[ddot,fill=white,dphase dimensions]
\tikzstyle{green phase ddot}=[ddot,fill=green,dphase dimensions]

\tikzstyle{white rect ddot}=[draw=black,fill=white,doubled,minimum size=5mm,font=\footnotesize,rectangle,rounded corners=2.5mm,inner sep=0.2mm]
\tikzstyle{gray rect ddot}=[draw=black,fill=gray!40!white,doubled,minimum size=6mm,font=\footnotesize,rectangle,rounded corners=3mm]

\tikzstyle{gray phase dot}=[dot,fill=gray!40!white,phase dimensions]
\tikzstyle{gray phase ddot}=[ddot,fill=gray!40!white,dphase dimensions]
\tikzstyle{red phase ddot}=[ddot,fill=red,dphase dimensions]

\tikzstyle{grey phase dot}=[gray phase dot]
\tikzstyle{grey phase ddot}=[gray phase ddot]

\tikzstyle{small phase dimensions}=[minimum size=4mm,font=\tiny,rectangle,rounded corners=2mm,inner sep=0.2mm,outer sep=-2mm]
\tikzstyle{small dphase dimensions}=[minimum size=4mm,font=\tiny,rectangle,rounded corners=2mm,inner sep=0.2mm,outer sep=-2mm]

\tikzstyle{small gray phase dot}=[dot,fill=gray!40!white,small phase dimensions]
\tikzstyle{small gray phase ddot}=[ddot,fill=gray!40!white,small dphase dimensions]

\tikzstyle{small map}=[draw,shape=rectangle,minimum height=4mm,minimum width=4mm,fill=white]

\tikzstyle{cnot}=[fill=white,shape=circle,inner sep=-1.4pt]

\tikzstyle{asym hadamard}=[fill=white,draw,shape=NEbox,inner sep=0.6mm,font=\footnotesize,minimum height=4mm]
\tikzstyle{asym hadamard conj}=[fill=white,draw,shape=NWbox,inner sep=0.6mm,font=\footnotesize,minimum height=4mm]
\tikzstyle{asym hadamard dag}=[fill=white,draw,shape=SEbox,inner sep=0.6mm,font=\footnotesize,minimum height=4mm]

\tikzstyle{hadamard}=[fill=white,draw,inner sep=0.6mm,font=\footnotesize,minimum height=4mm,minimum width=4mm]
\tikzstyle{small hadamard}=[fill=white,draw,inner sep=0.6mm,minimum height=1.5mm,minimum width=1.5mm]
\tikzstyle{dhadamard}=[hadamard,doubled]
\tikzstyle{small dhadamard}=[small hadamard,doubled]
\tikzstyle{small dhadamard rotate}=[small hadamard,doubled,rotate=45]
\tikzstyle{antipode}=[white dot,inner sep=0.3mm,font=\footnotesize]

\tikzstyle{scalar}=[diamond,draw,inner sep=0.5pt,font=\small]
\tikzstyle{dscalar}=[diamond,doubled, draw,inner sep=0.5pt,font=\small]

\tikzstyle{small box}=[rectangle,inline text,fill=white,draw,minimum height=5mm,yshift=-0.5mm,minimum width=5mm,font=\small]
\tikzstyle{small gray box}=[small box,fill=gray!30]
\tikzstyle{medium box}=[rectangle,inline text,fill=white,draw,minimum height=5mm,yshift=-0.5mm,minimum width=10mm,font=\small]
\tikzstyle{square box}=[small box] % for backwards-compatibility
\tikzstyle{medium gray box}=[small box,fill=gray!30]
\tikzstyle{semilarge box}=[rectangle,inline text,fill=white,draw,minimum height=5mm,yshift=-0.5mm,minimum width=12.5mm,font=\small]
\tikzstyle{large box}=[rectangle,inline text,fill=white,draw,minimum height=5mm,yshift=-0.5mm,minimum width=15mm,font=\small]
\tikzstyle{large gray box}=[small box,fill=gray!30]

\tikzstyle{Bayes box}=[rectangle,fill=black,draw, minimum height=3mm, minimum width=3mm]

\tikzstyle{gray square point}=[small box,fill=gray!50]

\tikzstyle{dphase box white}=[dhadamard]
\tikzstyle{dphase box gray}=[dhadamard,fill=gray!50!white]

\tikzstyle{point}=[regular polygon,regular polygon sides=3,draw,scale=0.75,inner sep=-0.5pt,minimum width=9mm,fill=white,regular polygon rotate=180]
\tikzstyle{copoint}=[regular polygon,regular polygon sides=3,draw,scale=0.75,inner sep=-0.5pt,minimum width=9mm,fill=white]
\tikzstyle{dpoint}=[point,doubled]
\tikzstyle{dcopoint}=[copoint,doubled]

\tikzstyle{wide copoint}=[fill=white,draw,shape=isosceles triangle,shape border rotate=90,isosceles triangle stretches=true,inner sep=0pt,minimum width=1.5cm,minimum height=6.12mm]
\tikzstyle{wide point}=[fill=white,draw,shape=isosceles triangle,shape border rotate=-90,isosceles triangle stretches=true,inner sep=0pt,minimum width=1.5cm,minimum height=6.12mm,yshift=-0.0mm]
\tikzstyle{wide point plus}=[fill=white,draw,shape=isosceles triangle,shape border rotate=-90,isosceles triangle stretches=true,inner sep=0pt,minimum width=1.74cm,minimum height=7mm,yshift=-0.0mm]

\tikzstyle{wide dpoint}=[fill=white,doubled,draw,shape=isosceles triangle,shape border rotate=-90,isosceles triangle stretches=true,inner sep=0pt,minimum width=1.5cm,minimum height=6.12mm,yshift=-0.0mm]
\tikzstyle{wide dcopoint}=[fill=white,doubled,draw,shape=isosceles triangle,shape border rotate=90,isosceles triangle stretches=true,inner sep=0pt,minimum width=1.5cm,minimum height=6.12mm,yshift=-0.0mm]

\tikzstyle{tinypoint}=[regular polygon,regular polygon sides=3,draw,scale=0.55,inner sep=-0.15pt,minimum width=6mm,fill=white,regular polygon rotate=180]

\tikzstyle{white point}=[point]
\tikzstyle{white dpoint}=[dpoint]
\tikzstyle{green point}=[white point] % for backwards-compatibility
\tikzstyle{white copoint}=[copoint]
\tikzstyle{gray point}=[point,fill=gray!40!white]
\tikzstyle{gray dpoint}=[gray point,doubled]
\tikzstyle{red point}=[gray point] % for backwards-compatibility
\tikzstyle{gray copoint}=[copoint,fill=gray!40!white]
\tikzstyle{gray dcopoint}=[gray copoint,doubled]

\tikzstyle{white point guide}=[regular polygon,regular polygon sides=3,font=\scriptsize,draw,scale=0.65,inner sep=-0.5pt,minimum width=9mm,fill=white,regular polygon rotate=180]

\tikzstyle{black point}=[point,fill=black,font=\color{white}]
\tikzstyle{black copoint}=[copoint,fill=black,font=\color{white}]

\tikzstyle{tiny gray point}=[tinypoint,fill=gray!40!white]

\tikzstyle{diredge}=[->]
\tikzstyle{ddiredge}=[<->]
\tikzstyle{rdiredge}=[<-]
\tikzstyle{thickdiredge}=[->, very thick]
\tikzstyle{pointer edge}=[->,very thick,gray]
\tikzstyle{pointer edge part}=[very thick,gray]
\tikzstyle{dashed edge}=[dashed]
\tikzstyle{thick dashed edge}=[very thick,dashed]
\tikzstyle{thick gray dashed edge}=[thick dashed edge,gray!40]
\tikzstyle{thick map edge}=[very thick,|->]

\makeatletter
\newcommand{\boxshape}[3]{%
\pgfdeclareshape{#1}{
\inheritsavedanchors[from=rectangle] % this is nearly a rectangle
\inheritanchorborder[from=rectangle]
\inheritanchor[from=rectangle]{center}
\inheritanchor[from=rectangle]{north}
\inheritanchor[from=rectangle]{south}
\inheritanchor[from=rectangle]{west}
\inheritanchor[from=rectangle]{east}
\backgroundpath{% this is new
\southwest \pgf@xa=\pgf@x \pgf@ya=\pgf@y
\northeast \pgf@xb=\pgf@x \pgf@yb=\pgf@y

\@tempdima=#2
\@tempdimb=#3

\pgfpathmoveto{\pgfpoint{\pgf@xa - 5pt + \@tempdima}{\pgf@ya}}
\pgfpathlineto{\pgfpoint{\pgf@xa - 5pt - \@tempdima}{\pgf@yb}}
\pgfpathlineto{\pgfpoint{\pgf@xb + 5pt + \@tempdimb}{\pgf@yb}}
\pgfpathlineto{\pgfpoint{\pgf@xb + 5pt - \@tempdimb}{\pgf@ya}}
\pgfpathlineto{\pgfpoint{\pgf@xa - 5pt + \@tempdima}{\pgf@ya}}
\pgfpathclose
}
}}

\boxshape{NEbox}{0pt}{5pt}
\boxshape{SEbox}{0pt}{-5pt}
\boxshape{NWbox}{5pt}{0pt}
\boxshape{SWbox}{-5pt}{0pt}
\boxshape{EBox}{-3pt}{3pt}
\boxshape{WBox}{3pt}{-3pt}
\makeatother

\tikzstyle{cloud}=[shape=cloud,draw,minimum width=1.5cm,minimum height=1.5cm]

\tikzstyle{map}=[draw,shape=NEbox,inner sep=2pt,minimum height=6mm,fill=white]
\tikzstyle{dashedmap}=[draw,dashed,shape=NEbox,inner sep=2pt,minimum height=6mm,fill=white]
\tikzstyle{mapdag}=[draw,shape=SEbox,inner sep=2pt,minimum height=6mm,fill=white]
\tikzstyle{mapadj}=[draw,shape=SEbox,inner sep=2pt,minimum height=6mm,fill=white]
\tikzstyle{maptrans}=[draw,shape=SWbox,inner sep=2pt,minimum height=6mm,fill=white]
\tikzstyle{mapconj}=[draw,shape=NWbox,inner sep=2pt,minimum height=6mm,fill=white]

\tikzstyle{langmap}=[draw,shape=NEbox,inner sep=2pt,minimum height=2.4mm,minimum width=3.2mm,fill=white]
\tikzstyle{langmaptrans}=[draw,shape=SWbox,inner sep=2pt,minimum height=2.4mm,minimum width=3.2mm,fill=white]

\tikzstyle{medium map}=[draw,shape=NEbox,inner sep=2pt,minimum height=6mm,fill=white,minimum width=7mm]
\tikzstyle{medium map dag}=[draw,shape=SEbox,inner sep=2pt,minimum height=6mm,fill=white,minimum width=7mm]
\tikzstyle{medium map adj}=[draw,shape=SEbox,inner sep=2pt,minimum height=6mm,fill=white,minimum width=7mm]
\tikzstyle{medium map trans}=[draw,shape=SWbox,inner sep=2pt,minimum height=6mm,fill=white,minimum width=7mm]
\tikzstyle{medium map conj}=[draw,shape=NWbox,inner sep=2pt,minimum height=6mm,fill=white,minimum width=7mm]
\tikzstyle{semilarge map}=[draw,shape=NEbox,inner sep=2pt,minimum height=6mm,fill=white,minimum width=9.5mm]
\tikzstyle{semilarge map trans}=[draw,shape=SWbox,inner sep=2pt,minimum height=6mm,fill=white,minimum width=9.5mm]
\tikzstyle{semilarge map adj}=[draw,shape=SEbox,inner sep=2pt,minimum height=6mm,fill=white,minimum width=9.5mm]
\tikzstyle{semilarge map dag}=[draw,shape=SEbox,inner sep=2pt,minimum height=6mm,fill=white,minimum width=9.5mm]
\tikzstyle{semilarge map conj}=[draw,shape=NWbox,inner sep=2pt,minimum height=6mm,fill=white,minimum width=9.5mm]
\tikzstyle{large map}=[draw,shape=NEbox,inner sep=2pt,minimum height=6mm,fill=white,minimum width=12mm]
\tikzstyle{large map conj}=[draw,shape=NWbox,inner sep=2pt,minimum height=6mm,fill=white,minimum width=12mm]
\tikzstyle{very large map}=[draw,shape=NEbox,inner sep=2pt,minimum height=6mm,fill=white,minimum width=17mm]

\tikzstyle{medium dmap}=[draw,doubled,shape=NEbox,inner sep=2pt,minimum height=6mm,fill=white,minimum width=7mm]
\tikzstyle{medium dmap dag}=[draw,doubled,shape=SEbox,inner sep=2pt,minimum height=6mm,fill=white,minimum width=7mm]
\tikzstyle{medium dmap adj}=[draw,doubled,shape=SEbox,inner sep=2pt,minimum height=6mm,fill=white,minimum width=7mm]
\tikzstyle{medium dmap trans}=[draw,doubled,shape=SWbox,inner sep=2pt,minimum height=6mm,fill=white,minimum width=7mm]
\tikzstyle{medium dmap conj}=[draw,doubled,shape=NWbox,inner sep=2pt,minimum height=6mm,fill=white,minimum width=7mm]
\tikzstyle{semilarge dmap}=[draw,doubled,shape=NEbox,inner sep=2pt,minimum height=6mm,fill=white,minimum width=9.5mm]
\tikzstyle{semilarge dmap trans}=[draw,doubled,shape=SWbox,inner sep=2pt,minimum height=6mm,fill=white,minimum width=9.5mm]
\tikzstyle{semilarge dmap adj}=[draw,doubled,shape=SEbox,inner sep=2pt,minimum height=6mm,fill=white,minimum width=9.5mm]
\tikzstyle{semilarge dmap dag}=[draw,doubled,shape=SEbox,inner sep=2pt,minimum height=6mm,fill=white,minimum width=9.5mm]
\tikzstyle{semilarge dmap conj}=[draw,doubled,shape=NWbox,inner sep=2pt,minimum height=6mm,fill=white,minimum width=9.5mm]
\tikzstyle{large dmap}=[draw,doubled,shape=NEbox,inner sep=2pt,minimum height=6mm,fill=white,minimum width=12mm]
\tikzstyle{large dmap conj}=[draw,doubled,shape=NWbox,inner sep=2pt,minimum height=6mm,fill=white,minimum width=12mm]
\tikzstyle{large dmap trans}=[draw,doubled,shape=SWbox,inner sep=2pt,minimum height=6mm,fill=white,minimum width=12mm]
\tikzstyle{large dmap adj}=[draw,doubled,shape=SEbox,inner sep=2pt,minimum height=6mm,fill=white,minimum width=12mm]
\tikzstyle{large dmap dag}=[draw,doubled,shape=SEbox,inner sep=2pt,minimum height=6mm,fill=white,minimum width=12mm]
\tikzstyle{very large dmap}=[draw,doubled,shape=NEbox,inner sep=2pt,minimum height=6mm,fill=white,minimum width=19.5mm]

\tikzstyle{muxbox}=[draw,shape=rectangle,minimum height=3mm,minimum width=3mm,fill=white]
\tikzstyle{dmuxbox}=[muxbox,doubled]

\tikzstyle{box}=[draw,shape=rectangle,inner sep=2pt,minimum height=6mm,minimum width=6mm,fill=white]
\tikzstyle{dbox}=[draw,doubled,shape=rectangle,inner sep=2pt,minimum height=6mm,minimum width=6mm,fill=white]
\tikzstyle{dmap}=[draw,doubled,shape=NEbox,inner sep=2pt,minimum height=6mm,fill=white]
\tikzstyle{dmapdag}=[draw,doubled,shape=SEbox,inner sep=2pt,minimum height=6mm,fill=white]
\tikzstyle{dmapadj}=[draw,doubled,shape=SEbox,inner sep=2pt,minimum height=6mm,fill=white]
\tikzstyle{dmaptrans}=[draw,doubled,shape=SWbox,inner sep=2pt,minimum height=6mm,fill=white]
\tikzstyle{dmapconj}=[draw,doubled,shape=NWbox,inner sep=2pt,minimum height=6mm,fill=white]

\tikzstyle{ddmap}=[draw,doubled,dashed,shape=NEbox,inner sep=2pt,minimum height=6mm,fill=white]
\tikzstyle{ddmapdag}=[draw,doubled,dashed,shape=SEbox,inner sep=2pt,minimum height=6mm,fill=white]
\tikzstyle{ddmapadj}=[draw,doubled,dashed,shape=SEbox,inner sep=2pt,minimum height=6mm,fill=white]
\tikzstyle{ddmaptrans}=[draw,doubled,dashed,shape=SWbox,inner sep=2pt,minimum height=6mm,fill=white]
\tikzstyle{ddmapconj}=[draw,doubled,dashed,shape=NWbox,inner sep=2pt,minimum height=6mm,fill=white]

\boxshape{sNEbox}{0pt}{3pt}
\boxshape{sSEbox}{0pt}{-3pt}
\boxshape{sNWbox}{3pt}{0pt}
\boxshape{sSWbox}{-3pt}{0pt}
\tikzstyle{smap}=[draw,shape=sNEbox,fill=white]
\tikzstyle{smapdag}=[draw,shape=sSEbox,fill=white]
\tikzstyle{smapadj}=[draw,shape=sSEbox,fill=white]
\tikzstyle{smaptrans}=[draw,shape=sSWbox,fill=white]
\tikzstyle{smapconj}=[draw,shape=sNWbox,fill=white]

\tikzstyle{dsmap}=[draw,dashed,shape=sNEbox,fill=white]
\tikzstyle{dsmapdag}=[draw,dashed,shape=sSEbox,fill=white]
\tikzstyle{dsmaptrans}=[draw,dashed,shape=sSWbox,fill=white]
\tikzstyle{dsmapconj}=[draw,dashed,shape=sNWbox,fill=white]

\boxshape{mNEbox}{0pt}{10pt}
\boxshape{mSEbox}{0pt}{-10pt}
\boxshape{mNWbox}{10pt}{0pt}
\boxshape{mSWbox}{-10pt}{0pt}
\tikzstyle{mmap}=[draw,shape=mNEbox]
\tikzstyle{mmapdag}=[draw,shape=mSEbox]
\tikzstyle{mmaptrans}=[draw,shape=mSWbox]
\tikzstyle{mmapconj}=[draw,shape=mNWbox]

\tikzstyle{mmapgray}=[draw,fill=gray!40!white,shape=mNEbox]
\tikzstyle{smapgray}=[draw,fill=gray!40!white,shape=sNEbox]

\makeatletter
\pgfdeclareshape{cornerpoint}{
\inheritsavedanchors[from=rectangle] % this is nearly a rectangle
\inheritanchorborder[from=rectangle]
\inheritanchor[from=rectangle]{center}
\inheritanchor[from=rectangle]{north}
\inheritanchor[from=rectangle]{south}
\inheritanchor[from=rectangle]{west}
\inheritanchor[from=rectangle]{east}
\backgroundpath{% this is new
\southwest \pgf@xa=\pgf@x \pgf@ya=\pgf@y
\northeast \pgf@xb=\pgf@x \pgf@yb=\pgf@y

\pgfmathsetmacro{\pgf@shorten@left}{\pgfkeysvalueof{/tikz/shorten left}}
\pgfmathsetmacro{\pgf@shorten@right}{\pgfkeysvalueof{/tikz/shorten right}}

\pgfpathmoveto{\pgfpoint{0.5 * (\pgf@xa + \pgf@xb)}{\pgf@ya - 5pt}}
\pgfpathlineto{\pgfpoint{\pgf@xa - 8pt + \pgf@shorten@left}{\pgf@yb - 1.5 * \pgf@shorten@left}}
\pgfpathlineto{\pgfpoint{\pgf@xa - 8pt + \pgf@shorten@left}{\pgf@yb}}
\pgfpathlineto{\pgfpoint{\pgf@xb + 8pt - \pgf@shorten@right}{\pgf@yb}}
\pgfpathlineto{\pgfpoint{\pgf@xb + 8pt - \pgf@shorten@right}{\pgf@yb - 1.5 * \pgf@shorten@right}}
\pgfpathclose
}
}

\pgfdeclareshape{cornercopoint}{
\inheritsavedanchors[from=rectangle] % this is nearly a rectangle
\inheritanchorborder[from=rectangle]
\inheritanchor[from=rectangle]{center}
\inheritanchor[from=rectangle]{north}
\inheritanchor[from=rectangle]{south}
\inheritanchor[from=rectangle]{west}
\inheritanchor[from=rectangle]{east}
\backgroundpath{% this is new
\southwest \pgf@xa=\pgf@x \pgf@ya=\pgf@y
\northeast \pgf@xb=\pgf@x \pgf@yb=\pgf@y

\pgfmathsetmacro{\pgf@shorten@left}{\pgfkeysvalueof{/tikz/shorten left}}
\pgfmathsetmacro{\pgf@shorten@right}{\pgfkeysvalueof{/tikz/shorten right}}

\pgfpathmoveto{\pgfpoint{0.5 * (\pgf@xa + \pgf@xb)}{\pgf@yb + 5pt}}
\pgfpathlineto{\pgfpoint{\pgf@xa - 8pt + \pgf@shorten@left}{\pgf@ya + 1.5 * \pgf@shorten@left}}
\pgfpathlineto{\pgfpoint{\pgf@xa - 8pt + \pgf@shorten@left}{\pgf@ya}}
\pgfpathlineto{\pgfpoint{\pgf@xb + 8pt - \pgf@shorten@right}{\pgf@ya}}
\pgfpathlineto{\pgfpoint{\pgf@xb + 8pt - \pgf@shorten@right}{\pgf@ya + 1.5 * \pgf@shorten@right}}
\pgfpathclose
}
}

\pgfdeclareshape{langpoint}{
\inheritsavedanchors[from=rectangle] % this is nearly a rectangle
\inheritanchorborder[from=rectangle]
\inheritanchor[from=rectangle]{center}
\inheritanchor[from=rectangle]{north}
\inheritanchor[from=rectangle]{south}
\inheritanchor[from=rectangle]{west}
\inheritanchor[from=rectangle]{east}
\backgroundpath{% this is new
\southwest \pgf@xa=\pgf@x \pgf@ya=\pgf@y
\northeast \pgf@xb=\pgf@x \pgf@yb=\pgf@y

\pgfmathsetmacro{\pgf@shorten@left}{\pgfkeysvalueof{/tikz/shorten left}}
\pgfmathsetmacro{\pgf@shorten@right}{\pgfkeysvalueof{/tikz/shorten right}}

\pgfpathmoveto{\pgfpoint{0.5 * (\pgf@xa + \pgf@xb)}{\pgf@ya - 2pt}}
\pgfpathlineto{\pgfpoint{\pgf@xa - 8pt}{\pgf@yb - 3 * \pgf@shorten@left + 5pt}} 
\pgfpathlineto{\pgfpoint{\pgf@xa - 8pt}{\pgf@yb -1pt}}
\pgfpathlineto{\pgfpoint{\pgf@xb + 8pt}{\pgf@yb -1pt}}
\pgfpathlineto{\pgfpoint{\pgf@xb + 8pt}{\pgf@yb - 3 * \pgf@shorten@left + 5pt}}%NEW POINT
\pgfpathclose
}
}

\pgfdeclareshape{langcopoint}{
\inheritsavedanchors[from=rectangle] % this is nearly a rectangle
\inheritanchorborder[from=rectangle]
\inheritanchor[from=rectangle]{center}
\inheritanchor[from=rectangle]{north}
\inheritanchor[from=rectangle]{south}
\inheritanchor[from=rectangle]{west}
\inheritanchor[from=rectangle]{east}
\backgroundpath{% this is new
\southwest \pgf@xa=\pgf@x \pgf@ya=\pgf@y
\northeast \pgf@xb=\pgf@x \pgf@yb=\pgf@y

\pgfmathsetmacro{\pgf@shorten@left}{\pgfkeysvalueof{/tikz/shorten left}}
\pgfmathsetmacro{\pgf@shorten@right}{\pgfkeysvalueof{/tikz/shorten right}}

\pgfpathmoveto{\pgfpoint{0.5 * (\pgf@xa + \pgf@xb)}{\pgf@yb +0pt}}
\pgfpathlineto{\pgfpoint{\pgf@xa - 8pt}{\pgf@ya + 3 * \pgf@shorten@left - 5pt}} 
\pgfpathlineto{\pgfpoint{\pgf@xa - 8pt}{\pgf@ya + 1pt}}
\pgfpathlineto{\pgfpoint{\pgf@xb + 8pt}{\pgf@ya + 1pt}}
\pgfpathlineto{\pgfpoint{\pgf@xb + 8pt}{\pgf@ya + 3 * \pgf@shorten@left - 5pt}}%NEW POINT
\pgfpathclose
}
}

\pgfdeclareshape{langrect}{
\inheritsavedanchors[from=rectangle] % this is nearly a rectangle
\inheritanchorborder[from=rectangle]
\inheritanchor[from=rectangle]{center}
\inheritanchor[from=rectangle]{north}
\inheritanchor[from=rectangle]{south}
\inheritanchor[from=rectangle]{west}
\inheritanchor[from=rectangle]{east}
\backgroundpath{% this is new
\southwest \pgf@xa=\pgf@x \pgf@ya=\pgf@y
\northeast \pgf@xb=\pgf@x \pgf@yb=\pgf@y

\pgfmathsetmacro{\pgf@shorten@left}{\pgfkeysvalueof{/tikz/shorten left}}
\pgfmathsetmacro{\pgf@shorten@right}{\pgfkeysvalueof{/tikz/shorten right}}

\pgfpathmoveto{\pgfpoint{\pgf@xa - 8pt}{\pgf@ya + 3 * \pgf@shorten@left - 5pt}} 
\pgfpathlineto{\pgfpoint{\pgf@xa - 8pt}{\pgf@ya + 1pt}}
\pgfpathlineto{\pgfpoint{\pgf@xb + 8pt}{\pgf@ya + 1pt}}
\pgfpathlineto{\pgfpoint{\pgf@xb + 8pt}{\pgf@ya + 3 * \pgf@shorten@left - 5pt}}%NEW POINT
\pgfpathclose
}
}
\makeatother

\pgfkeyssetvalue{/tikz/shorten left}{0pt}
\pgfkeyssetvalue{/tikz/shorten right}{0pt}

\tikzstyle{kpoint common}=[draw,fill=white,inner sep=1pt,minimum height=4mm]

\tikzstyle{langstate}=[shape=langcopoint,shorten left=5pt,kpoint common,font=\footnotesize]
\tikzstyle{langeffect}=[shape=langpoint,shorten left=5pt,kpoint common,font=\footnotesize]
\tikzstyle{langstatedash}=[shape=langcopoint,dashed, shorten left=5pt,kpoint common,font=\footnotesize]
\tikzstyle{langeffectdash}=[shape=langpoint,dashed, shorten left=5pt,kpoint common,font=\footnotesize]
\tikzstyle{langbox}=[shape=langrect,shorten left=5pt,kpoint common,font=\footnotesize] 
\tikzstyle{kpoint}=[shape=cornerpoint,shorten left=5pt,kpoint common]
\tikzstyle{kpoint adjoint}=[shape=cornercopoint,shorten left=5pt,kpoint common]

\tikzstyle{kpoint conjugate}=[shape=cornerpoint,shorten right=5pt,kpoint common]
\tikzstyle{kpoint transpose}=[shape=cornercopoint,shorten right=5pt,kpoint common]
\tikzstyle{kpoint symm}=[shape=cornerpoint,shorten left=5pt,shorten right=5pt,kpoint common]

\tikzstyle{black kpoint}=[shape=cornerpoint,shorten left=5pt,kpoint common,fill=black,font=\color{white}]
\tikzstyle{black kpoint adjoint}=[shape=cornercopoint,shorten left=5pt,kpoint common,fill=black,font=\color{white}]
\tikzstyle{black kpointadj}=[shape=cornercopoint,shorten left=5pt,kpoint common,fill=black,font=\color{white}]

\tikzstyle{black dkpoint}=[shape=cornerpoint,shorten left=5pt,kpoint common,fill=black, doubled,font=\color{white}]
\tikzstyle{black dkpoint adjoint}=[shape=cornercopoint,shorten left=5pt,kpoint common,fill=black, doubled,font=\color{white}]
\tikzstyle{black dkpointadj}=[shape=cornercopoint,shorten left=5pt,kpoint common,fill=black, doubled,font=\color{white}]

\tikzstyle{kpointdag}=[kpoint adjoint]
\tikzstyle{kpointadj}=[kpoint adjoint]
\tikzstyle{kpointconj}=[kpoint conjugate]
\tikzstyle{kpointtrans}=[kpoint transpose]

\tikzstyle{big kpoint}=[kpoint, minimum width=1.2 cm, minimum height=8mm, inner sep=4pt, text depth=3mm]

\tikzstyle{wide kpoint}=[kpoint, minimum width=1 cm, inner sep=2pt]%, text depth=-0.7 mm]
\tikzstyle{wide kpointdag}=[kpointdag, minimum width=1 cm, inner sep=2pt]%, text depth=0.7 mm]
\tikzstyle{wide kpointconj}=[kpointconj, minimum width=1 cm, inner sep=2pt]%, text depth=-0.7 mm]
\tikzstyle{wide kpointtrans}=[kpointtrans, minimum width=1 cm, inner sep=2pt]%, text depth=0.7 mm]

\tikzstyle{gray kpoint}=[kpoint,fill=gray!50!white]
\tikzstyle{gray kpointdag}=[kpointdag,fill=gray!50!white]
\tikzstyle{gray kpointadj}=[kpointadj,fill=gray!50!white]
\tikzstyle{gray kpointconj}=[kpointconj,fill=gray!50!white]
\tikzstyle{gray kpointtrans}=[kpointtrans,fill=gray!50!white]

\tikzstyle{gray dkpoint}=[kpoint,fill=gray!50!white,doubled]
\tikzstyle{gray dkpointdag}=[kpointdag,fill=gray!50!white,doubled]
\tikzstyle{gray dkpointadj}=[kpointadj,fill=gray!50!white,doubled]
\tikzstyle{gray dkpointconj}=[kpointconj,fill=gray!50!white,doubled]
\tikzstyle{gray dkpointtrans}=[kpointtrans,fill=gray!50!white,doubled]

\tikzstyle{white label}=[draw,fill=white,rectangle,inner sep=0.7 mm]
\tikzstyle{gray label}=[draw,fill=gray!50!white,rectangle,inner sep=0.7 mm]
\tikzstyle{black label}=[draw,fill=black,rectangle,inner sep=0.7 mm]

\tikzstyle{dkpoint}=[kpoint,doubled]
\tikzstyle{wide dkpoint}=[wide kpoint,doubled]
\tikzstyle{dkpointdag}=[kpoint adjoint,doubled]
\tikzstyle{wide dkpointdag}=[wide kpointdag,doubled]
\tikzstyle{dkcopoint}=[kpoint adjoint,doubled]
\tikzstyle{dkpointadj}=[kpoint adjoint,doubled]
\tikzstyle{dkpointconj}=[kpoint conjugate,doubled]
\tikzstyle{dkpointtrans}=[kpoint transpose,doubled]

\tikzstyle{kscalar}=[kpoint common, shape=EBox, inner xsep=-1pt, inner ysep=3pt,font=\small]
\tikzstyle{kscalarconj}=[kpoint common, shape=WBox, inner xsep=-1pt, inner ysep=3pt,font=\small]

 \tikzstyle{upground}=[circuit ee IEC,ground,rotate=90,scale=2.5]
 \tikzstyle{downground}=[circuit ee IEC,ground,rotate=-90,scale=2.5]
 \tikzstyle{bigground}=[regular polygon,regular polygon sides=3,draw=gray,scale=0.50,inner sep=-0.5pt,minimum width=10mm,fill=gray]
\tikzstyle{arrs}=[-latex,font=\small,auto]
\tikzstyle{arrow plain}=[arrs]
\tikzstyle{arrow dashed}=[dashed,arrs]
\tikzstyle{arrow bold}=[very thick,arrs]
\tikzstyle{arrow hide}=[draw=white!0,-]
\tikzstyle{arrow reverse}=[latex-]
\tikzstyle{cdnode}=[]

\tikzstyle{H}=[-, style=dashed]
\tikzstyle{K}=[-, line width=1pt]
\tikzstyle{Kv}=[-, line width=1pt, ->]
\tikzstyle{Kv<>}=[-,line width=1pt,{<->}]
\tikzstyle{gF}=[-, draw=none, fill={rgb,255: red,191; green,191; blue,191}]
\tikzstyle{KB}=[-, draw=blue, line width=1pt]
\tikzstyle{KO}=[-, draw={rgb,255: red,255; green,128; blue,0}, line width=1pt]
\tikzstyle{KL}=[-, draw={rgb,255: red,191; green,255; blue,0}, line width=1pt]
\tikzstyle{KHO}=[-, draw={rgb,255: red,255; green,128; blue,0}, style=dashed, line width=1pt]
\tikzstyle{KTG}=[-, draw={rgb,255: red,128; green,128; blue,128}, style=dotted, line width=1pt]
\tikzstyle{KTlG}=[-, draw={rgb,255: red,191; green,191; blue,191}, style=dotted, line width=1pt]
\tikzstyle{KBv}=[-, draw=blue, ->]
\tikzstyle{KOv}=[-, draw={rgb,255: red,255; green,128; blue,0}, line width=1pt, ->]
\tikzstyle{KLv}=[-, draw={rgb,255: red,191; green,255; blue,0}, ->]
\tikzstyle{T}=[-, style=dotted]
\tikzstyle{wF}=[-, fill=white, draw=none]
\tikzstyle{KH}=[-, style=dashed, line width=1pt]
\tikzstyle{Hv}=[->, style=dashed]
\tikzstyle{cv}=[-,right hook->]
\tikzstyle{vv}=[-,->>]
\tikzstyle{v}=[-,->]
\tikzstyle{<>}=[-,<->]
\tikzstyle{Hvv}=[-,->>,style=dashed]
\tikzstyle{Kvp}=[->]
\tikzstyle{KTB}=[-, draw=blue, style=dotted, line width=1pt]
\tikzstyle{KBgF}=[-, fill={rgb,255: red,191; green,191; blue,191}, draw=blue, line width=1 pt]
\tikzstyle{KBggF}=[-, fill={rgb,255: red,128; green,128; blue,128}, draw=blue, line width=1 pt]
\tikzstyle{b-wf}=[-, fill=white]
\tikzstyle{b-gf}=[-, fill={rgb,255: red,191; green,191; blue,191}, draw=black]
\tikzstyle{KHB}=[-, draw=blue, style=dashed, line width=1pt]

\tikzstyle{bigunit}=[dot,fill=white,text depth=-0.2mm]
\tikzstyle{smallblackdot}=[fill=black, inner sep=0mm,minimum width=1mm,minimum height=1mm,draw,shape=circle]
\tikzstyle{smallorangedot}=[fill={rgb,255: red,255; green,128; blue,0}, inner sep=0mm,minimum width=1mm,minimum height=1mm,draw,shape=circle]
\tikzstyle{smallbluedot}=[fill=blue, inner sep=0mm,minimum width=1mm,minimum height=1mm,draw,shape=circle] 

\definecolor{hexcolor0xa9a9a9}{rgb}{0.663,0.663,0.663} 
\tikzstyle{GrayLine}=[dashed,draw=hexcolor0xa9a9a9] 
\tikzstyle{gray}=[dashed,draw=hexcolor0xa9a9a9]

\theoremstyle{definition}
 
\newtheorem*{theorem*}{Theorem}

\newtheorem{example*}[theorem]{Example*}
\newtheorem{examples*}[theorem]{Examples*}

\newtheorem{remark*}[theorem]{Remark*}

\def\bR{\begin{color}{red}}  
\def\bB{\begin{color}{blue}}
\def\bM{\begin{color}{magenta}}  
\def\bC{\begin{color}{cyan}}
\def\bW{\begin{color}{white}}
\def\bBl{\begin{color}{black}}
\def\bG{\begin{color}{green}}
\def\bY{\begin{color}{yellow}}
\def\e{\end{color}\xspace}
\newcommand{\bit}{\begin{itemize}}
\newcommand{\eit}{\end{itemize}\par\noindent}
\newcommand{\ben}{\begin{enumerate}}
\newcommand{\een}{\end{enumerate}\par\noindent}
\newcommand{\beq}{\begin{equation}}
\newcommand{\eeq}{\end{equation}\par\noindent}
\newcommand{\beqa}{\begin{eqnarray*}}
\newcommand{\eeqa}{\end{eqnarray*}\par\noindent}
\newcommand{\beqn}{\begin{eqnarray}}
\newcommand{\eeqn}{\end{eqnarray}\par\noindent}

\title{Basic ZX-calculus for students and professionals}

\author{Bob Coecke\\ 
Quantinuum\\ 
17 Beaumont Street\\ OX1 2NA Oxford, UK\\  
bob.coecke@quantinuum.com}

\begin{document}    
\maketitle  
\begin{abstract}
These are the lecture notes of guest lectures for Artur Ekert’s course Introduction to Quantum Information at the Mathematical Institute of Oxford University, Hilary Term 2023. Some basic familiarity with Dirac notation is assumed.  %The idea is that you can read this in a couple of toilet visits, hence also avoiding the judgement of disapproving colleagues.  

For the readers of Quantum in Pictures who have some basic quantum background, these notes also constitute the shortest path to an explanation of how what they learn in QIP relates to the traditional quantum formalism.
\end{abstract}

\section{Introduction} 

\bM ZX-calculus\e  \cite{CD0, CD1, CD2}  has already been around as long as 2007, but it only has become  widely used in the past couple of years, and especially with the \bM emergence of quantum industry\e, given the new scientific and technological challenges this poses \cite{coecke2021kindergarden}.  Areas where it has become prominent include compilation \cite{cowtan2019phase, khesinGraphicalQuantumCliffordencoder2023}, circuit optimisation \cite{clifford-simp, KissingerTcount}, error-correction \cite{de2017zx, Gidney2019}, quantum natural language processing \cite{QNLP-foundations}, QML \cite{zhao2021analyzing, wangDifferentiatingIntegratingZX2022}, and also many issues surrounding photonic quantum computing \cite{defeliceQuantumLinearOptics2022, litinski2022active}.  

Another reason for this more recent rise to prominence is the availability of the book \bM Picturing Quantum Processes \cite{CKbook}\e, which presents ZX-calculus without any reference to category theory, which before was an obstacle to many.  Even more so, since very recently there also is the book \bM Quantum in Pictures \cite{QiP}\e which has no mathematical prerequisites whatsoever.  Hence, this book also establishes ZX-calculus as an \bM educational tool\e that can importantly contribute to making quantum more inclusive, while at the same time providing an entirely new perspective on it. 

\section{Process theories as wires and boxes}

ZX-calculus is an instance of what in \cite{CKbook} we call a \bM process theory\e, that is, a theory in which processes are first class citizens.  Historically, going 2500 years back there was a debate between pre-Socratics Heraclitus and Parmenides, where the former advocated that all there is are processes, with a world in constant flux, while Parmenides advocated that at any time the world is a static instance.  Plato, Aristotle, Kepler, Newton, etc.~all followed the Parmenides line of thinking, having a kinematic description of the world before dynamic evolution is discussed.  Whitehead was the 1st to bring Heraclitus’ process ideas back.  With the birth of category theory, the appropriate mathematics to do so is now also around, and can equivalently be described as pictures \cite{JS}.

The language of process theories consists of \bM wires\e and \bM boxes\e:   
\begin{center}
\epsfig{figure=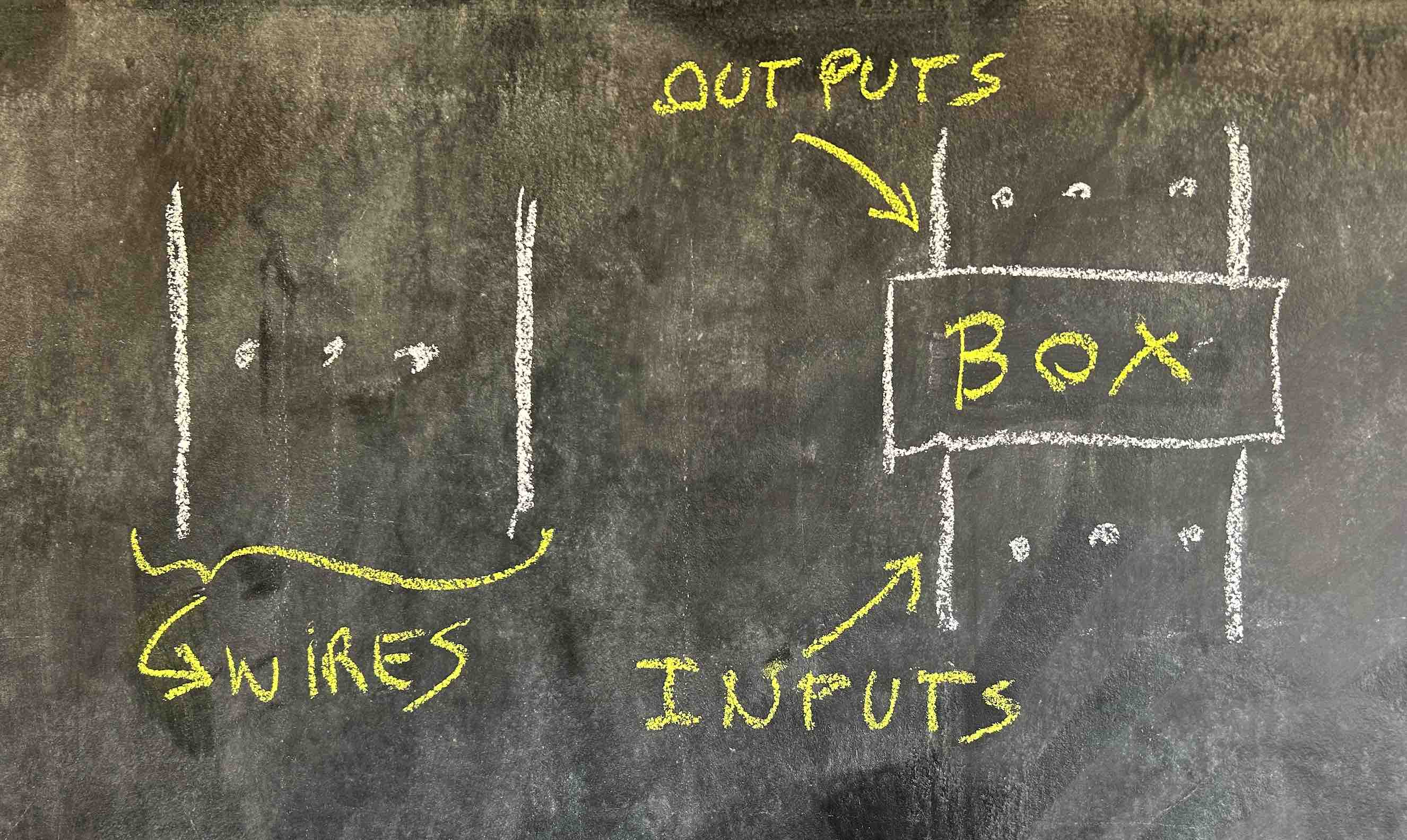,width=260pt} 
\end{center} 
where these boxes will have some \bM input and output wires\e.  Here the wires represent quantum systems, or classical systems — cf.~a place were quantum particles and classical data live/travel.  The boxes transform those particles and the data.  So for example,  the boxes could be a unitary gate, or a quantum measurement, or encoding classical data as a quantum state etc.  We take the convention that \bM time flows upwards\e, that is, inputs are at the bottom and outputs are at the top.

There are special kinds of boxes, called \bM states and effects (or tests)\e, and these respectively correspond to boxes without inputs and boxes without outputs:
\begin{center}
\epsfig{figure=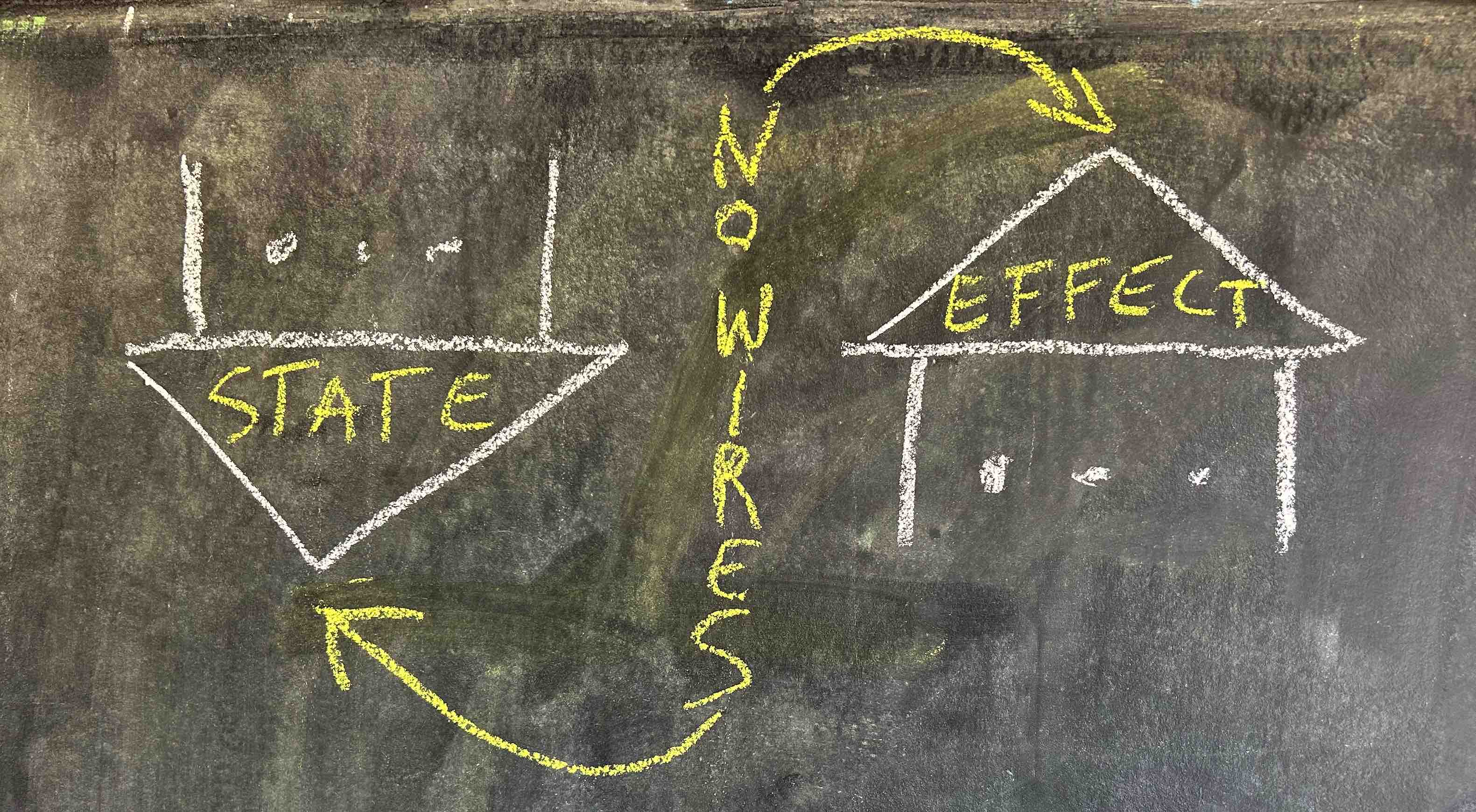,width=260pt}  
\end{center} 
We depicted them as triangles, since then they resemble Dirac notation as follows:
\begin{center}
\epsfig{figure=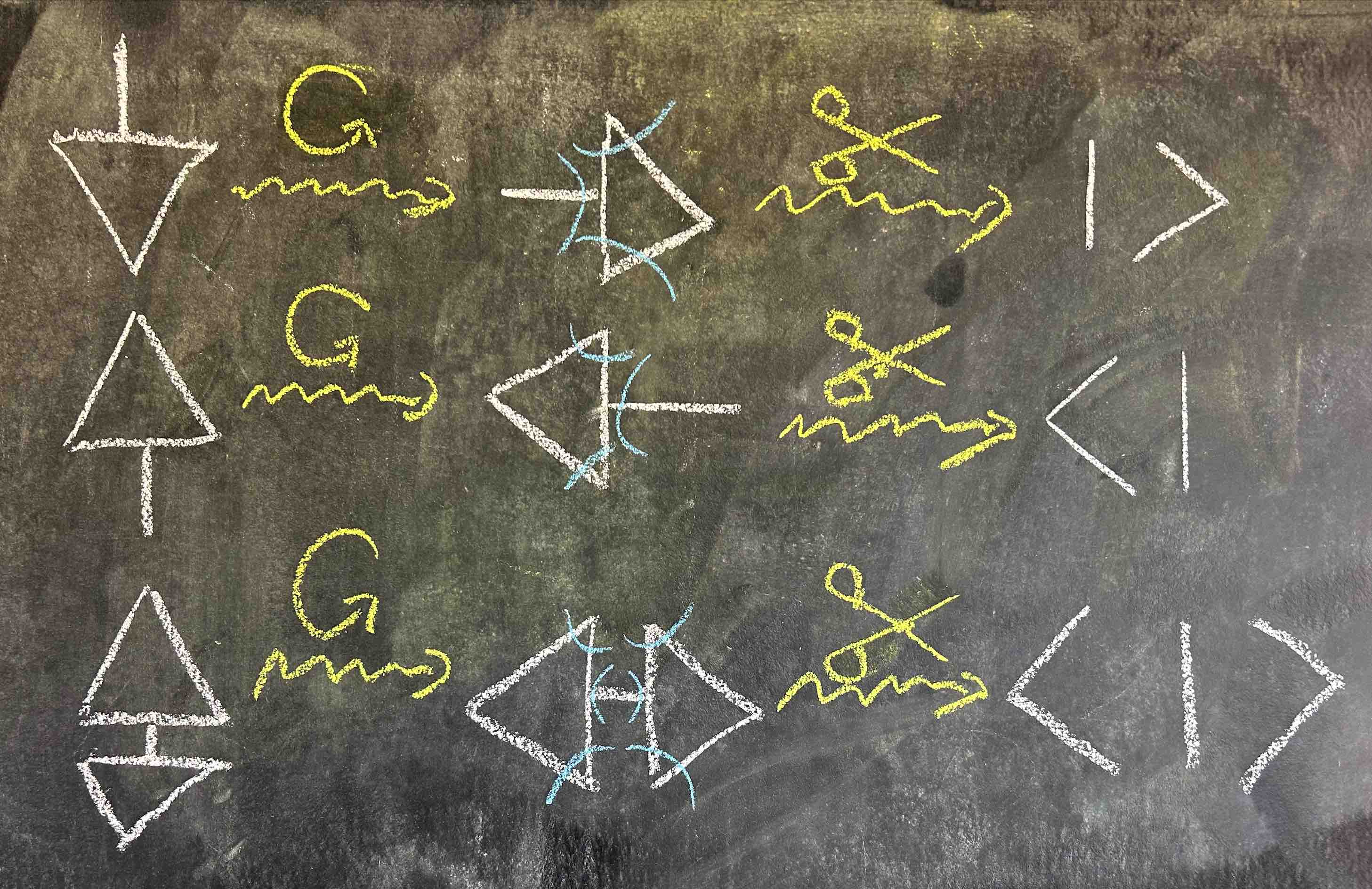,width=320pt} 
\end{center} 
We can indeed think of these process theories as a \bM 2D extension of Dirac notation\e, where putting two wires and/or boxes side-by-side corresponds to the tensor product:
\begin{center}
\epsfig{figure=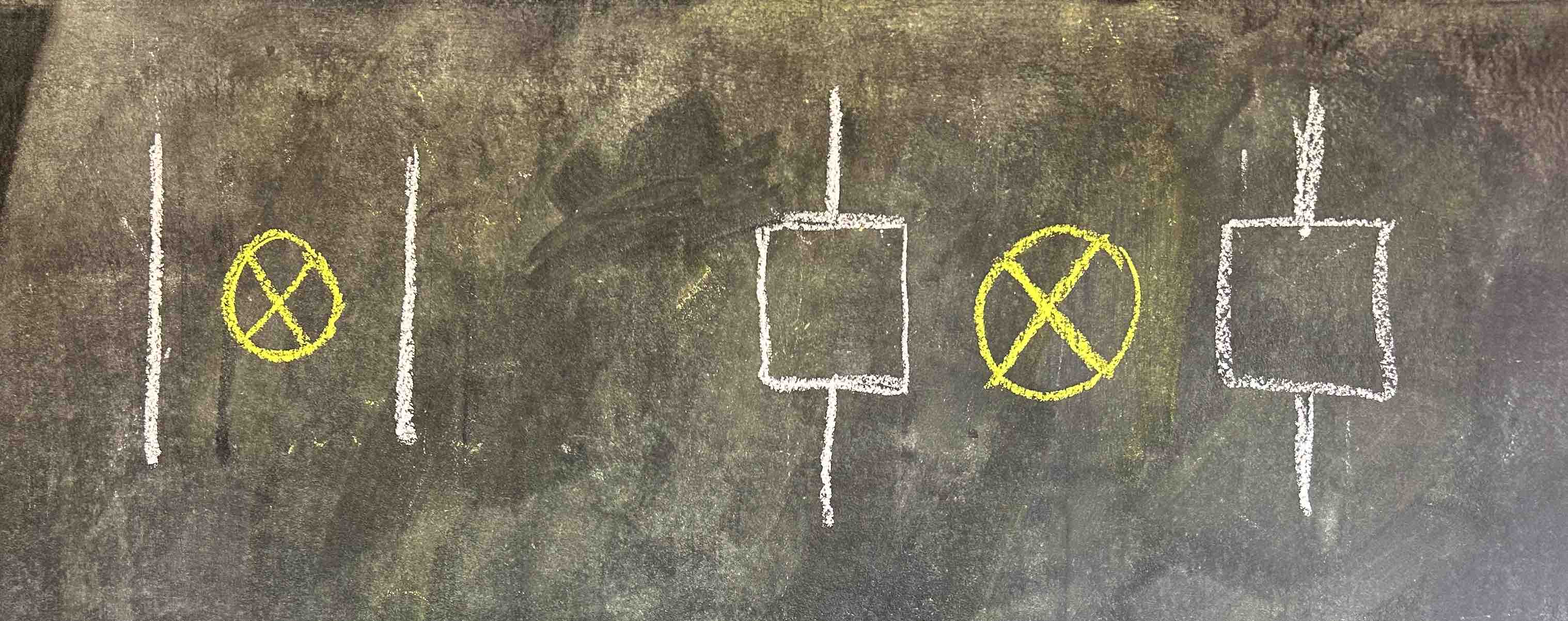,width=210pt} 
\end{center} 
We drew tensor symbols here, but in pictures, these tensor symbols are just empty space.  Note also that in process theories \bM numbers\e correspond to boxes without any input or output, denoted as a diamond, again because of Dirac notation.  Since there are no external wires constraining them, they can freely move about:
\begin{center}
\epsfig{figure=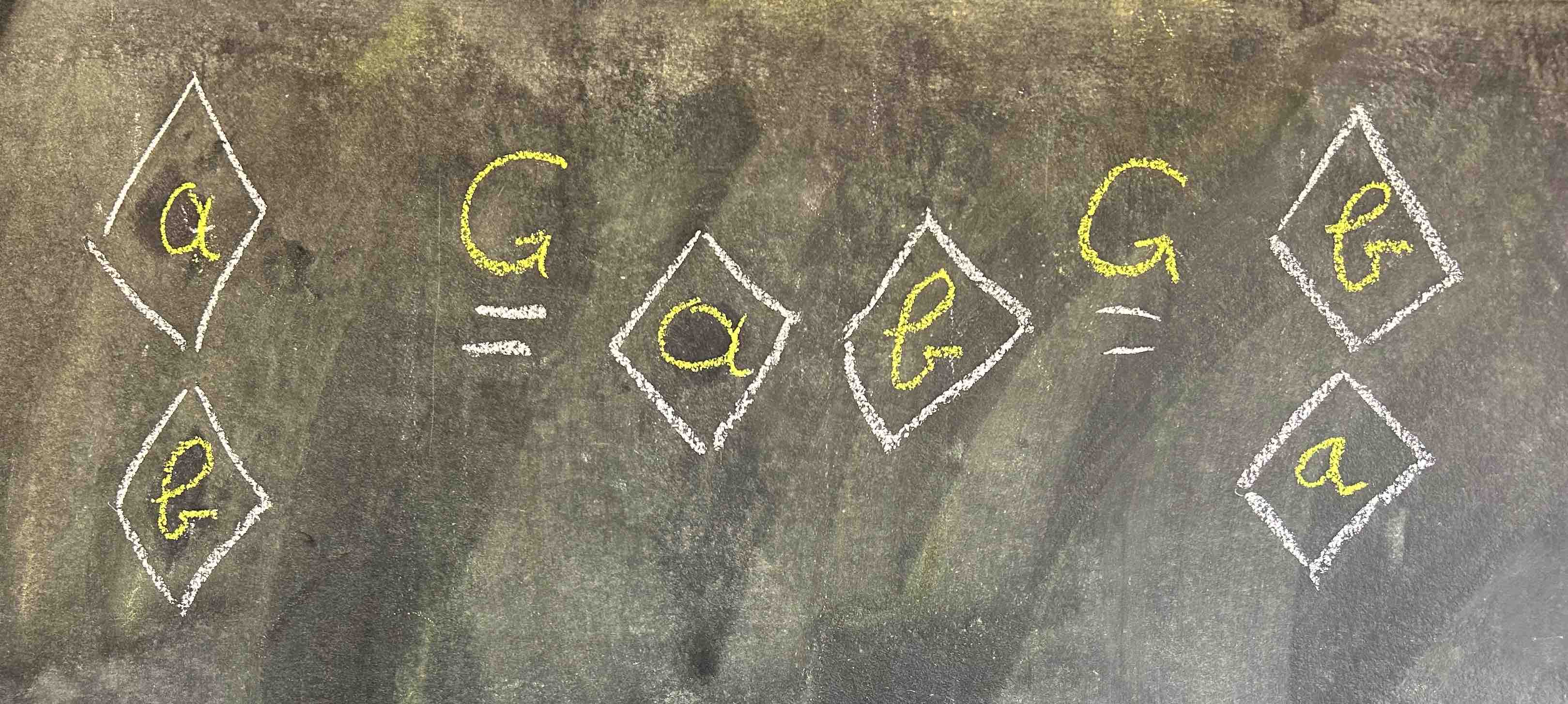,width=240pt} 
\end{center} 
We also saw how we can wire boxes together in order to form \bM diagrams\e.  Here are examples of a proper process, a state, and a number as diagrams:
\begin{center}
\epsfig{figure=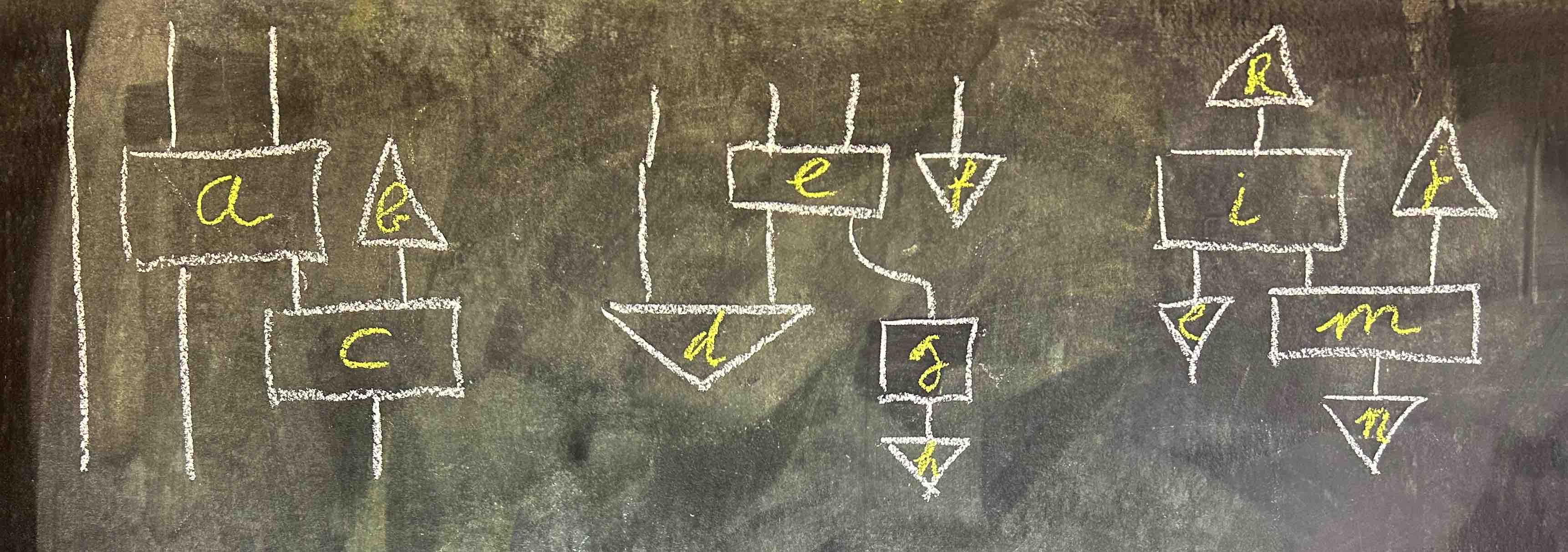,width=360pt} 
\end{center} 

\section{ZX-notation for linear maps}

Above the boxes are just abstract boxes, or, black boxes if you wish.  We will now fill them in with certain linear maps, which we call \bM ZX-notation\e.  

\subsection*{Plain spiders}

ZX-notation has mainly two symbols, called \bM Z-spiders\e (\bG green\e) and \bM X-spiders\e (\bR red\e):
\begin{center}
\epsfig{figure=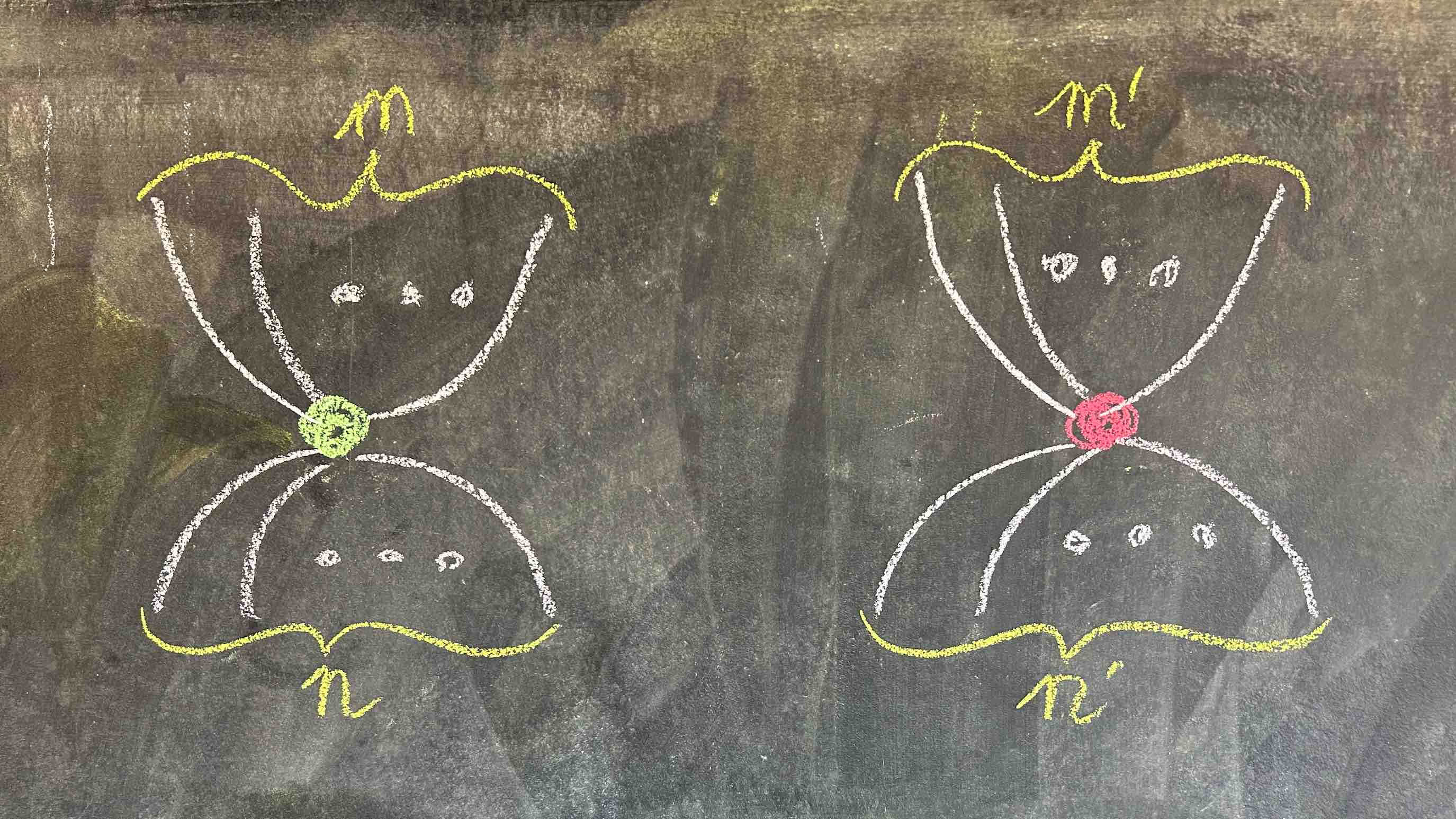,width=270pt} 
\end{center} 
They represent the following linear maps respectively:
\begin{center}
\epsfig{figure=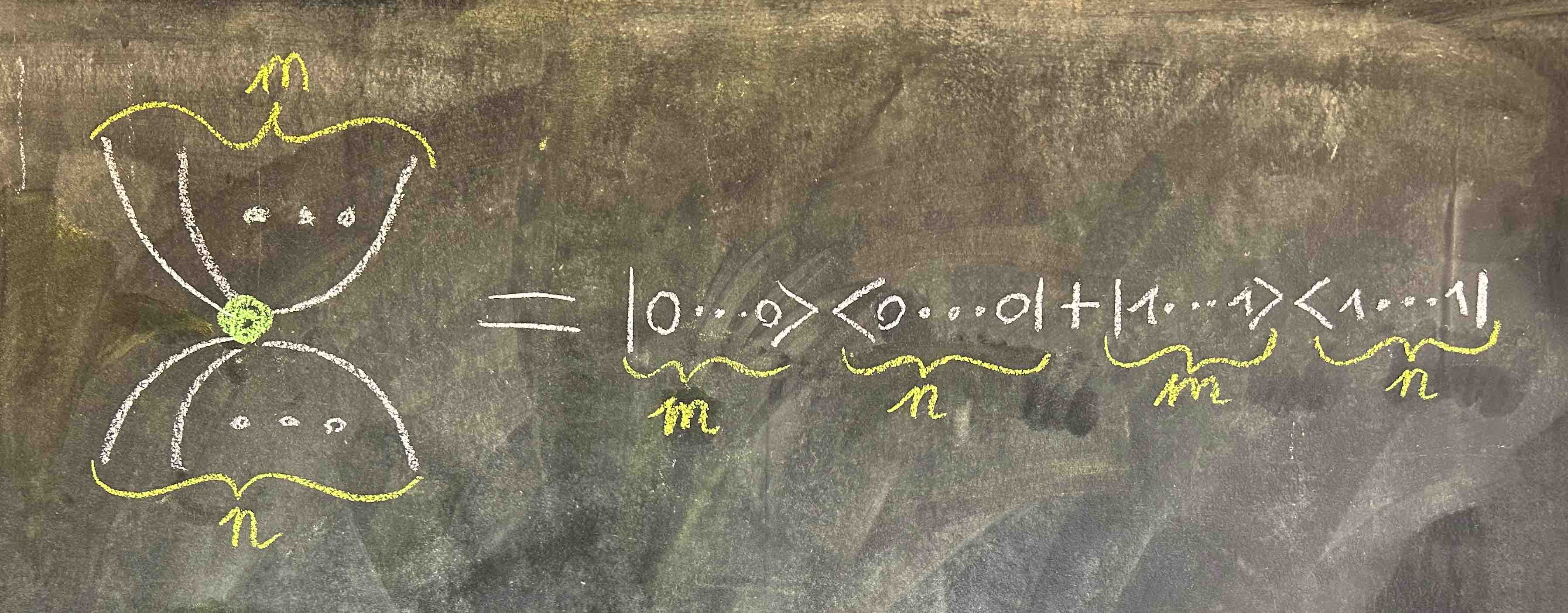,width=330pt} 
\end{center} 
and:
\begin{center}
\epsfig{figure=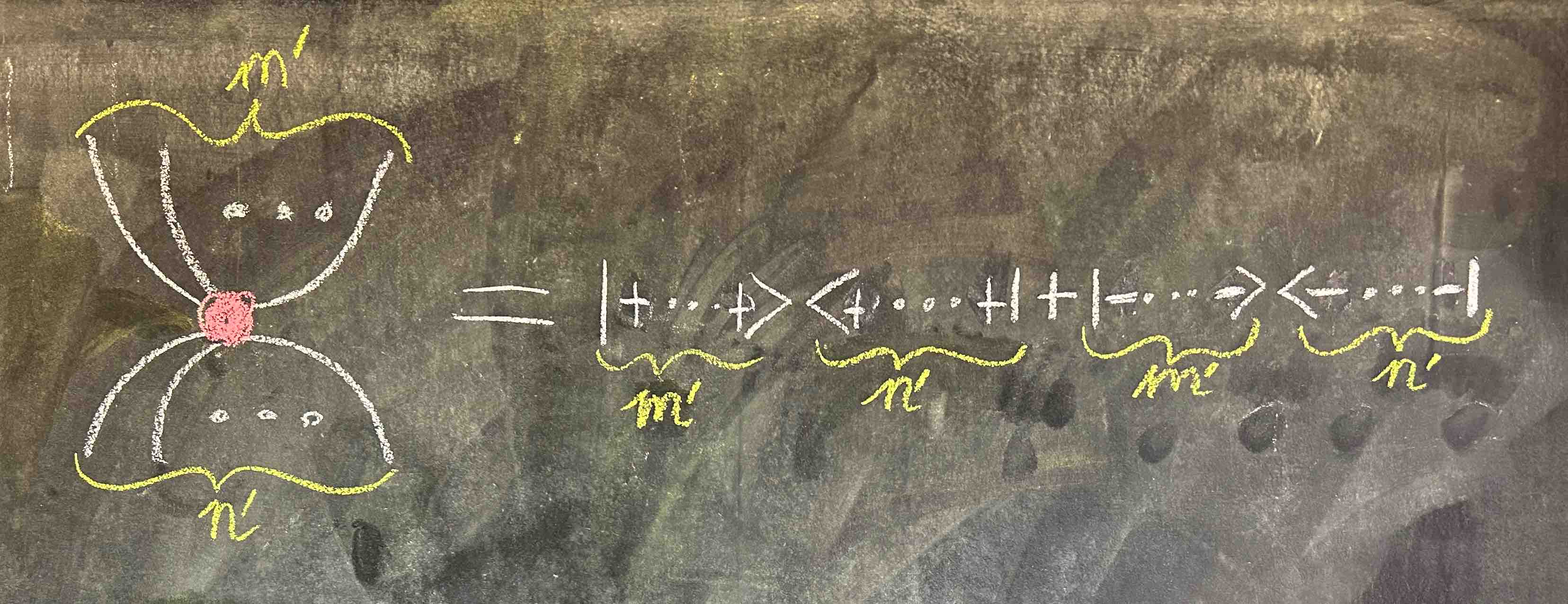,width=330pt}  
\end{center} 
So they only differ in the basis in which they are expressed, respectively:  
\[
\mbox{ \bM Z-basis\e} =   \{ |0\rangle, |1\rangle\}
\]
and:
\[
\mbox{\bM X-basis\e} = 
\left\{\, |+\rangle = {1\over\sqrt{2}} (|0\rangle + |1\rangle)\ ,\ 
         |-\rangle = {1\over\sqrt{2}} (|0\rangle - |1\rangle)\,\right\} 
\]
Here are some important examples of such spiders:
\[	
\{ |0\rangle, |1\rangle\} \mbox{-\bM copy/delete\e} = \raisebox{-30pt}{\epsfig{figure=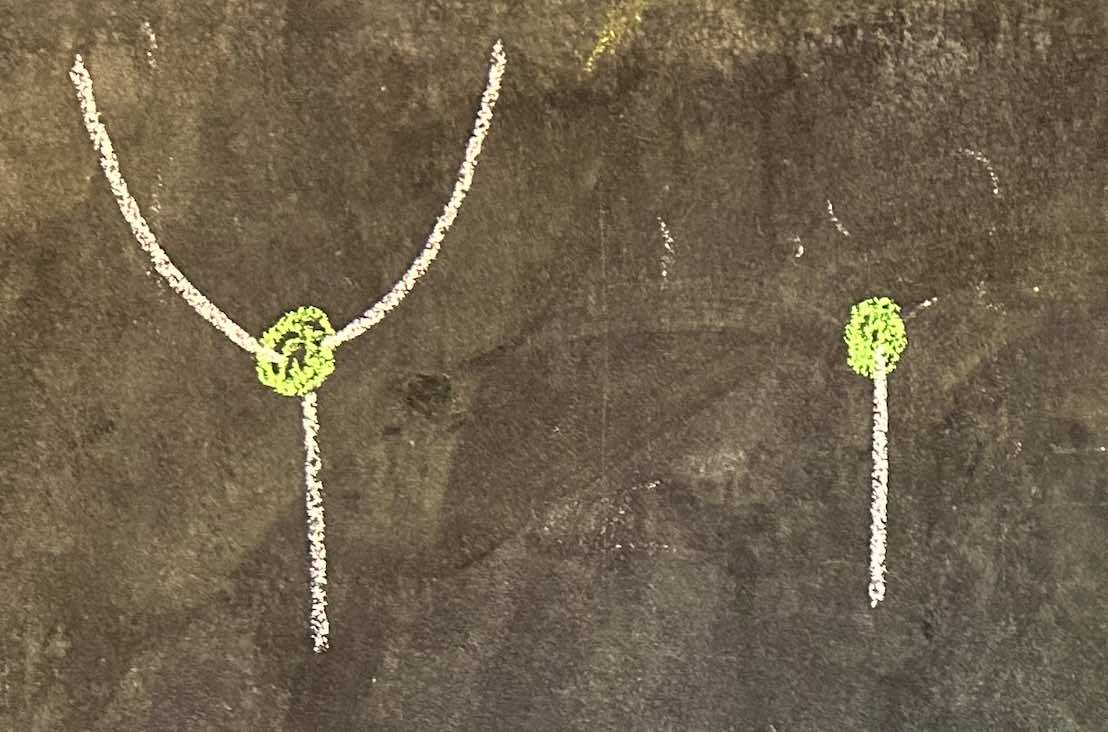,width=100pt}}\vspace{-1.6mm}
\]
\[
\{ |+\rangle, |-\rangle\} \mbox{-\bM copy/delete\e} = \raisebox{-32pt}{\epsfig{figure=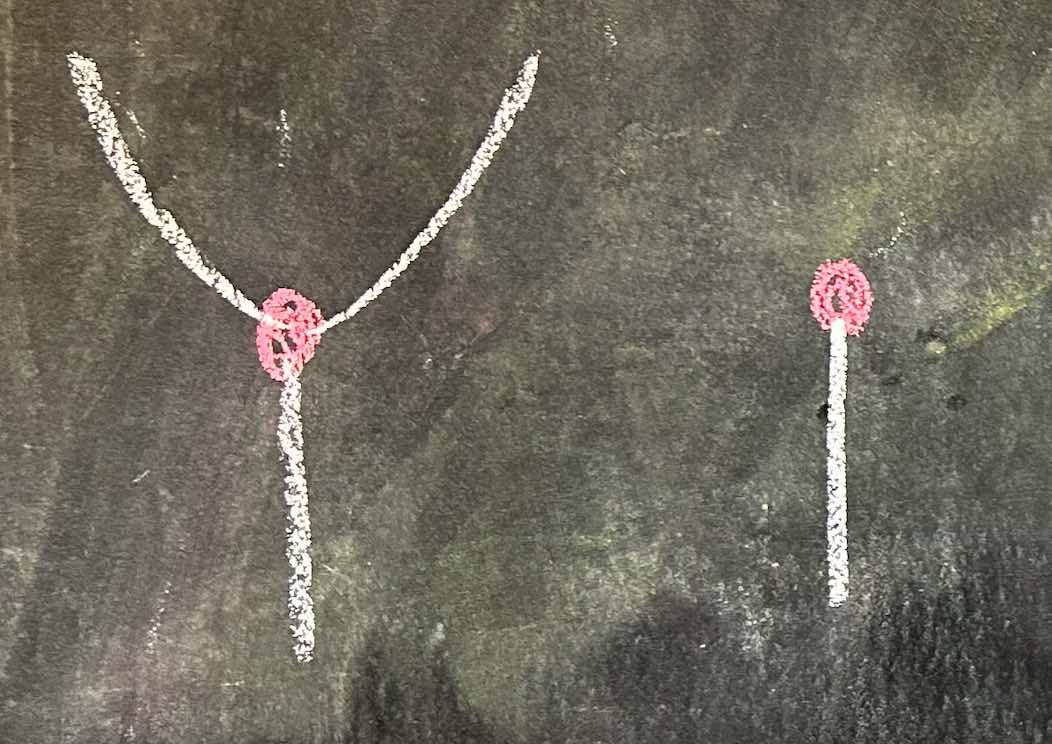,width=100pt}} 
\]
\[	
\mbox{\bM identity\e} =  \raisebox{-24pt}{\epsfig{figure=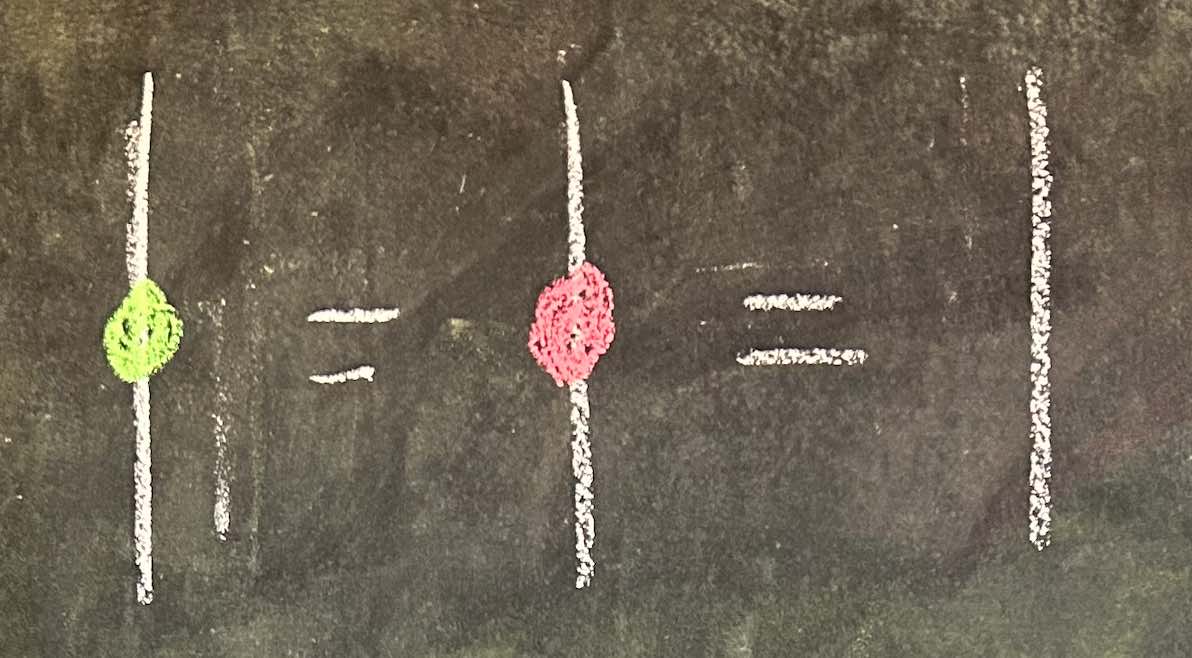,width=100pt}}
\]
\[	
\mbox{\bM Bell-state or cup\e} =  \raisebox{-18pt}{\epsfig{figure=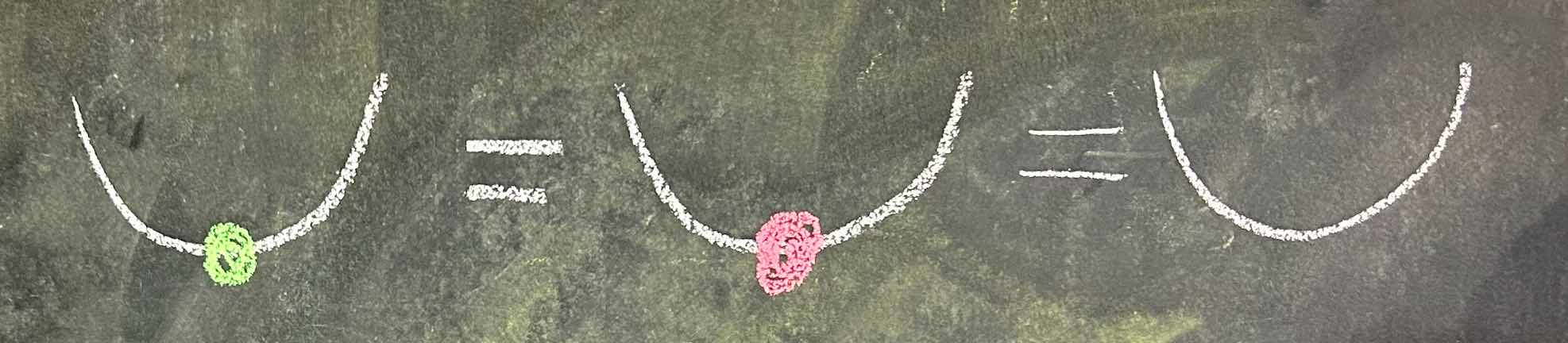,width=200pt}}
\]
\[	
\mbox{\bM Bell-effect or cap\e} =   \raisebox{-18pt}{\epsfig{figure=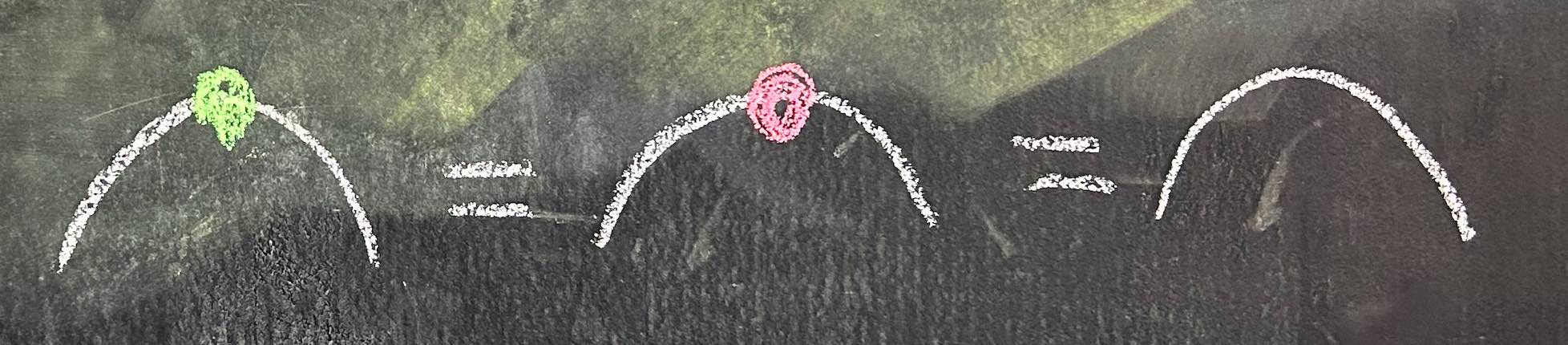,width=200pt}}
\]
\[	
\mbox{\bM GHZ-state\e} =   \raisebox{-24pt}{\epsfig{figure=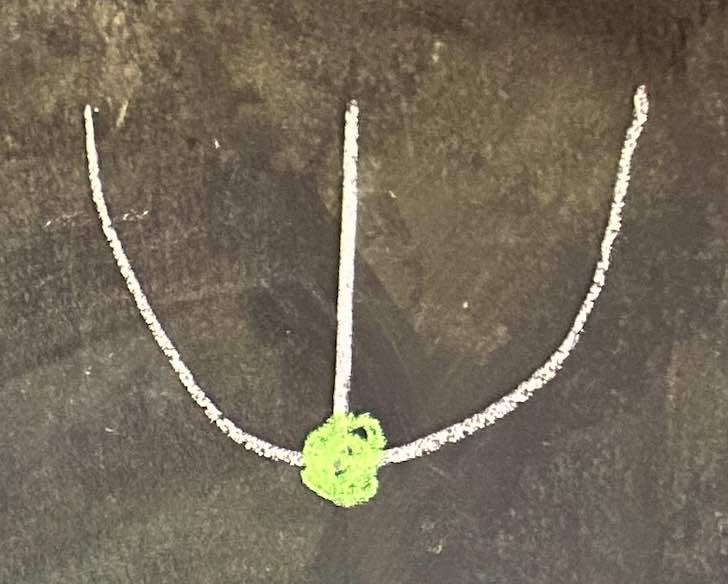,width=65pt}}
\]
Note first that the states here are not normalised, which has calculational advantages.  

\bigskip\noindent
{\bf Exercise.}  Show that the cup-spiders and cap-spiders are indeed the same for both colours.  Convince yourself that except for these, the plain wire, and the spider with no legs, no other spiders of different colours coincide.  

\bigskip\noindent
{\bf Exercise.}  Show that the spider with no legs is the number 2, and that it is equal to a circle made up of a cup and a cap.
 
\subsection*{Spiders with phases}

We will now also allow these spiders to carry \bM phases\e, that is, a number in $[0, 2\pi]$, which modify the linear maps as follows:
\begin{center}
\epsfig{figure=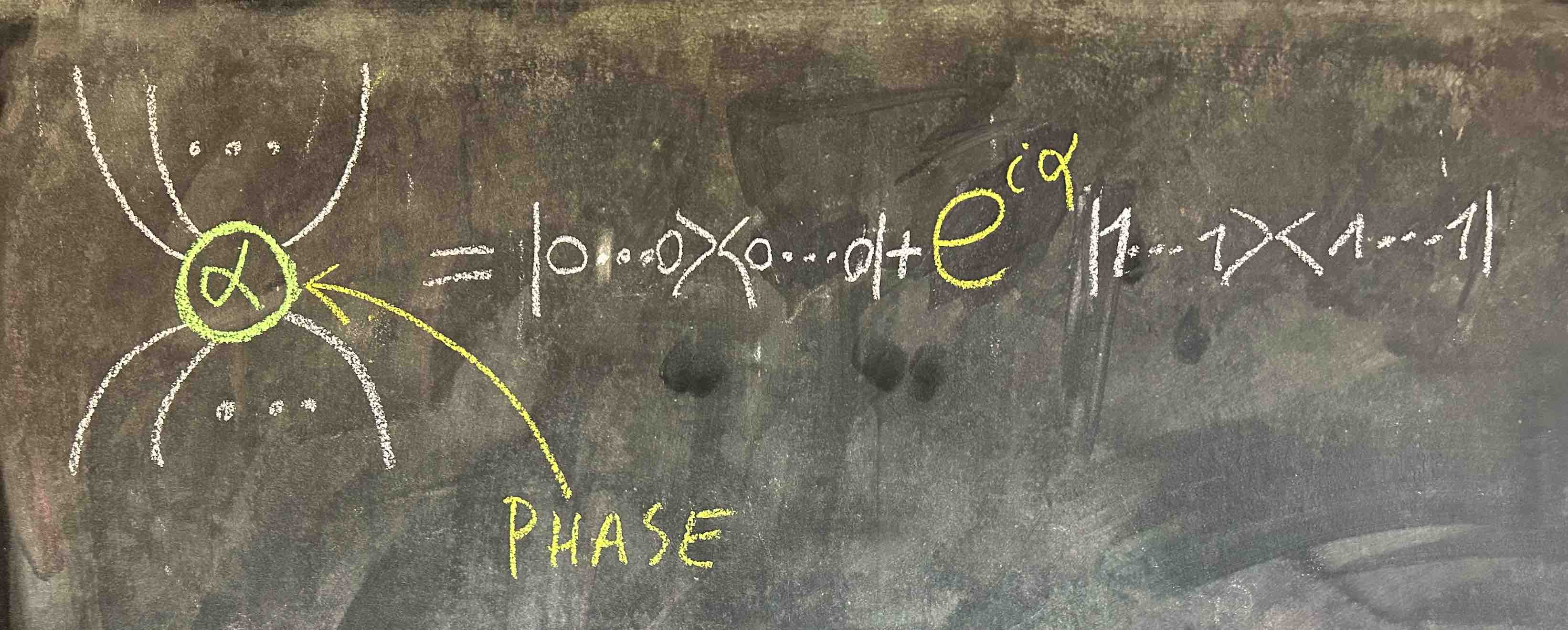,width=330pt} 
\end{center} 
and:
\begin{center}
\epsfig{figure=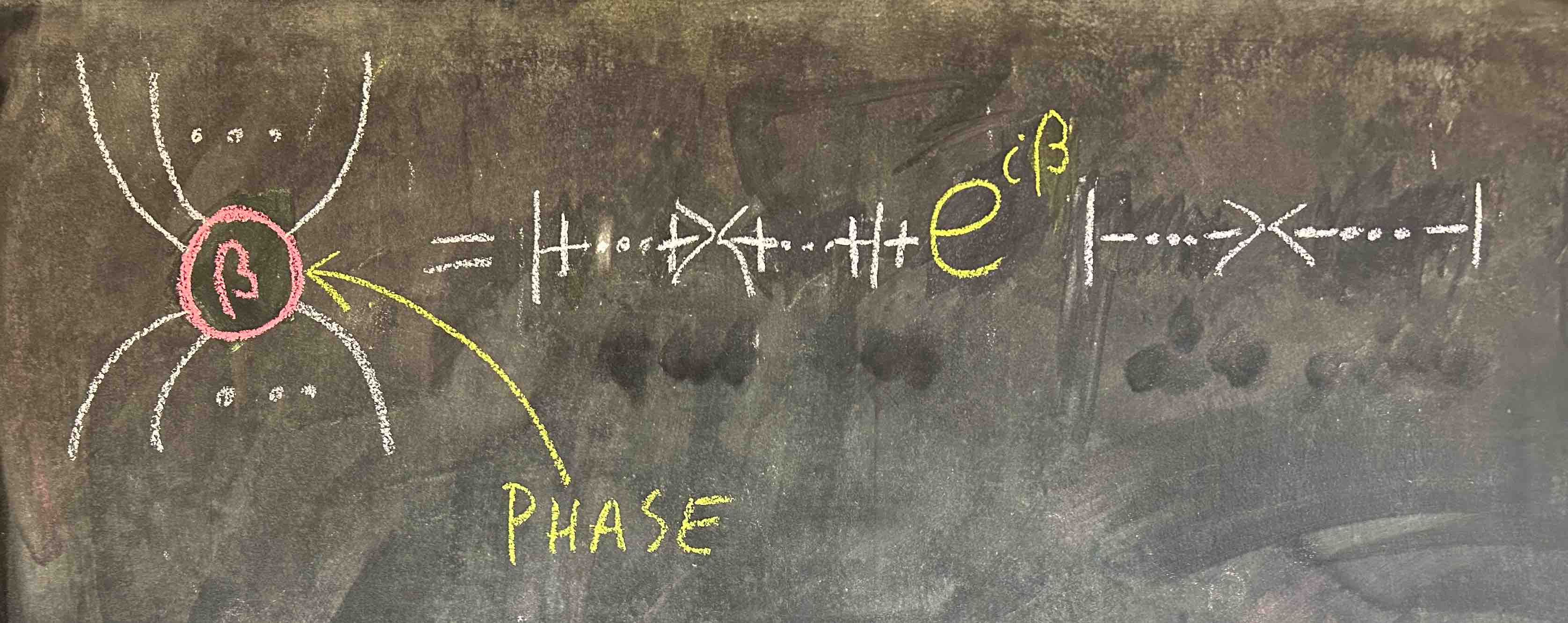,width=330pt} 
\end{center} 
We typically don't write the 0-phase:
\begin{center}
\epsfig{figure=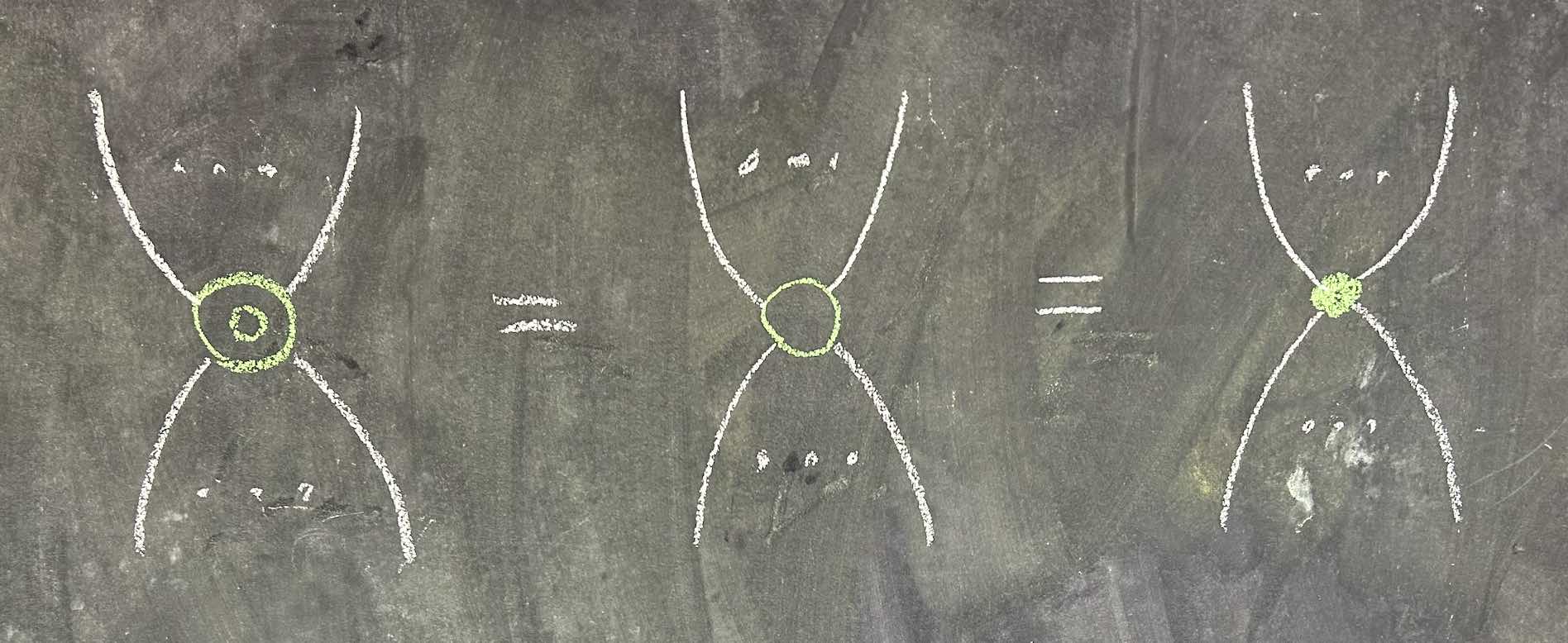,width=280pt} 
\end{center} 
Some examples here are:
\[	
\mbox{\bM phase-gate\e} = \raisebox{-39pt}{\epsfig{figure=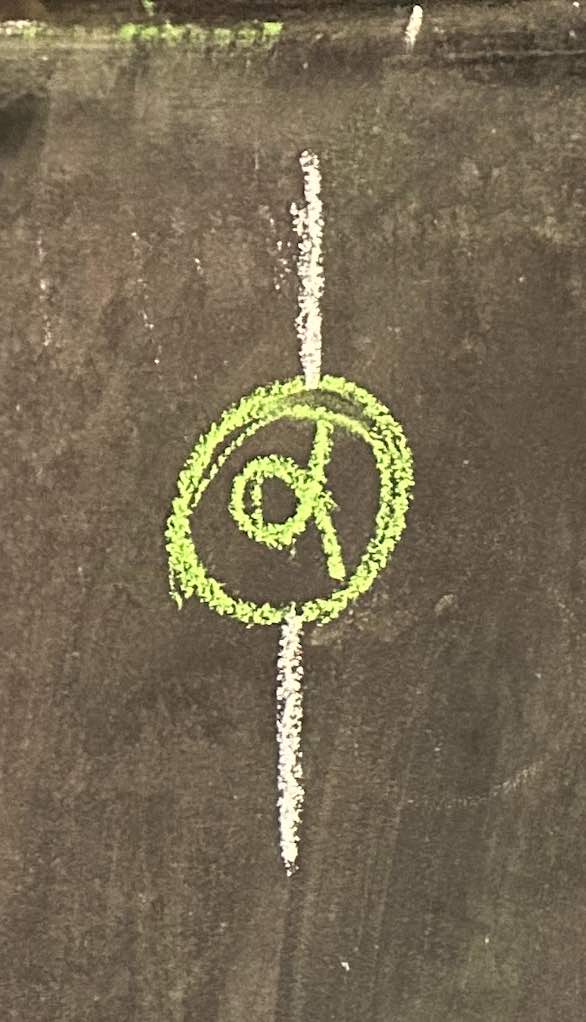,width=55pt}} 
\]
\[	
\mbox{\bM NOT-gate\e} = \raisebox{-39pt}{\epsfig{figure=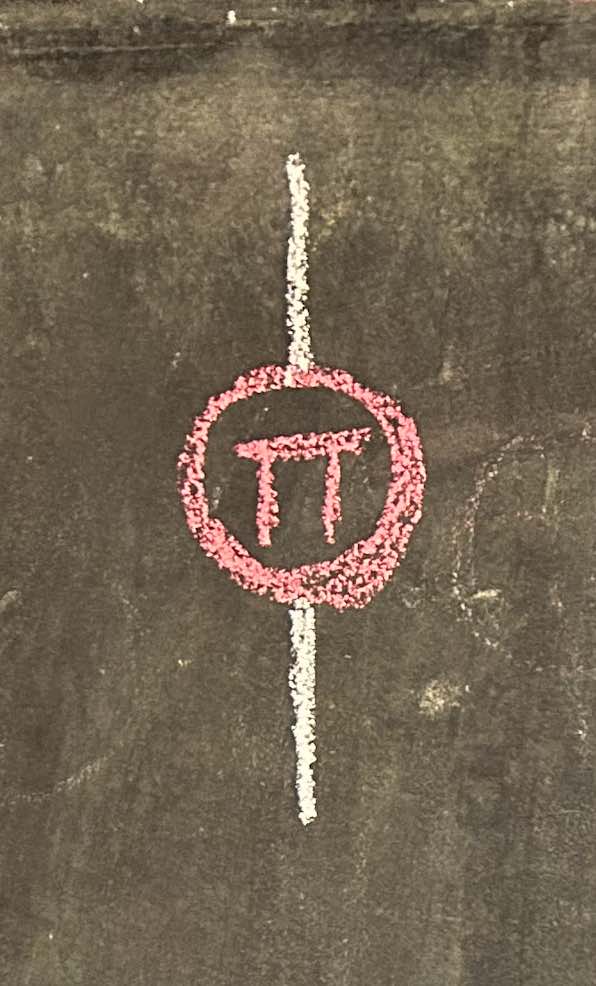,width=55pt}}
\]
A \bM ZX-diagram\e is a `web’ made up of Z-spiders and X-spiders: 
\begin{center}
\epsfig{figure=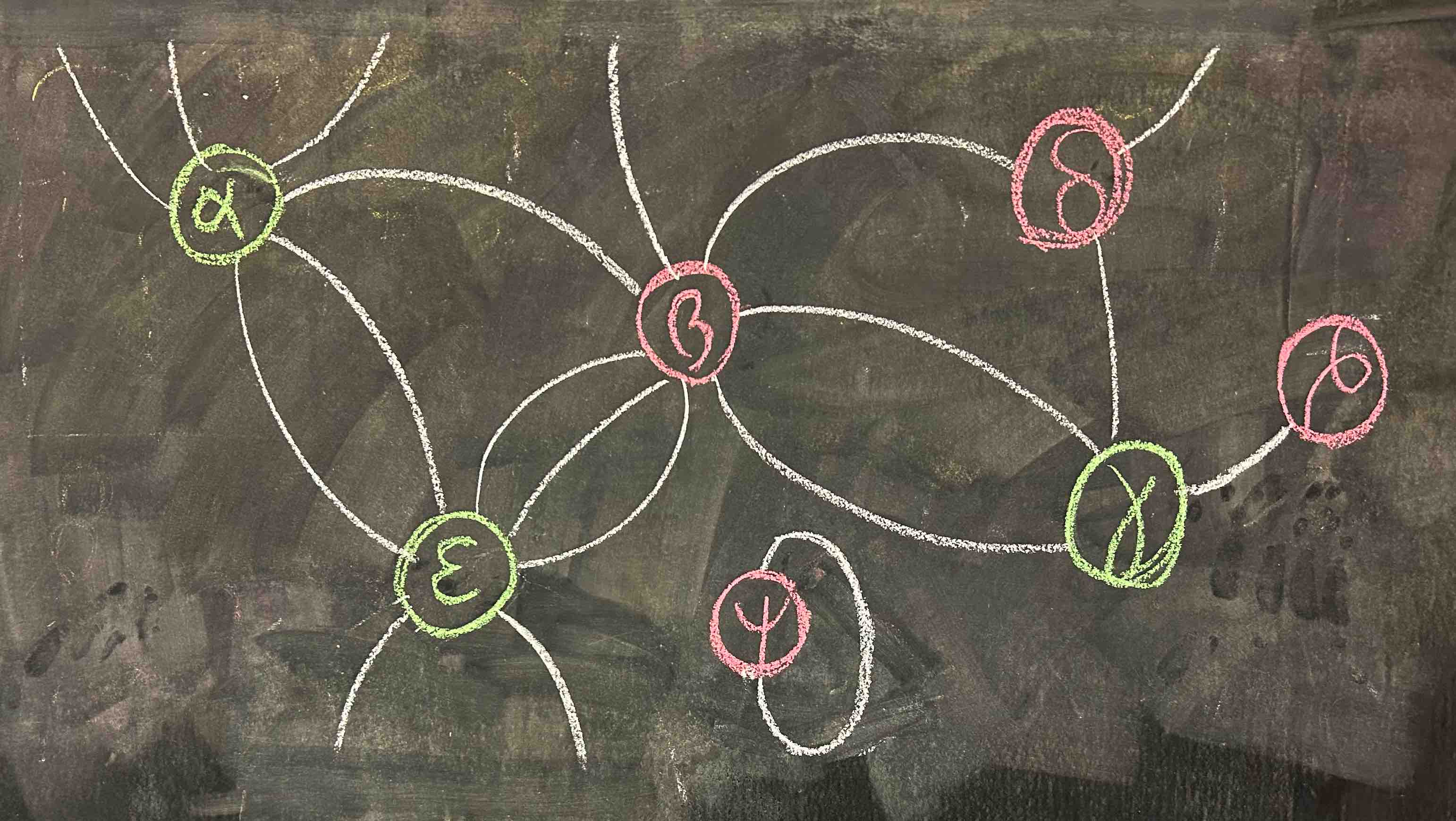,width=330pt} 
\end{center} 
When we combine spiders of two colours the most important example is: 
\[	
\mbox{\bM CNOT-gate\e} = \raisebox{-48pt}{\epsfig{figure=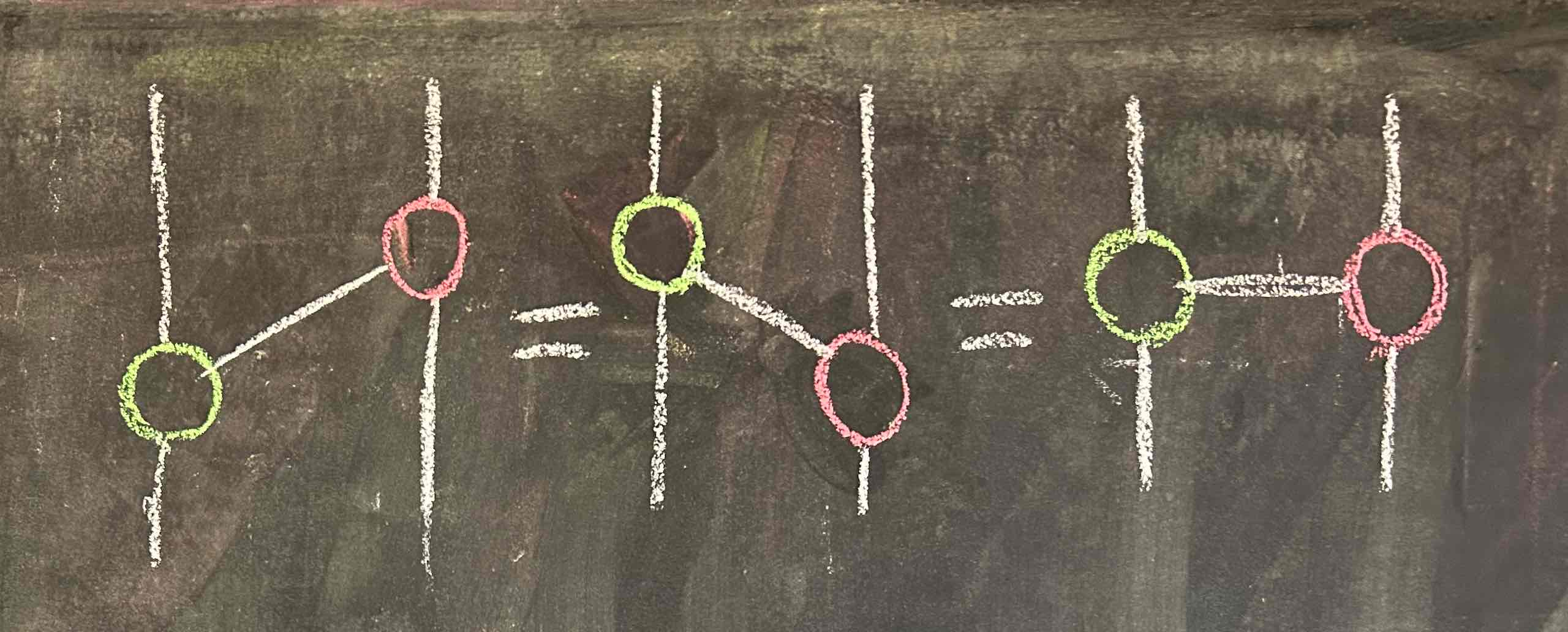,width=240pt}} 
\]

\bigskip\noindent
{\bf Exercise.}  Show that the two ways of defining the CNOT-gate are indeed the same, hence justifying the horizontal wire. 

\bigskip\noindent
Also, up to normalisation, we can write all basis states as spiders:
\begin{center}
\epsfig{figure=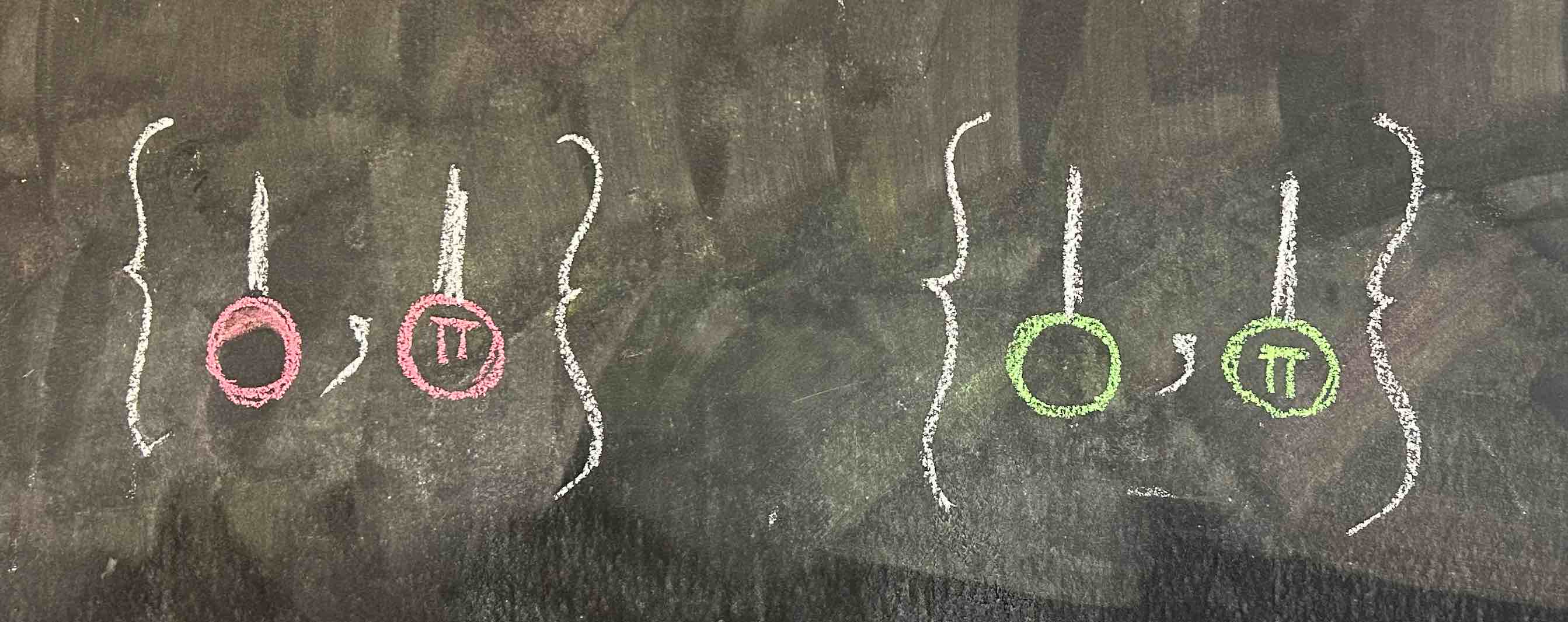,width=300pt} 
\end{center} 
Note here that while the green spiders are built in terms of the Z-basis, its basis vectors are expressed in terms of the X-spiders, and vice versa.

\bigskip\noindent
{\bf Exercise.}  Show that this is indeed the case.  

\section{Universality of ZX-notation} 

{\bf Theorem.}  Any linear map between qubits can be written in terms of spiders.  

\bigskip\noindent
In other words, \bM ZX-notation is universal\e  \cite{ContPhys, CD2, CKbook}.  We can prove this as follows.
It is well-known that any unitary map between qubits can be written in terms of CNOT-gates,  Z-phase gates and X-phase gates, and each of these can be written in terms of ZX-spiders as we have seen above.  This means that we can take a state of $n+m$ qubits expressible in terms of spiders, e.g.~$|0\ldots 0\rangle$, and produce any other $n+m$ qubit state by means of an appropriate unitary:
\begin{center}
\epsfig{figure=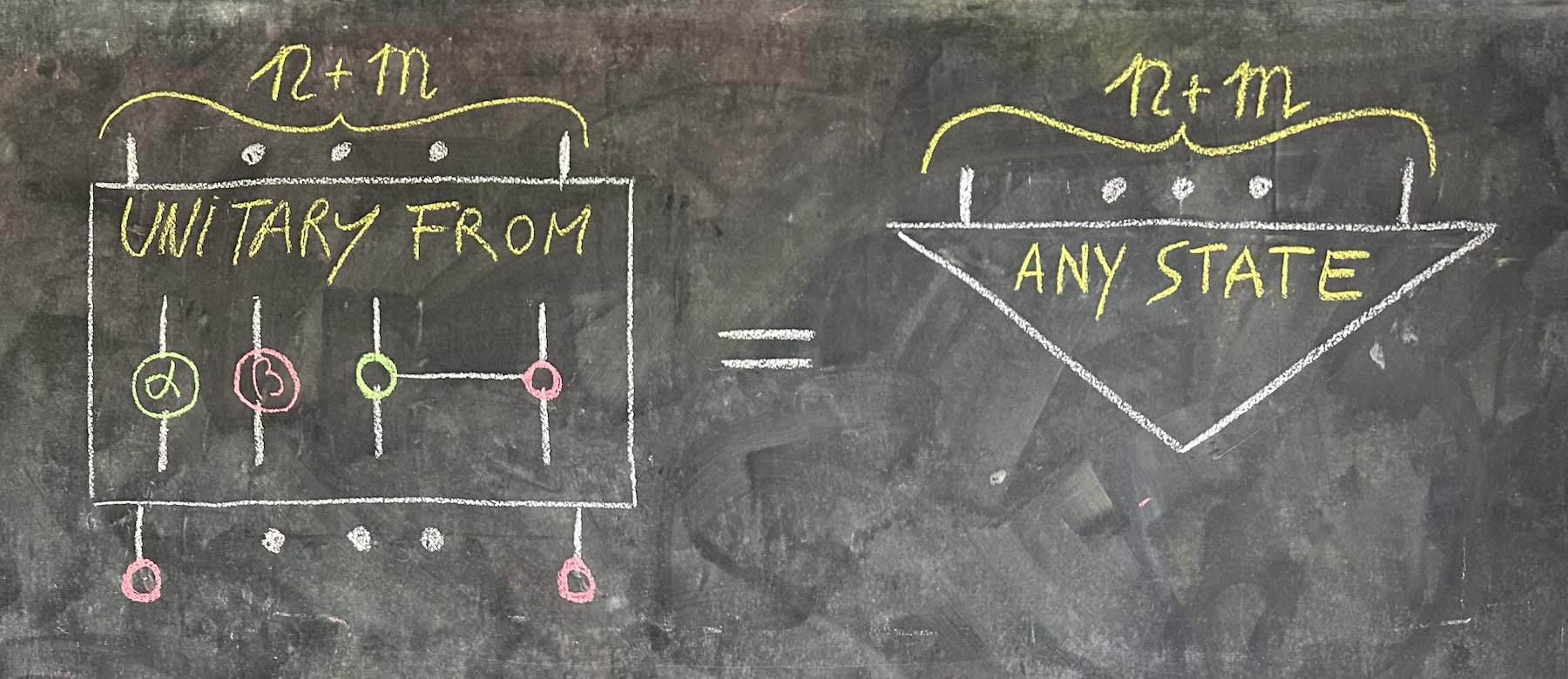,width=330pt} 
\end{center} 
Then we can use cups to turn that arbitrary $n+m$ qubit state into a $n$ qubit to $m$ qubit linear map:  
\begin{center}
\epsfig{figure=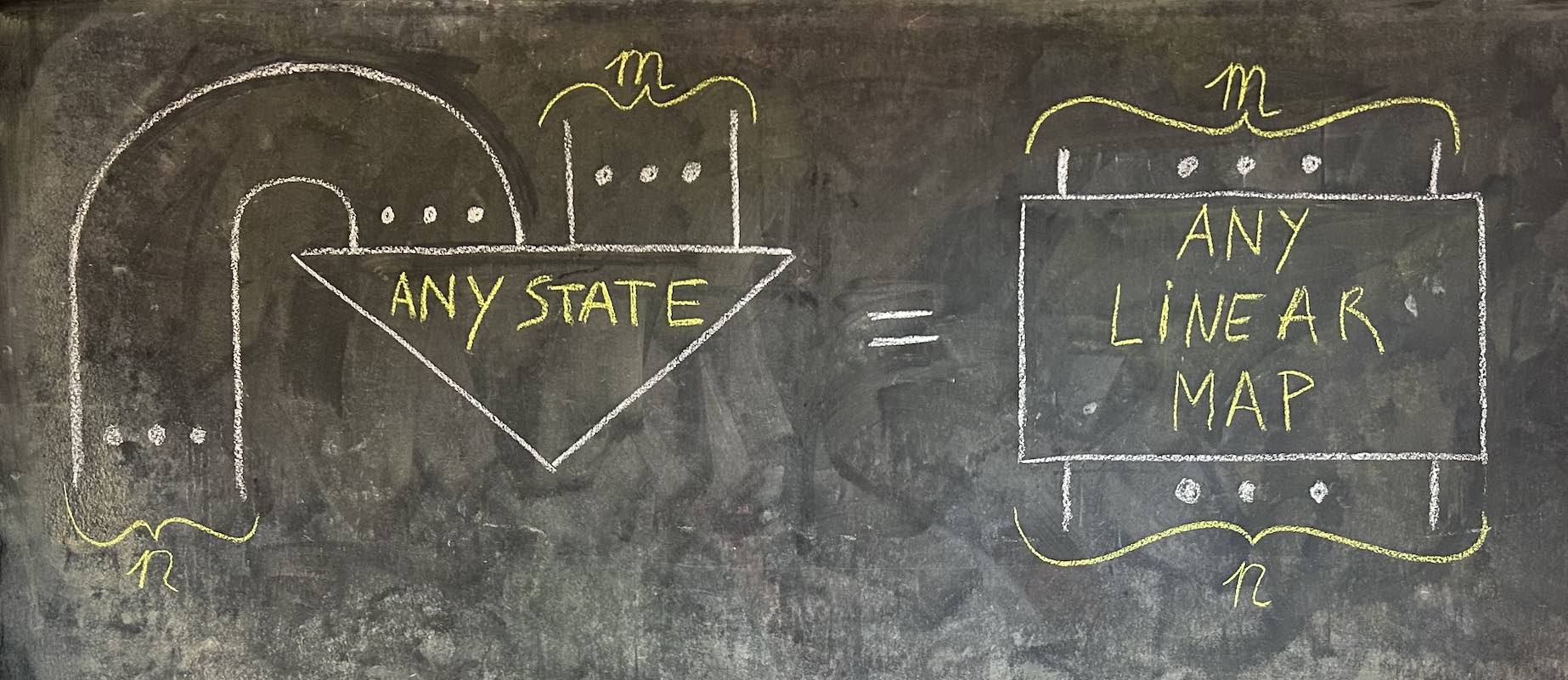,width=330pt} 
\end{center}   
By the isomorphism between $n+m$ qubit states and $n$ qubit to $m$ qubit linear maps, we can obtain any $n$ qubit to $m$ qubit linear map by using caps. %he isomorphism is also called Choi-Jamiolkowsi when generalised to mixed states.
A more detailed diagrammatic account on this isomorphism can be found in  \cite[\S 3.1]{CKpaperI}  and  \cite[\S 4.1.2]{CKbook}.

\section{Rules of ZX-calculus}

\subsection*{Spider-fusion}

What makes ZX-calculus truly useful are the simple rules that allow us to transform ZX-diagrams. For spiders of the same colour without phases we have the following rule:
\begin{center}
\epsfig{figure=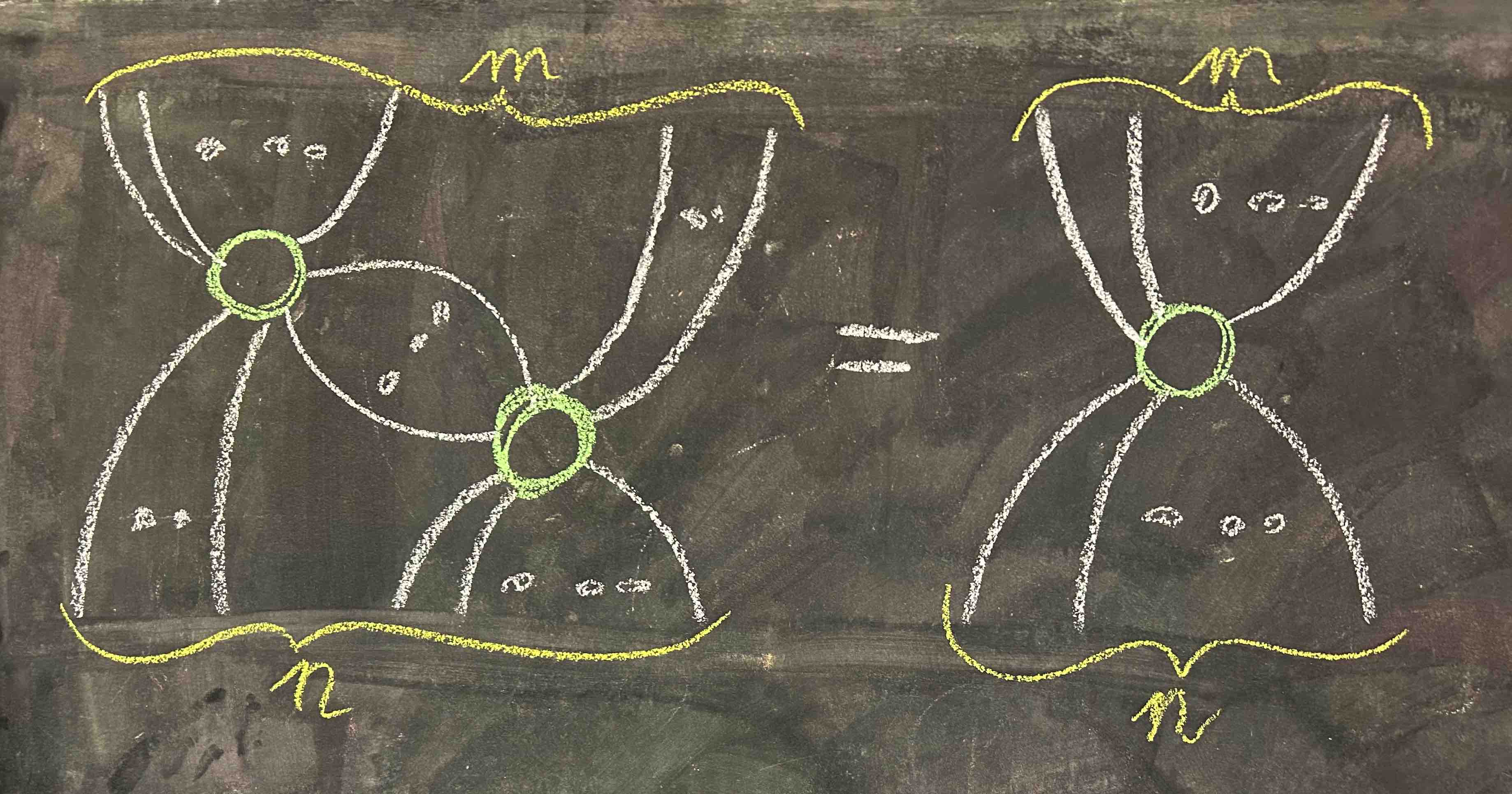,width=240pt} \qquad\epsfig{figure=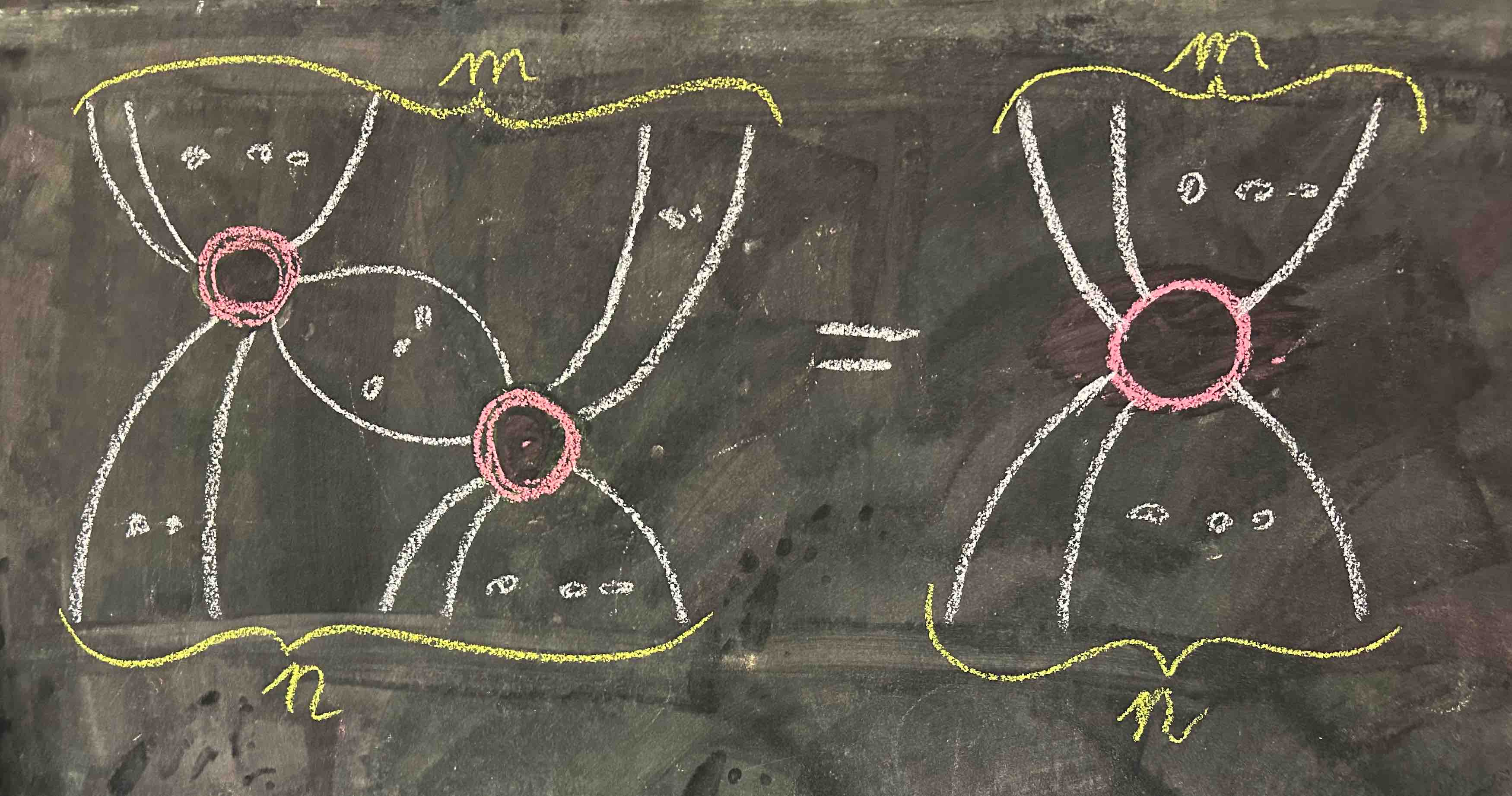,width=240pt} 
\end{center} 
which tells us that \bM spiders of the same colour  fuse together\e.  This also means that any \underline{connected} diagram of spiders of the same colour only depends on its number of inputs and outputs, since each such web is itself a spider.  
If we have spiders with phases than these phases add up when fusing:
\begin{center}
\epsfig{figure=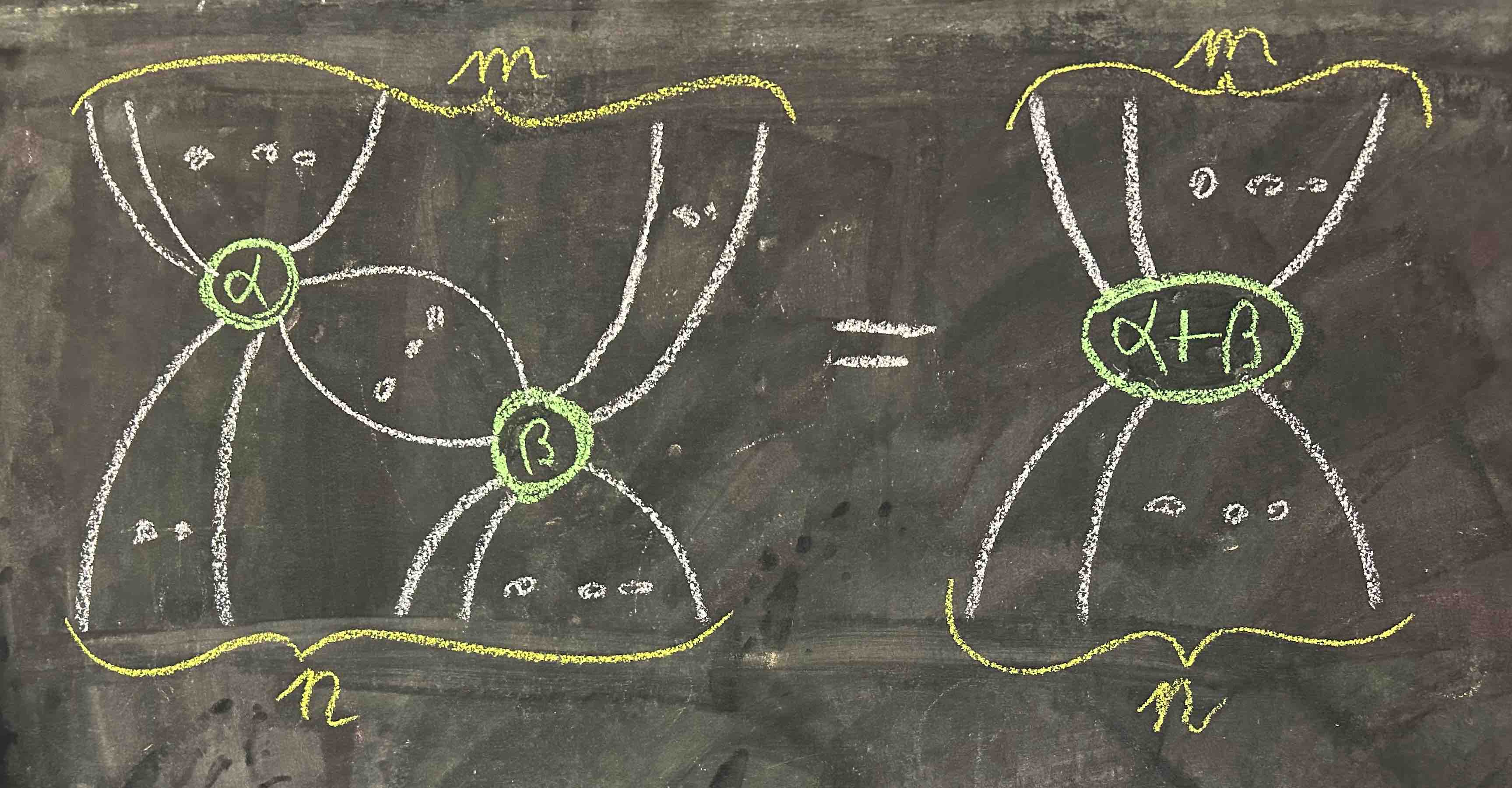,width=240pt}\qquad \epsfig{figure=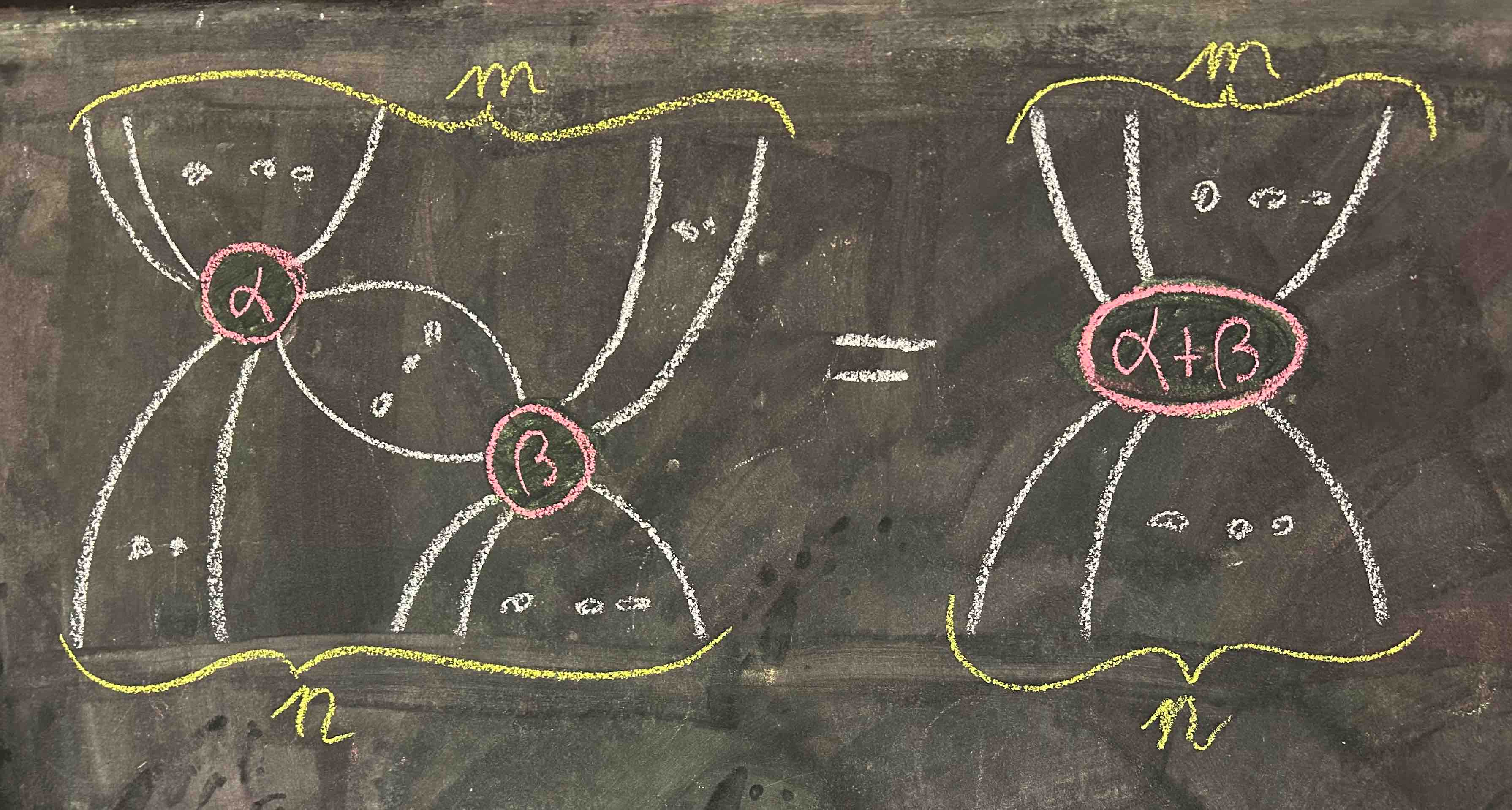,width=240pt} 
\end{center} 

\bigskip\noindent
{\bf Exercise.}  Prove the fusion rule for spiders with phases.

\bigskip\noindent
{\bf Remark.} From spider fusion it follows that inside a ZX-diagram one can always ignore what is an input and what is an output of a spider, as we can freely turn one into the other by  (un-)fusing, using either a cup or cap as follows:
\begin{center}
\epsfig{figure=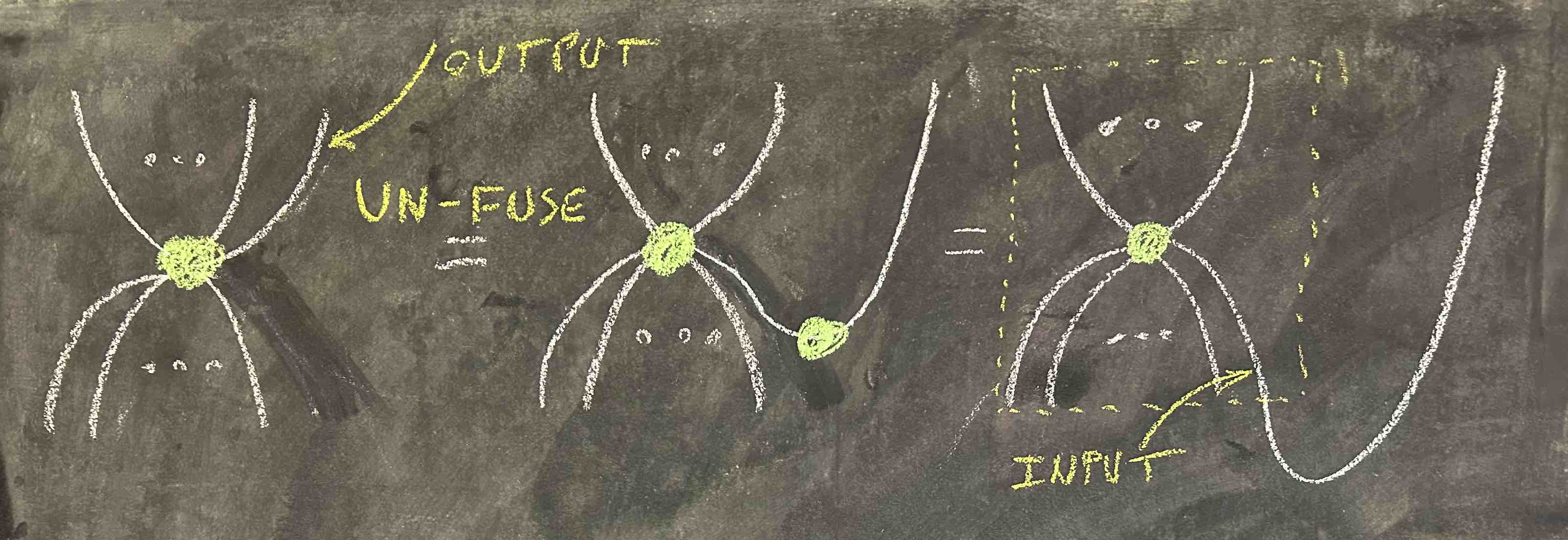,width=350pt}  
\end{center} 

\subsection*{Hopf-rule (or leg-chopping) and bialgebra-rule (or square-popping)}

When spiders of different colours share two legs, then these vanish:
\begin{center}
\epsfig{figure=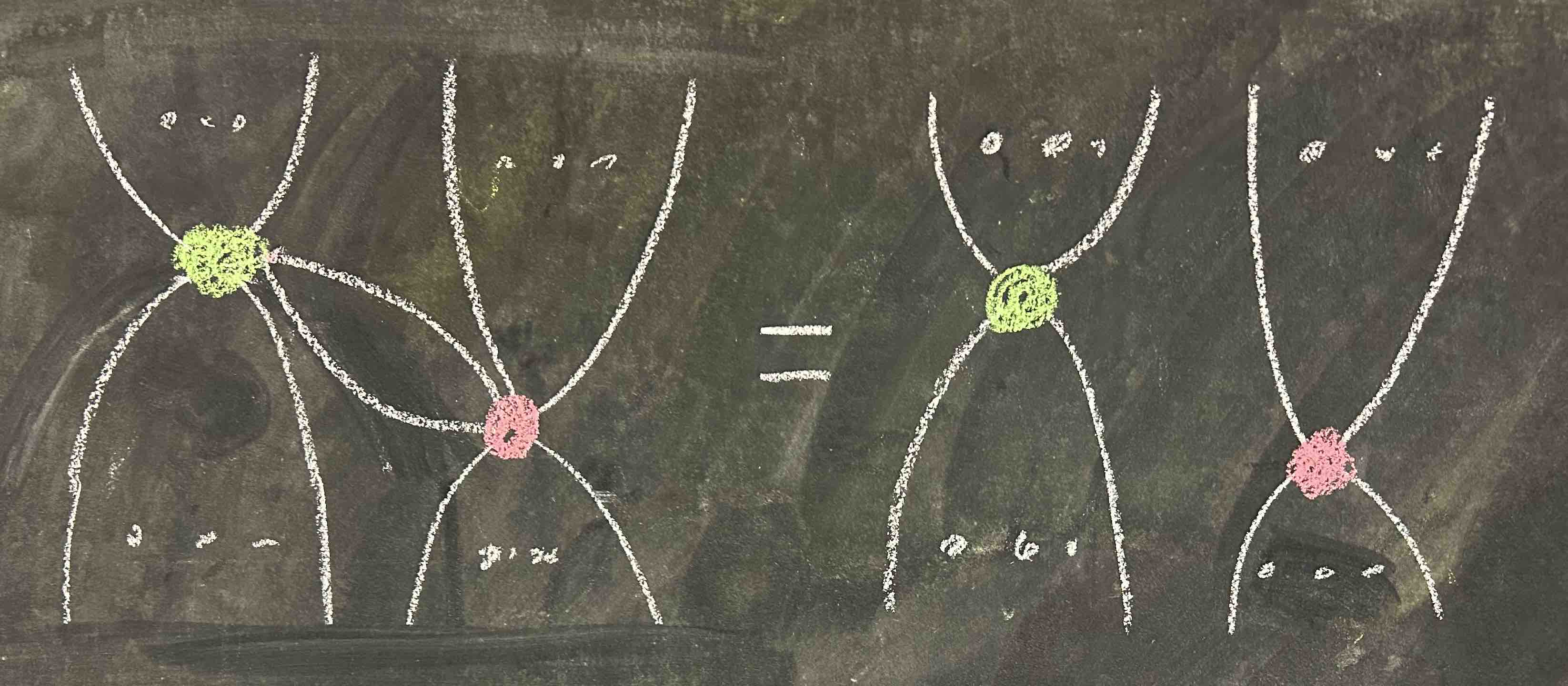,width=260pt} 
\end{center} 
which, thanks to spider-fusion, can be simplified as follows:
\begin{center}
\epsfig{figure=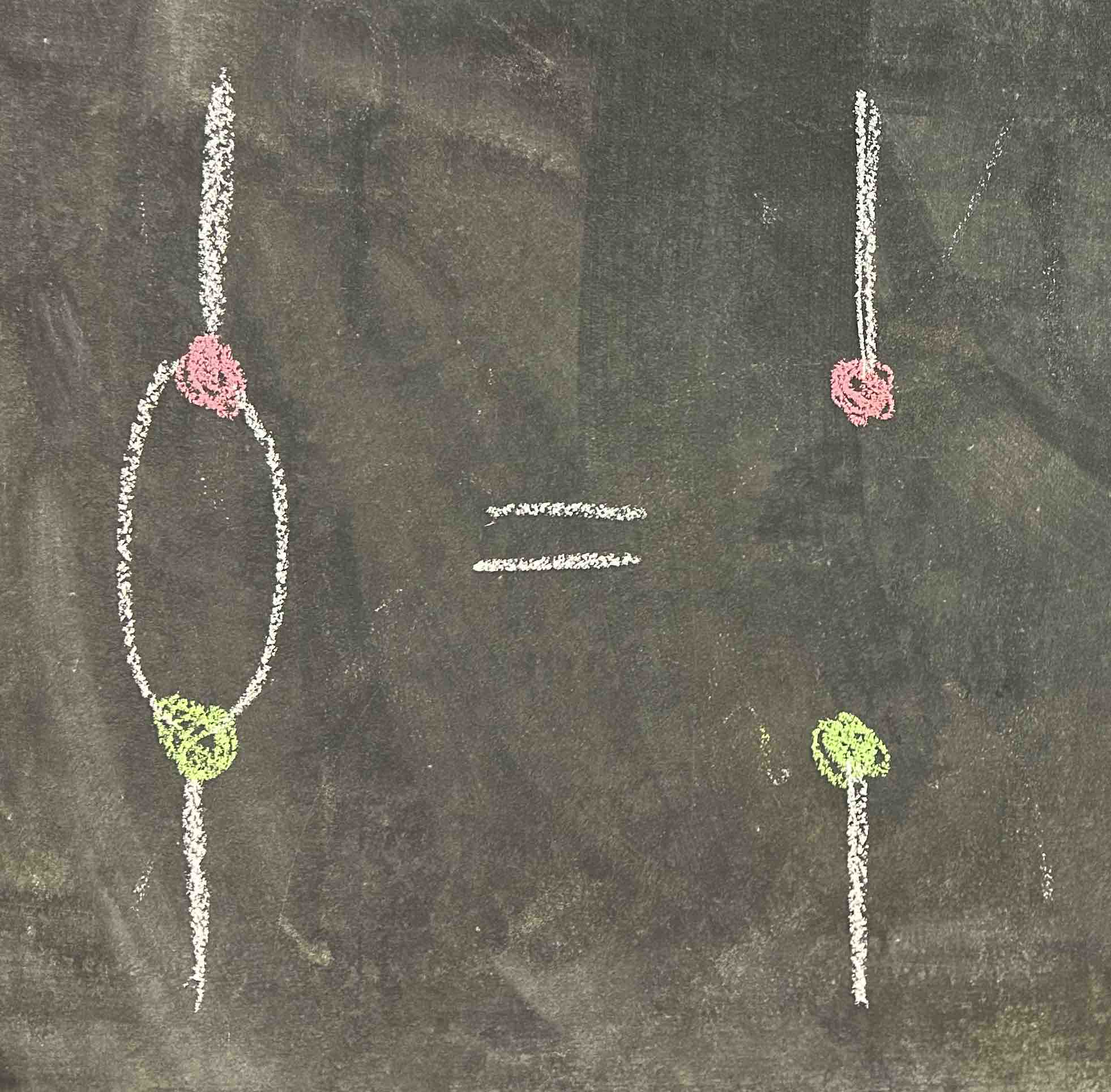,width=142pt} 
\end{center} 
For historical reasons we call this the \bM Hopf-rule\e. In \cite{QiP} we call it \bM leg-chopping\e .  
%This equation equivalent to mutual-unbiasedness of the bases of the corresponding bases, or complementarity \cite{CKbook}.

\bigskip\noindent
{\bf Remark}.  This equation is only true up to a number, but for the sake of the argument we won’t go into this issue here.  In fact, as long as you are not computing a probability, then you can typically always figure out at the end of the calculation what the normalising number should be \cite[\S 3.4.3]{CKbook}.  

\bigskip\noindent
{\bf Exercise.}  Prove the Hopf-rule.

\bigskip\noindent
There is a rule which is stronger than the Hopf-rule  and consists of three equations:
\begin{center}
\epsfig{figure=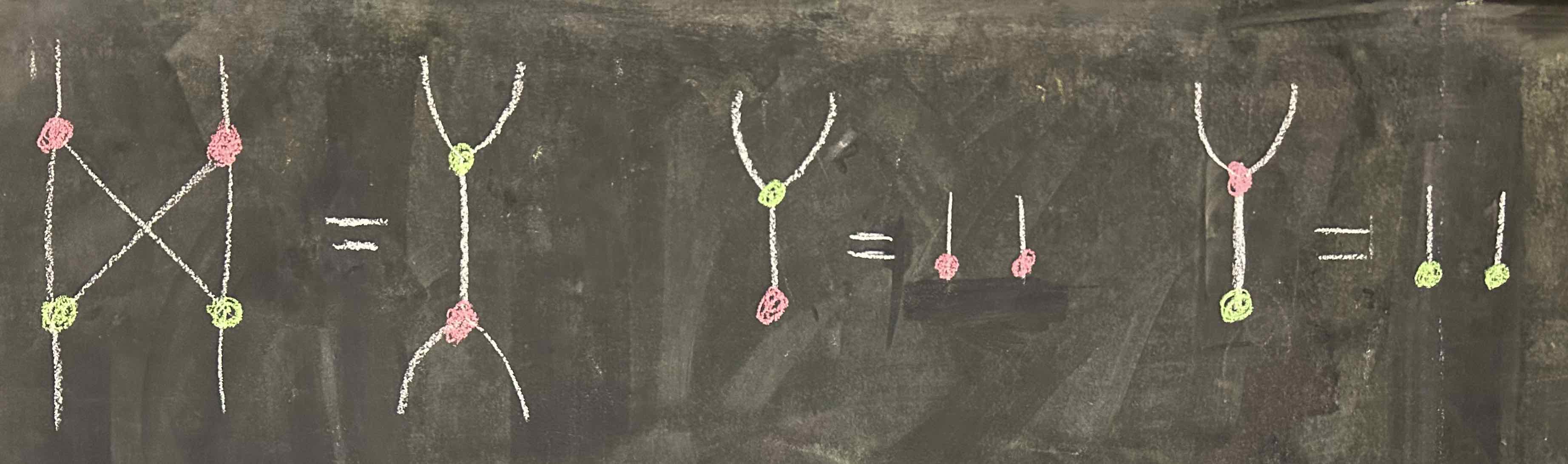,width=420pt} 
\end{center} 
For historical reasons we call the first of these the \bM bialgebra-rule\e.  The two two \bM copy-rules\e say that ``green copies red”, and that ``red copies green”.  In practice, the way the 1st one of these is more often used is in the following form, which gets rid of squares, while leg-chopping only gets rid of 2-cycles: 
\begin{center}
\epsfig{figure=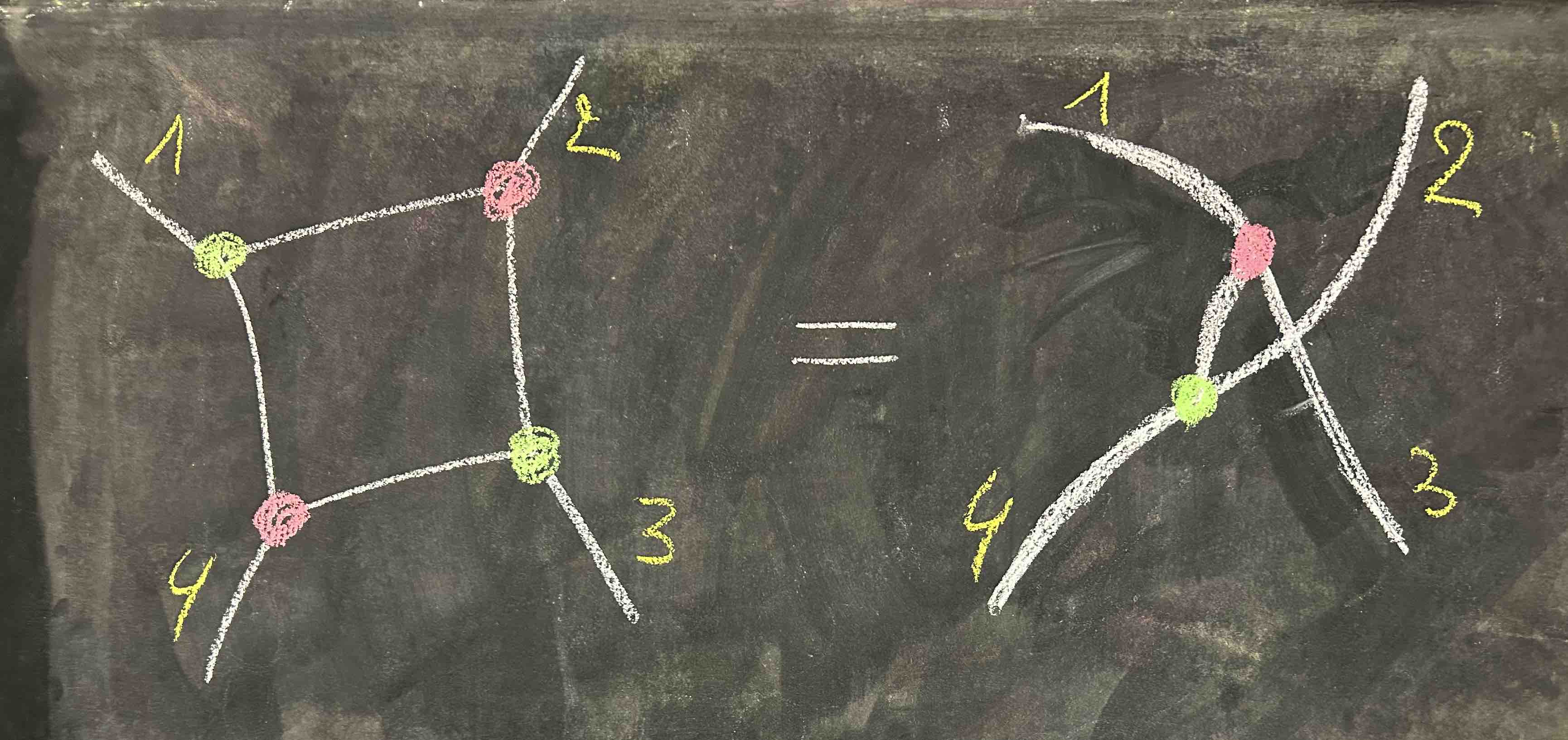,width=280pt} 
\end{center} 
Therefore  in \cite{QiP} it is called \bM square-popping\e.  One can derive the Hopf-rule from the bialgebra-rule and the two copy-rules:  
\begin{center}
\epsfig{figure=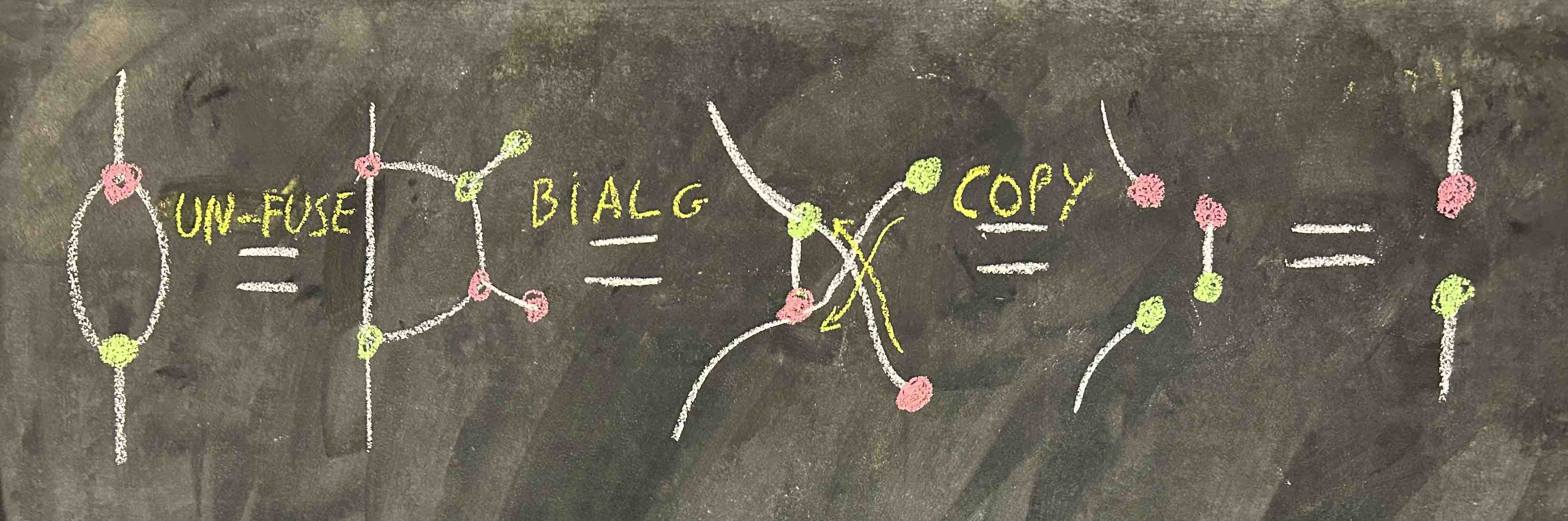,width=430pt} 
\end{center} 
Using some group theory, which we won't do here and you can find in \cite[\S 9.3.4]{CKbook}, one can also show that the other basis states are also copied by the opposite colour:
\begin{center}
\epsfig{figure=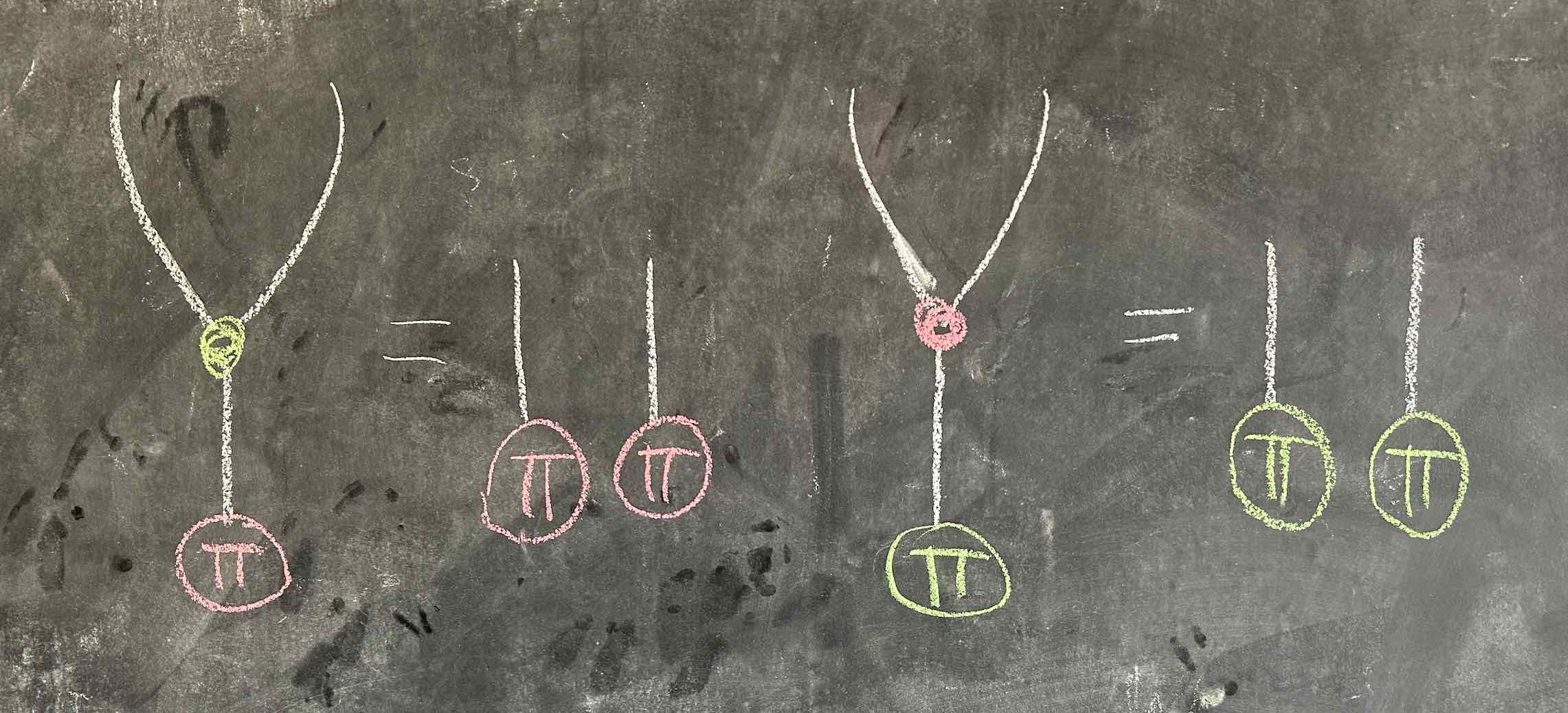,width=300pt} 
\end{center} 
The  rules listed above will get you a very long way already, including some cutting-edge new quantum computing techniques, as we will see further below.  But we will first present a larger set of rules about which we can prove something very important.

\subsection*{Colour-change}

It is useful to have a special notation for the Hadamard-gate which in terms of Euler angles can be written as follows in ZX-notation: 
\begin{center}
\epsfig{figure=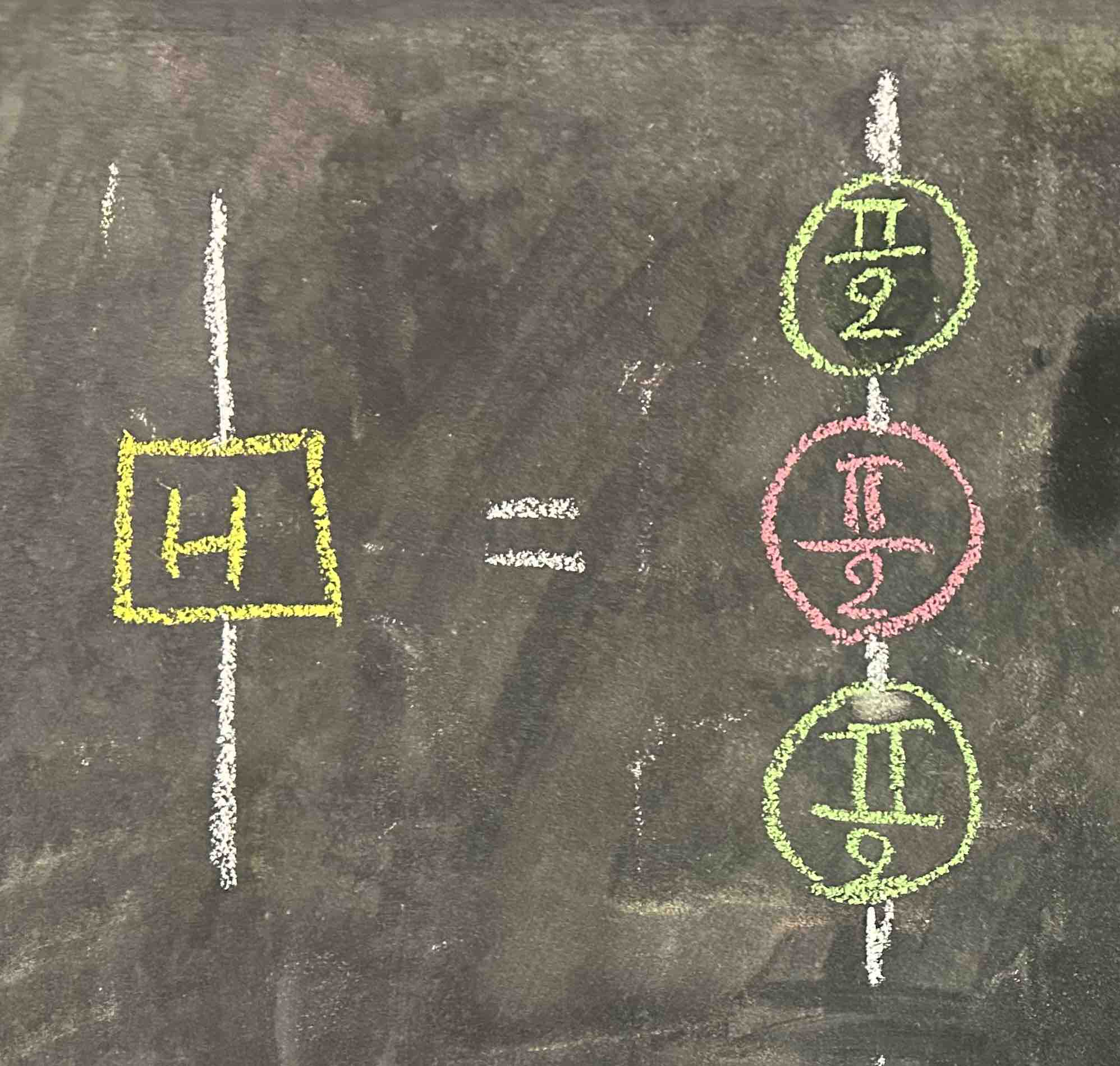,width=140pt} 
\end{center} 
 One extra rule then is called \bM colour-change\e:
\begin{center}
\epsfig{figure=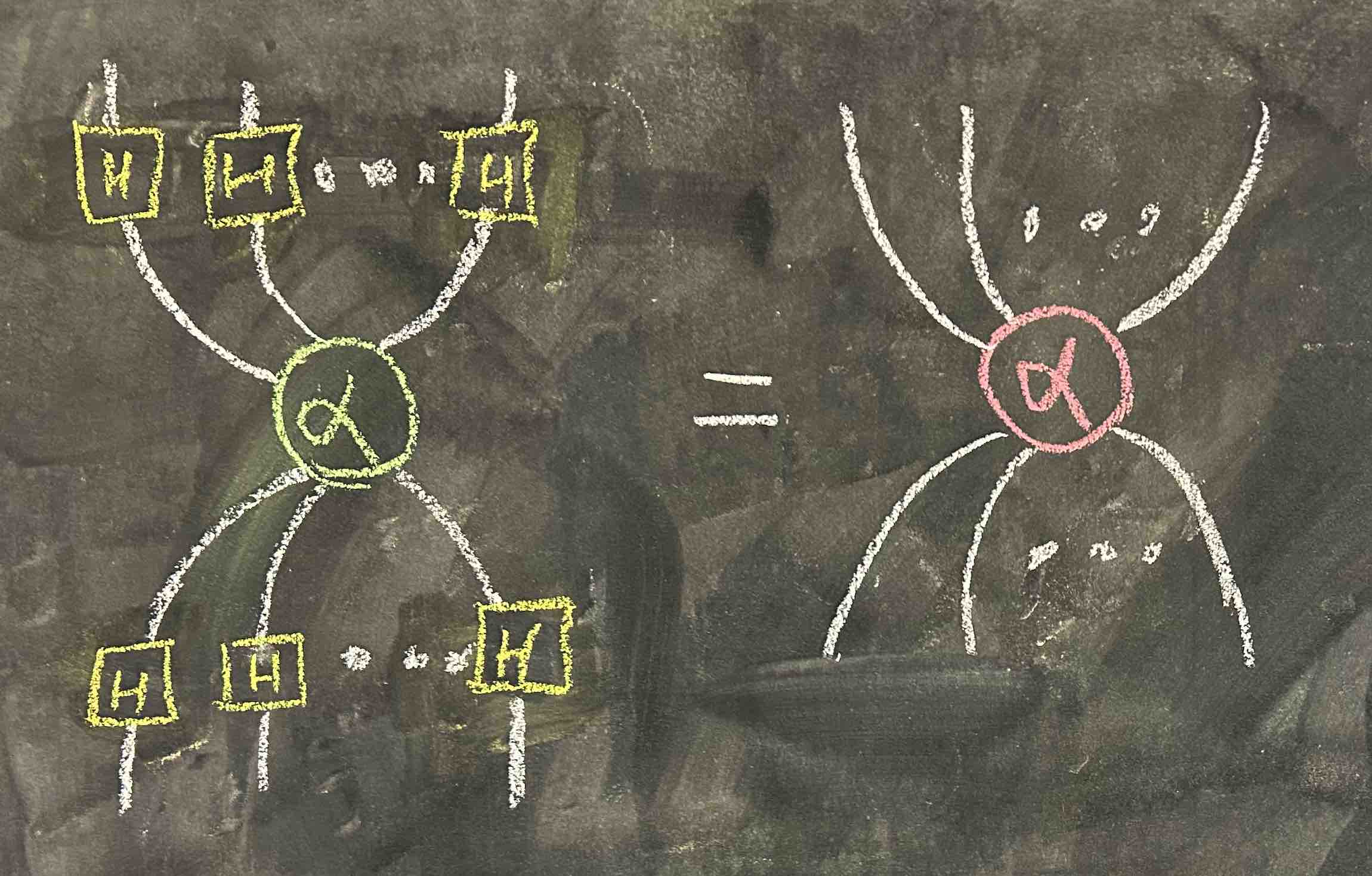,width=220pt} 
\end{center} 
Equivalently, an $H$-box can be copied through a spider provided its colour changes: 
\begin{center}
\epsfig{figure=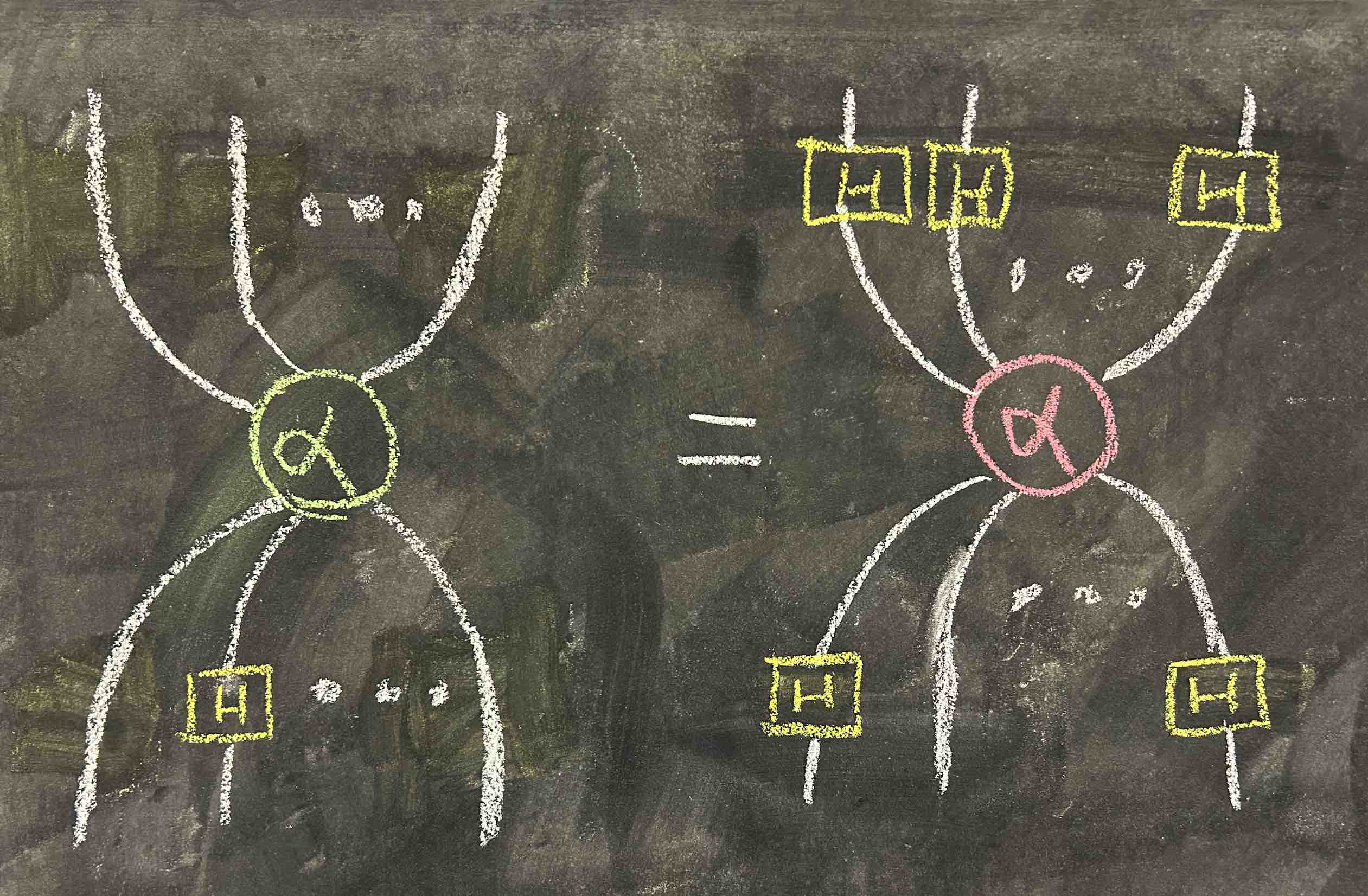,width=220pt}    
\end{center} 

\bigskip\noindent
{\bf Exercise.}  Prove  the colour-change rule.

\subsection*{Phase-colour swap}

A final rule takes the shape:
\begin{center}
\epsfig{figure=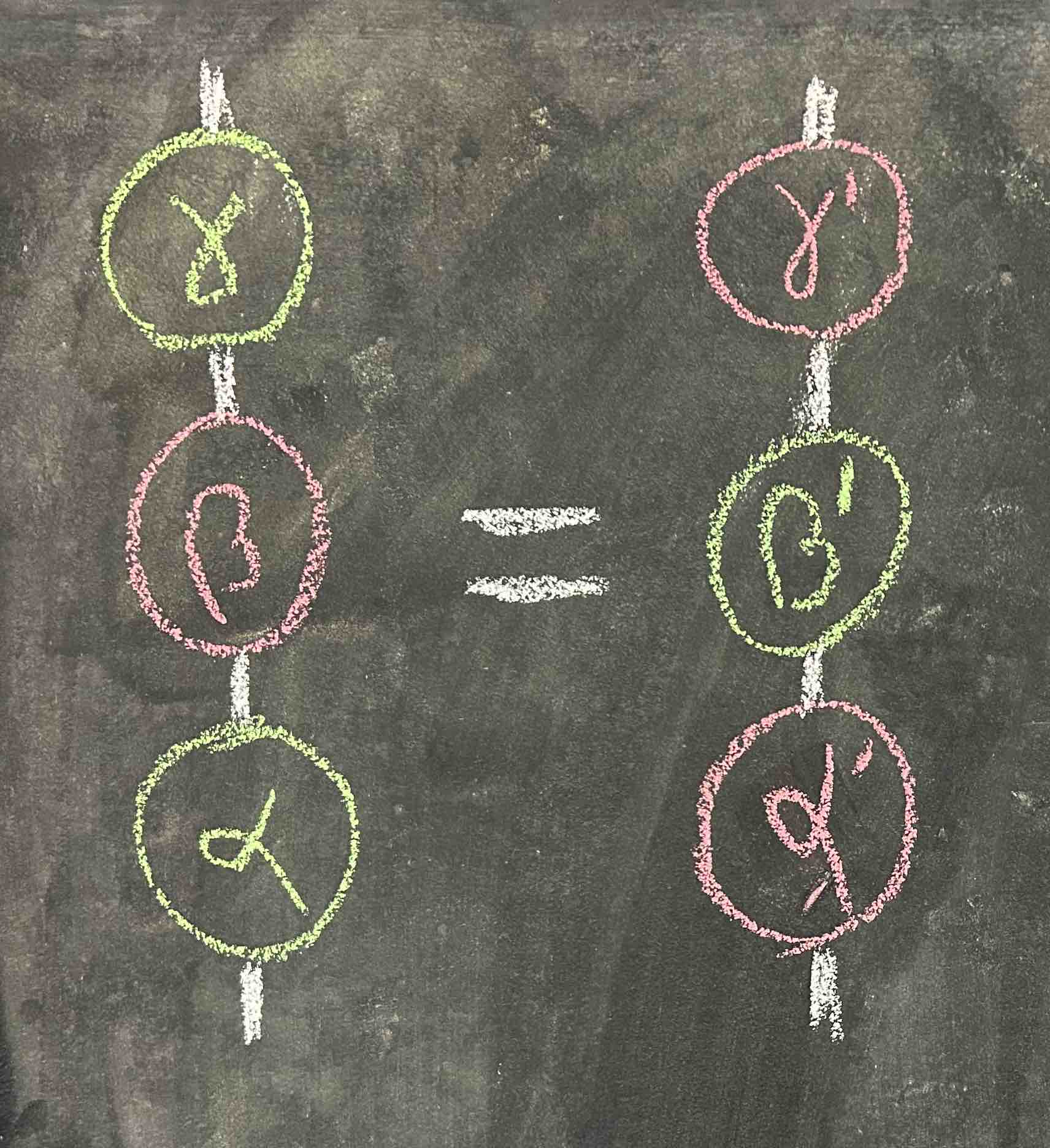,width=120pt} 
\end{center} 
and we call it \bM phase-colour swap\e \cite{DBLP:conf/rc/CoeckeW18, vilmart2018near}, but we won’t specify here the relation between the angles as we won’t use it.  In fact, it is not very often used in practical applications.  
	
\section{Completeness of ZX-calculus}

{\bf Theorem.}  Any equation that holds for linear maps between qubits and compositions thereof can be derived in ZX-calculus.  

\bigskip\noindent
In other words, \bM ZX-calculus is  complete for linear maps\e.  The proof here is  hard, and it took a string of papers to get to this point \cite{Backens, Backens2, Amar, jeandel2018complete, hadzihasanovic2018two, vilmart2018near}.    

\section{Doing stuff with ZX-calculus part I}  

We will now see ZX-calculus in action.  The point of the protocols described below is not to give a detailed discussion of them, but to illustrate how ZX-calculus helps to derive or verify them.  We provide links to the original papers so you can compare descriptions.

\subsection{Stuff with quantum circuits}

We can now use ZX-calculus to simplify this quantum circuit:
\begin{center}
\epsfig{figure=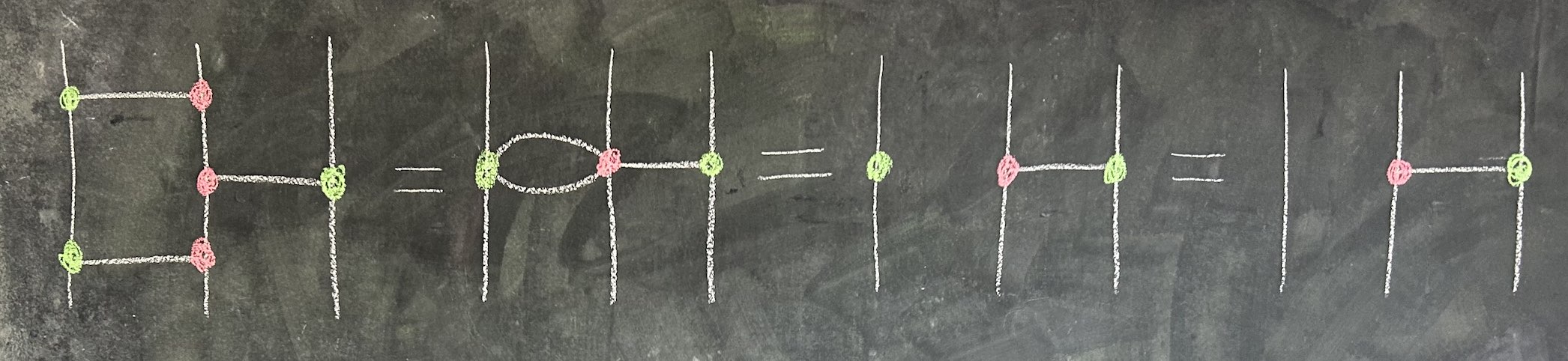,width=460pt} 
\end{center} 
Or this quantum circuit:
\begin{center}
\epsfig{figure=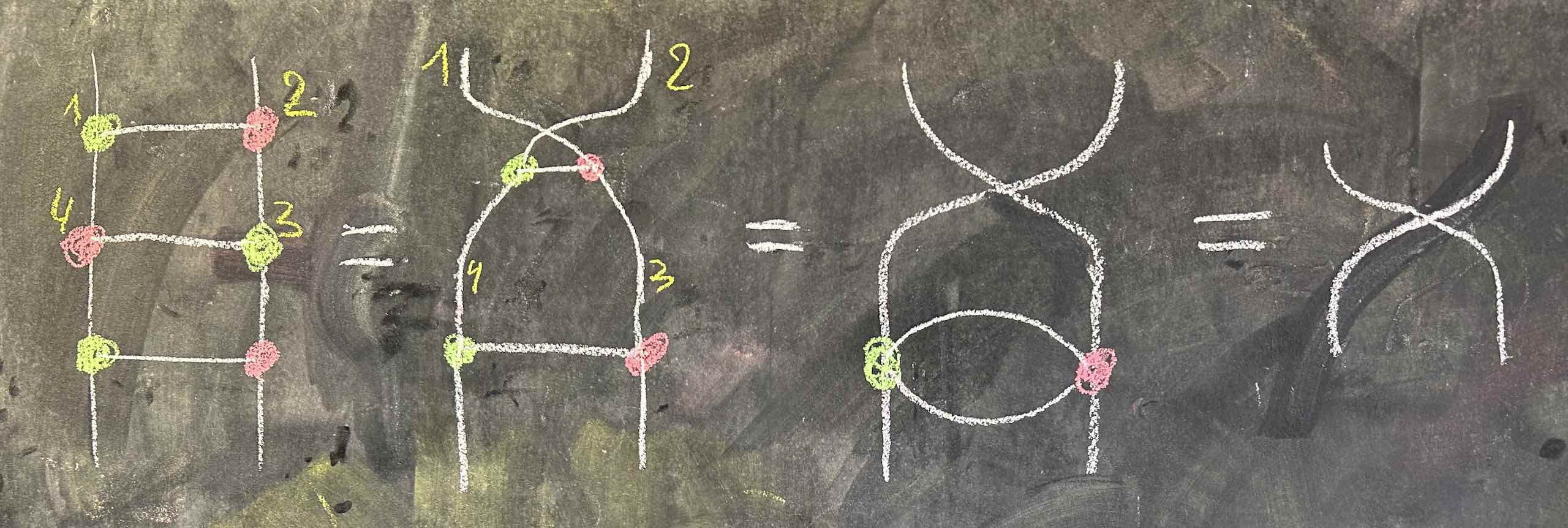,width=360pt} 
\end{center} 
Or we can use it to show that a CNOT-gate indeed does what a CNOT-gate should do:
\begin{center}
\epsfig{figure=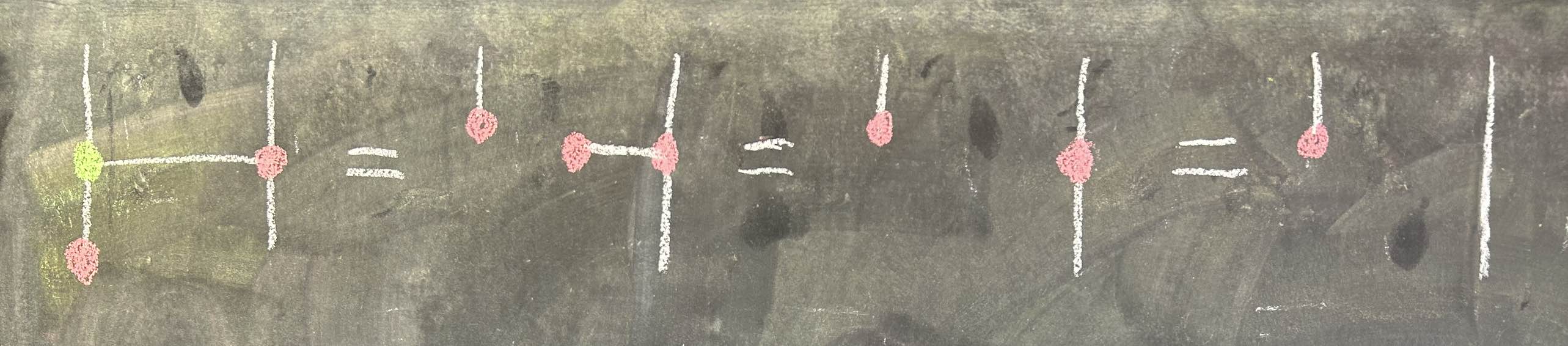,width=360pt} 
\end{center} 
and:
\begin{center}
\epsfig{figure=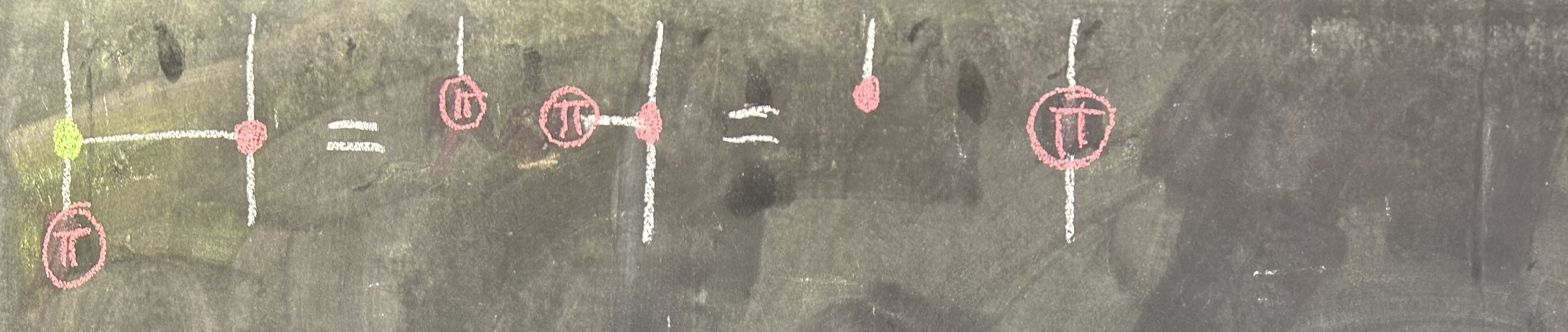,width=360pt} 
\end{center} 
These were very simple examples, and now we will show a generally applicable technique that is used in quantum circuit optimisation, e.g.~within the tket-compiler \cite{cowtan2019phase}.  The following circuit fragment has a `hidden' square and we can simplify it:
\begin{center}
\epsfig{figure=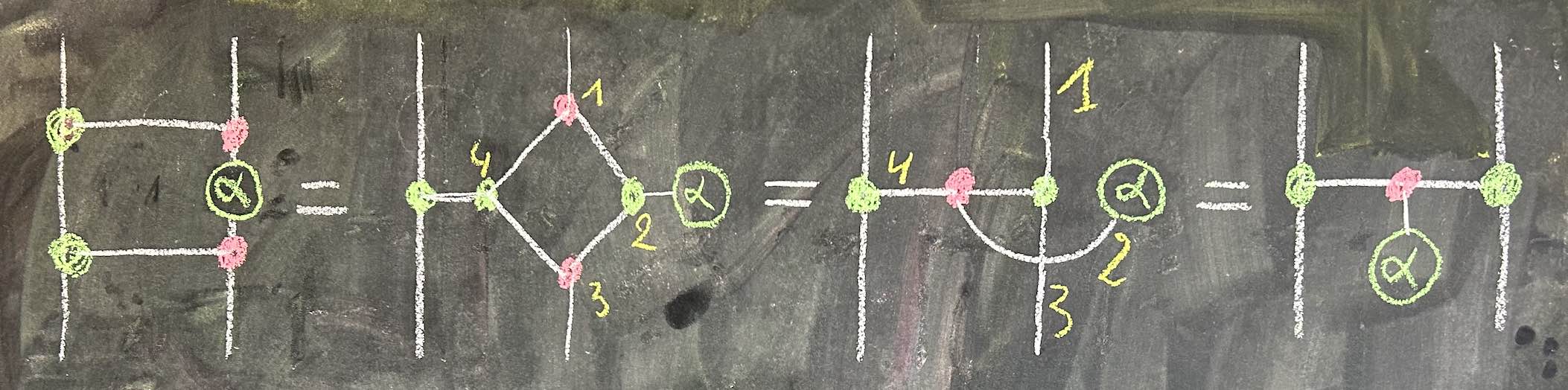,width=400pt} 
\end{center} 
We now only have green spiders on the vertical wires, which allows to, for example, simplify the following circuit:
\begin{center}
\epsfig{figure=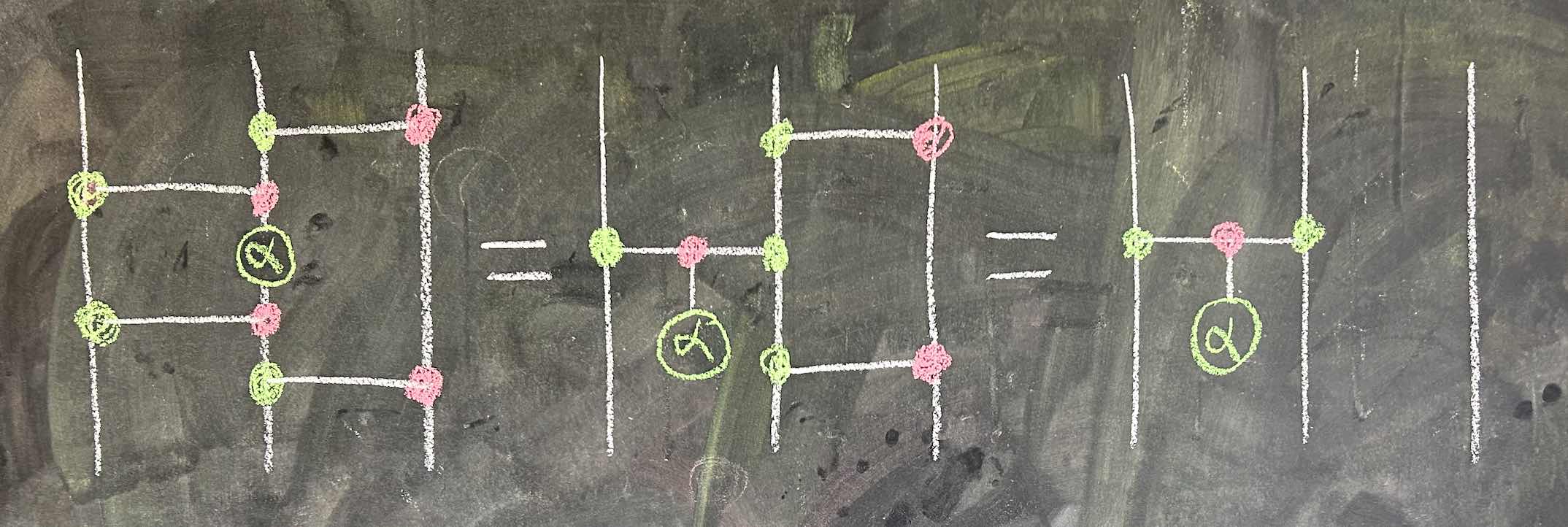,width=400pt}   
\end{center} 
Similar tricks moreover scale up, cf.:
\begin{center}
\epsfig{figure=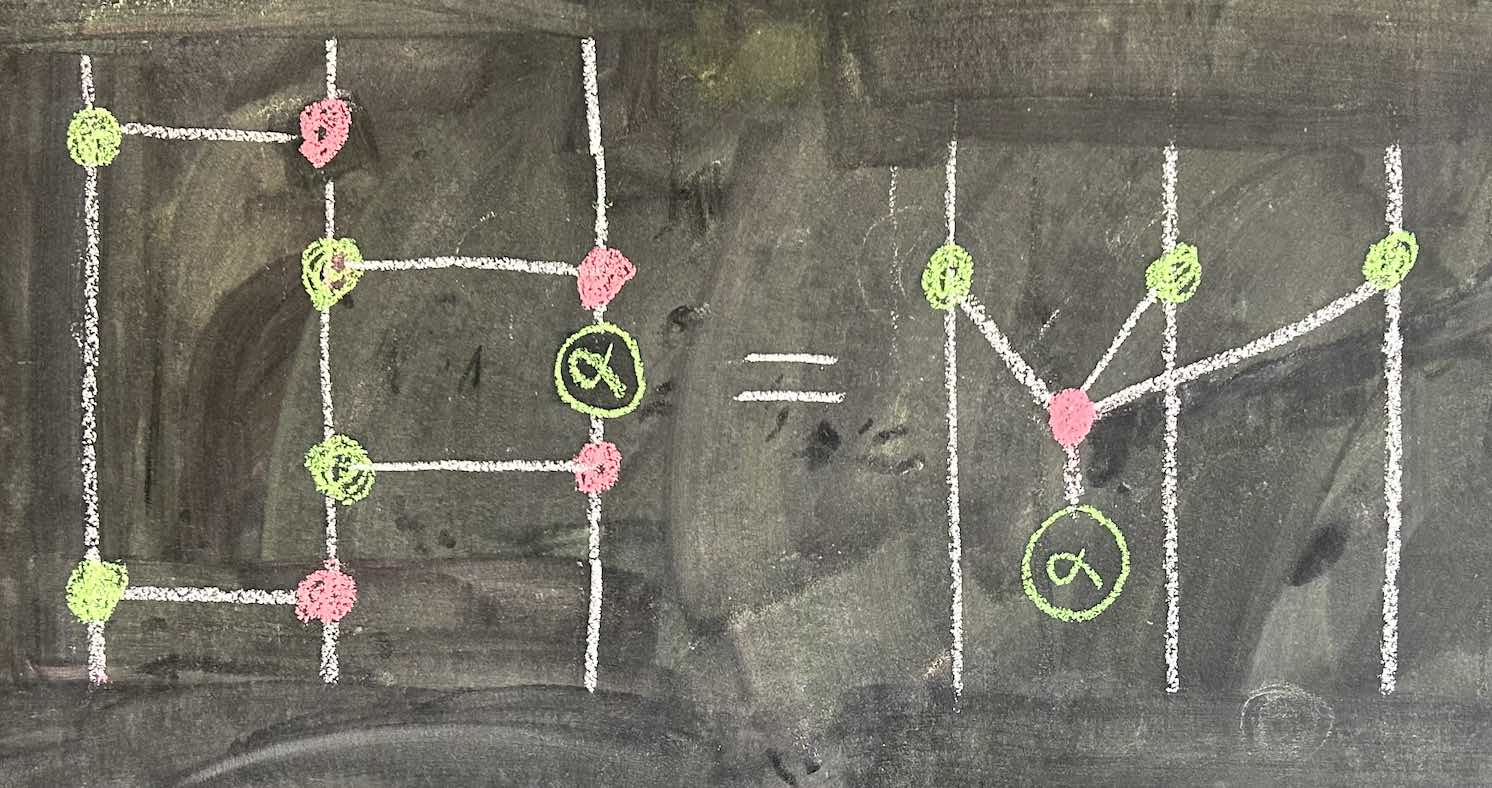,width=235pt}  
\end{center} 
and:
\begin{center}
\epsfig{figure=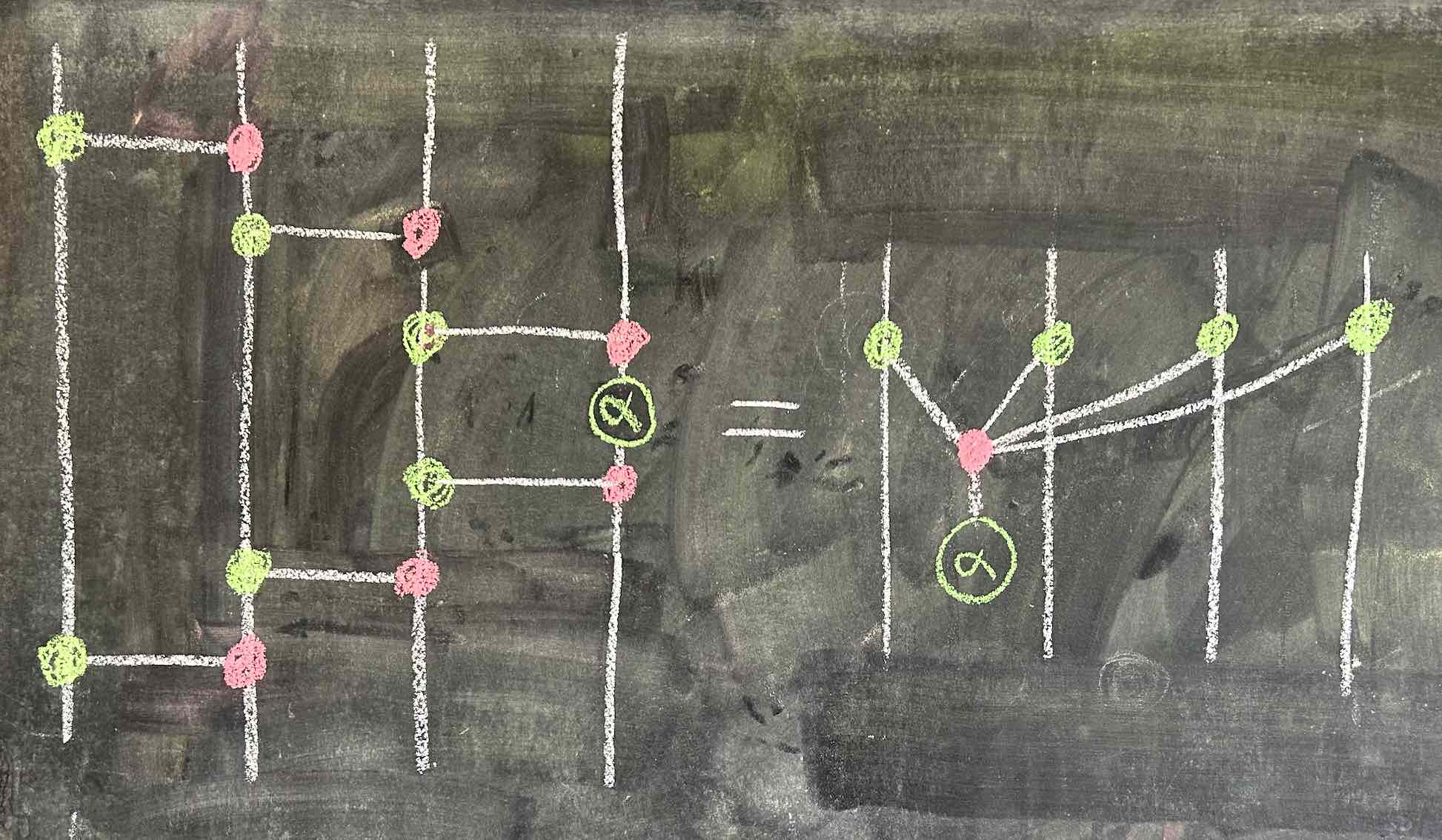,width=310pt} 
\end{center} 
and so on.

\subsection{Quantum teleportation}

We can represent the Bell-basis as follows:
\begin{center}
\epsfig{figure=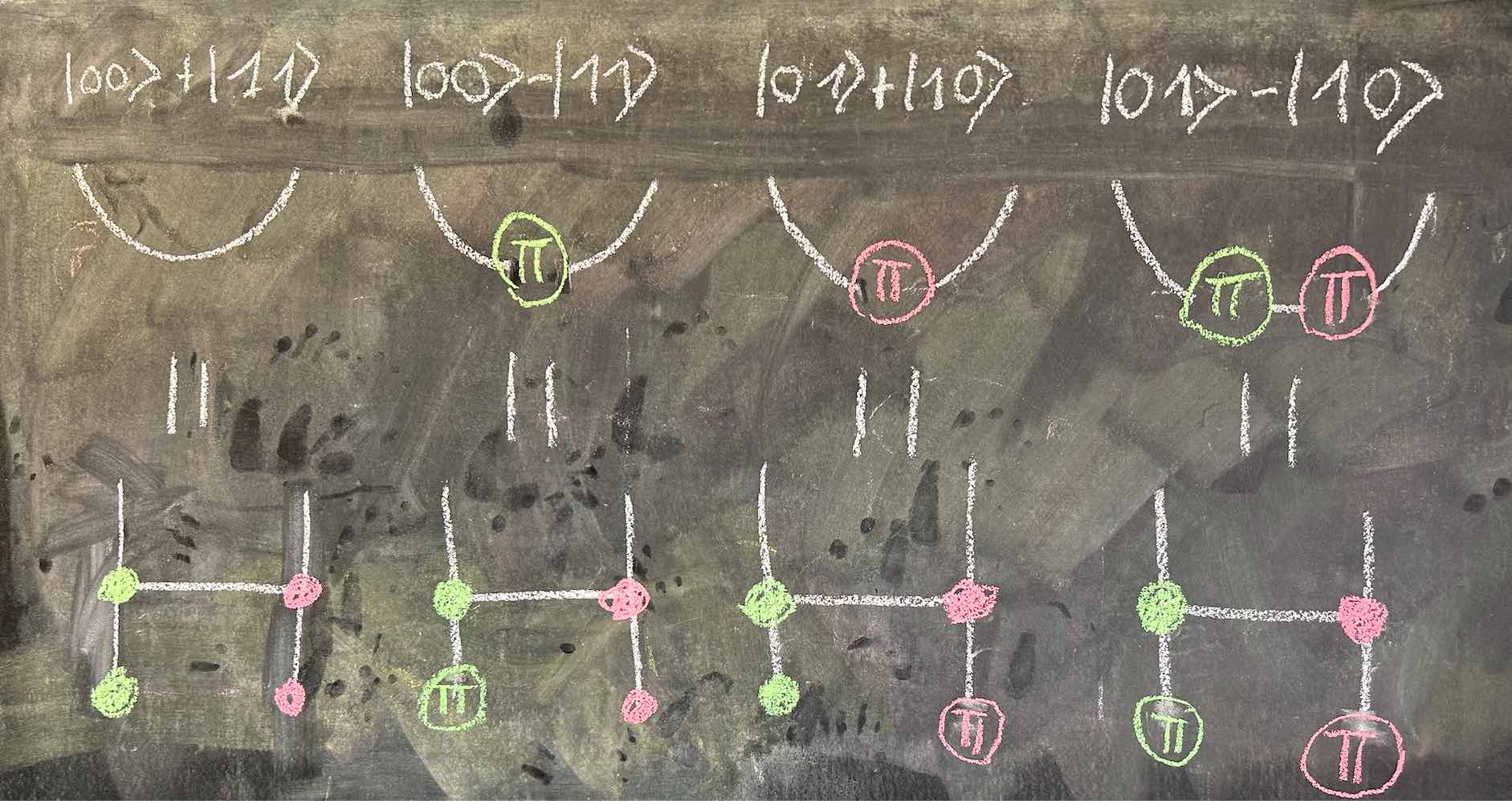,width=280pt} 
\end{center} 

\bigskip\noindent
{\bf Exercise.}  Verify that this is true.  
 
\bigskip\noindent
Now, suppose Alice and Bob share a Bell-state,  Alice  has a qubit in state $|\psi\rangle$, and Bob needs that state.  How can that state be passed on without effectively sending the qubit:
\begin{center}
\epsfig{figure=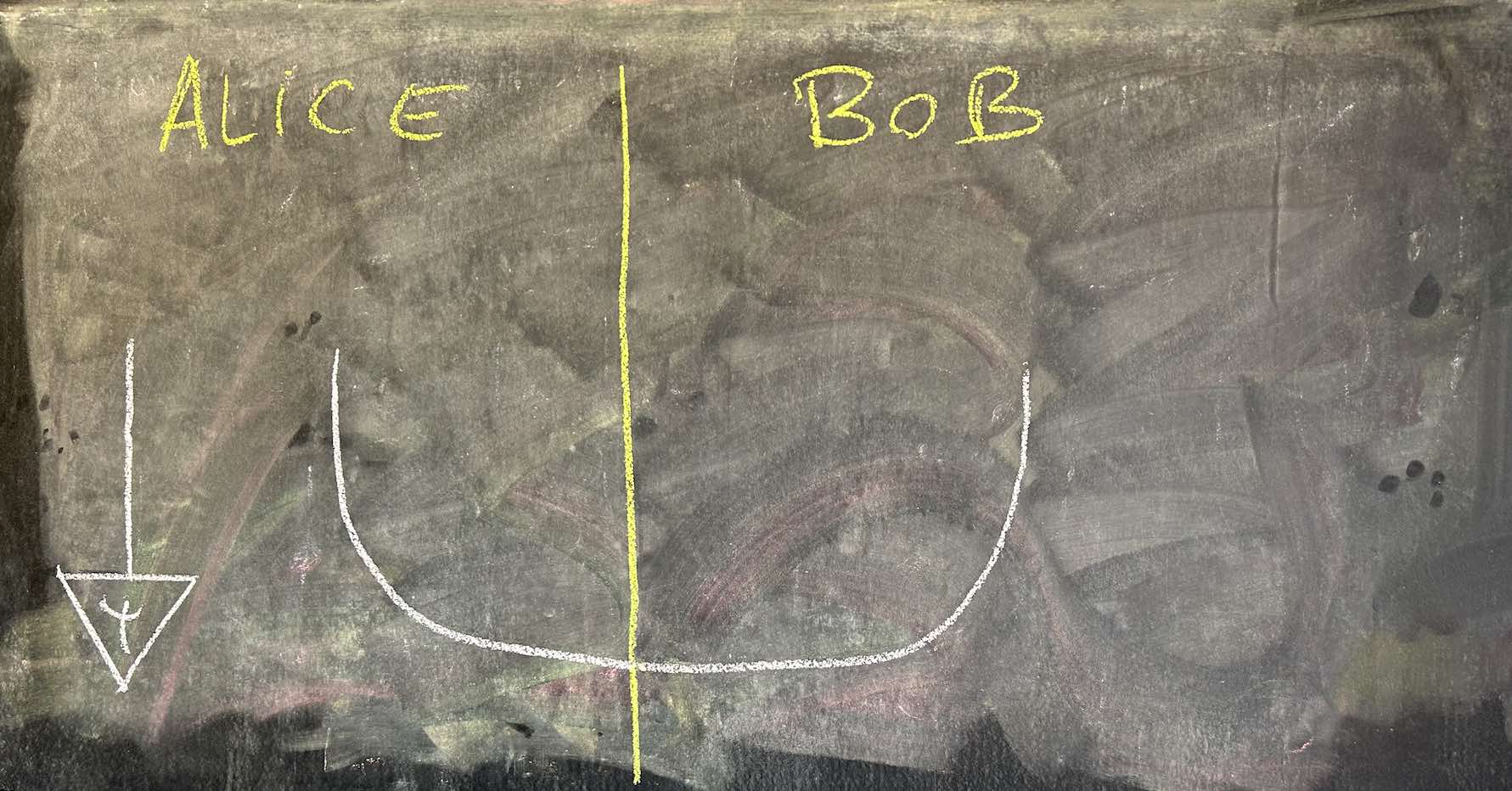,width=260pt} 
\end{center} 
An obvious solution now presents itself: Alice applies a Bell-effect:  
\begin{center}
\epsfig{figure=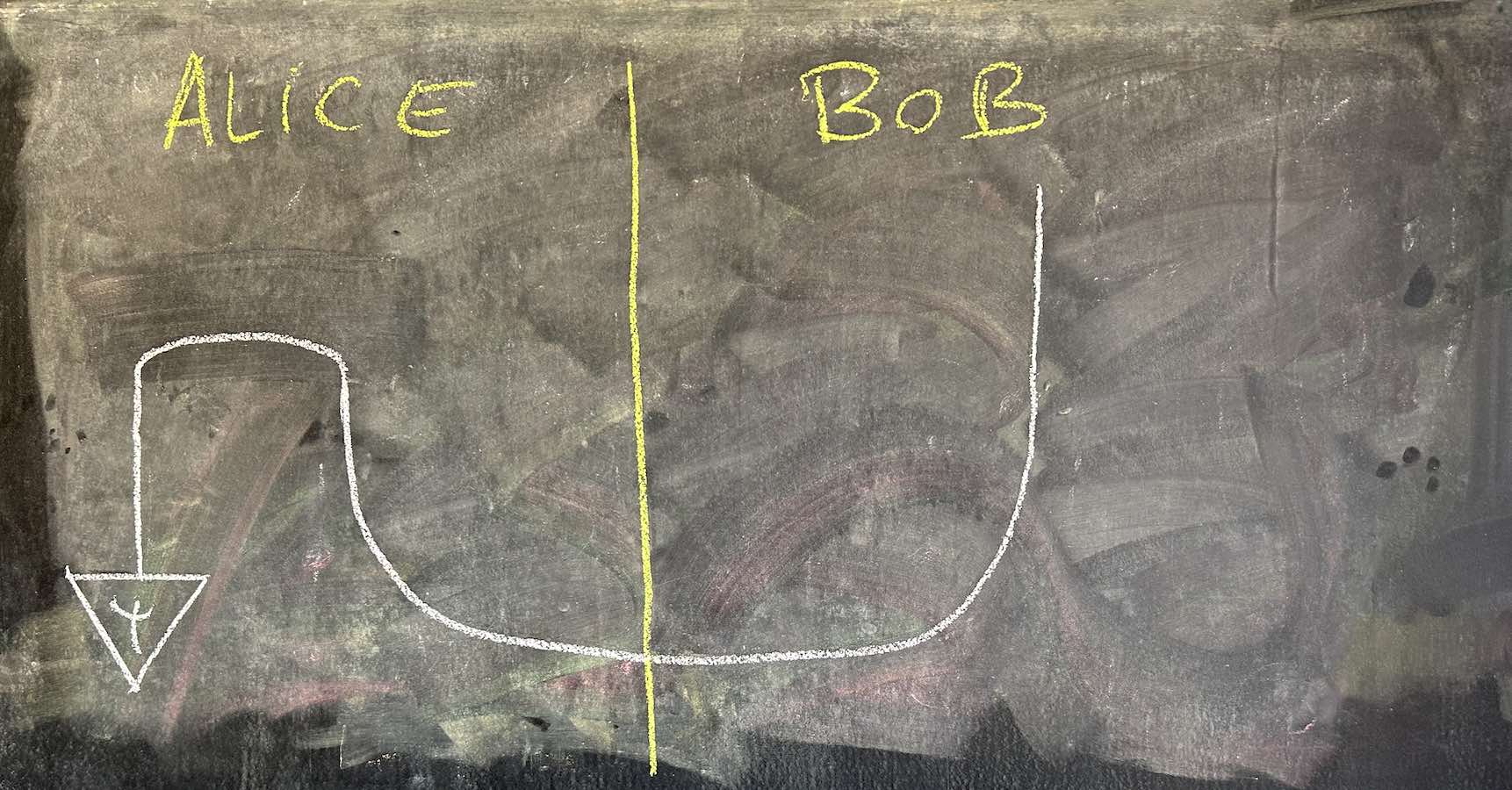,width=260pt} 
\end{center} 
We now indeed can slide the state along the wire:
\begin{center}
\epsfig{figure=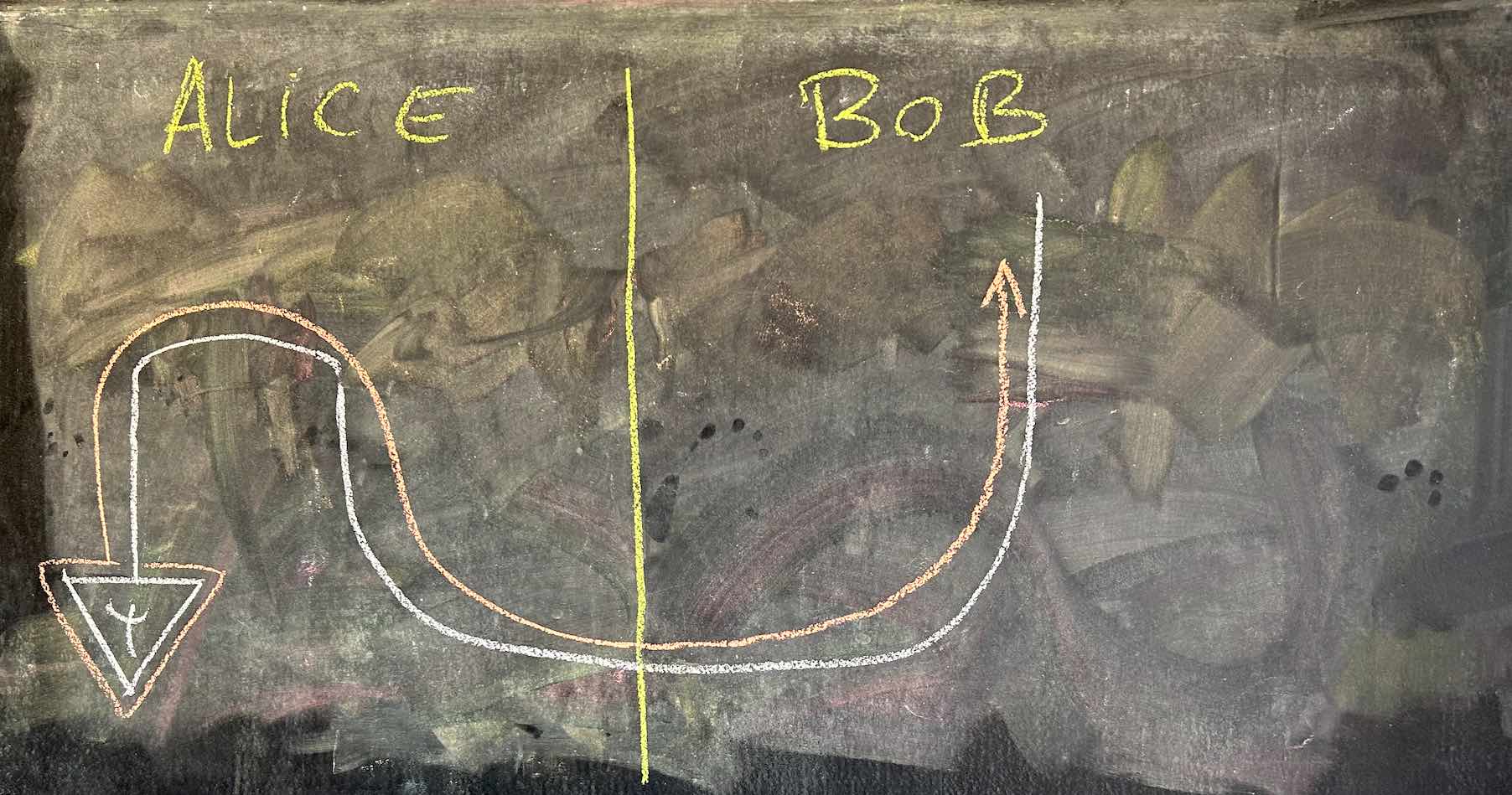,width=260pt} 
\end{center} 
so that it ends up with Bob:
\begin{center}
\epsfig{figure=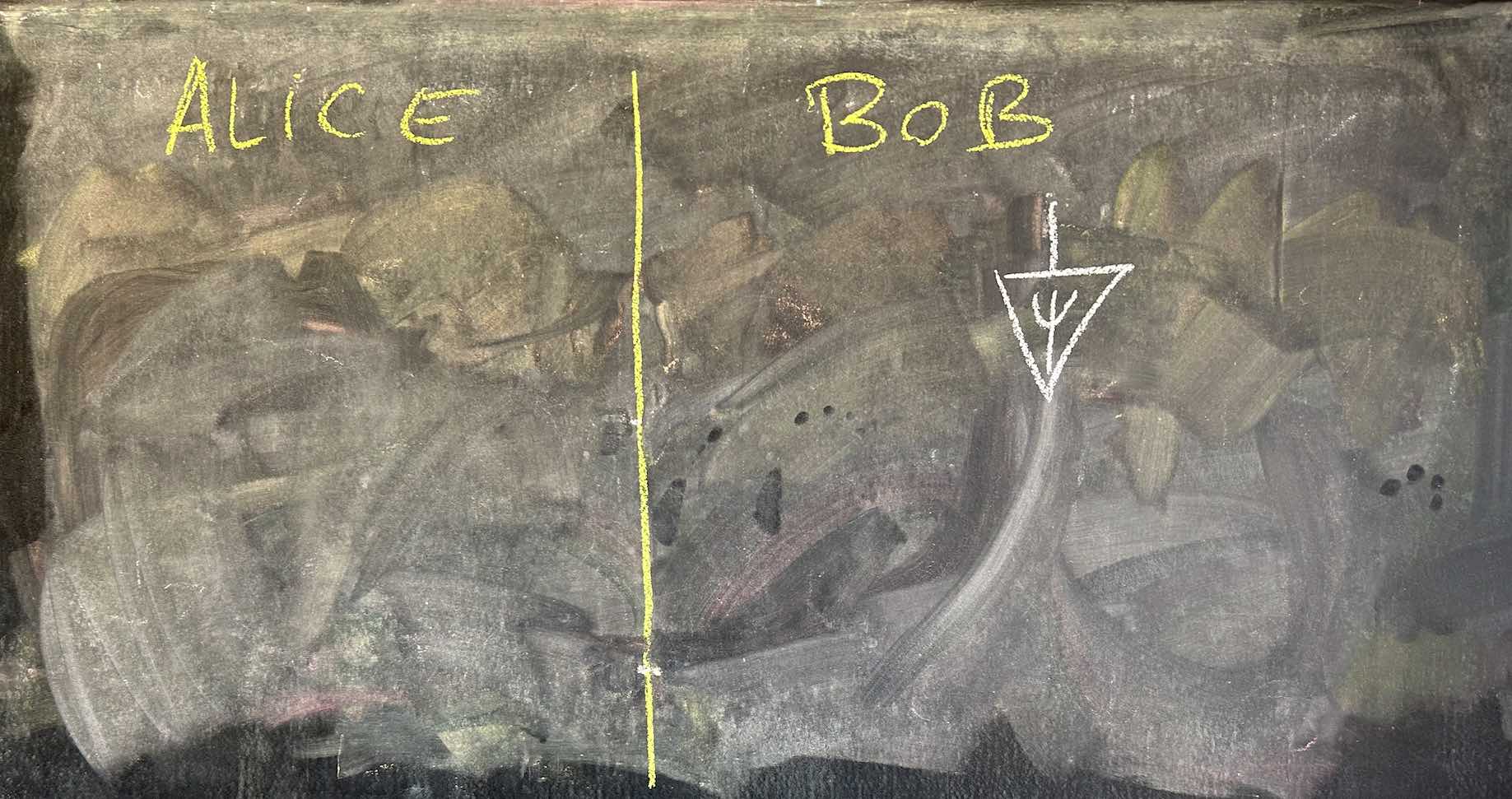,width=260pt}  
\end{center} 
But we can only achieve a Bell-effect as part of a measurement against the Bell basis, which would yield one of the four Bell-effects:
\begin{center}
\epsfig{figure=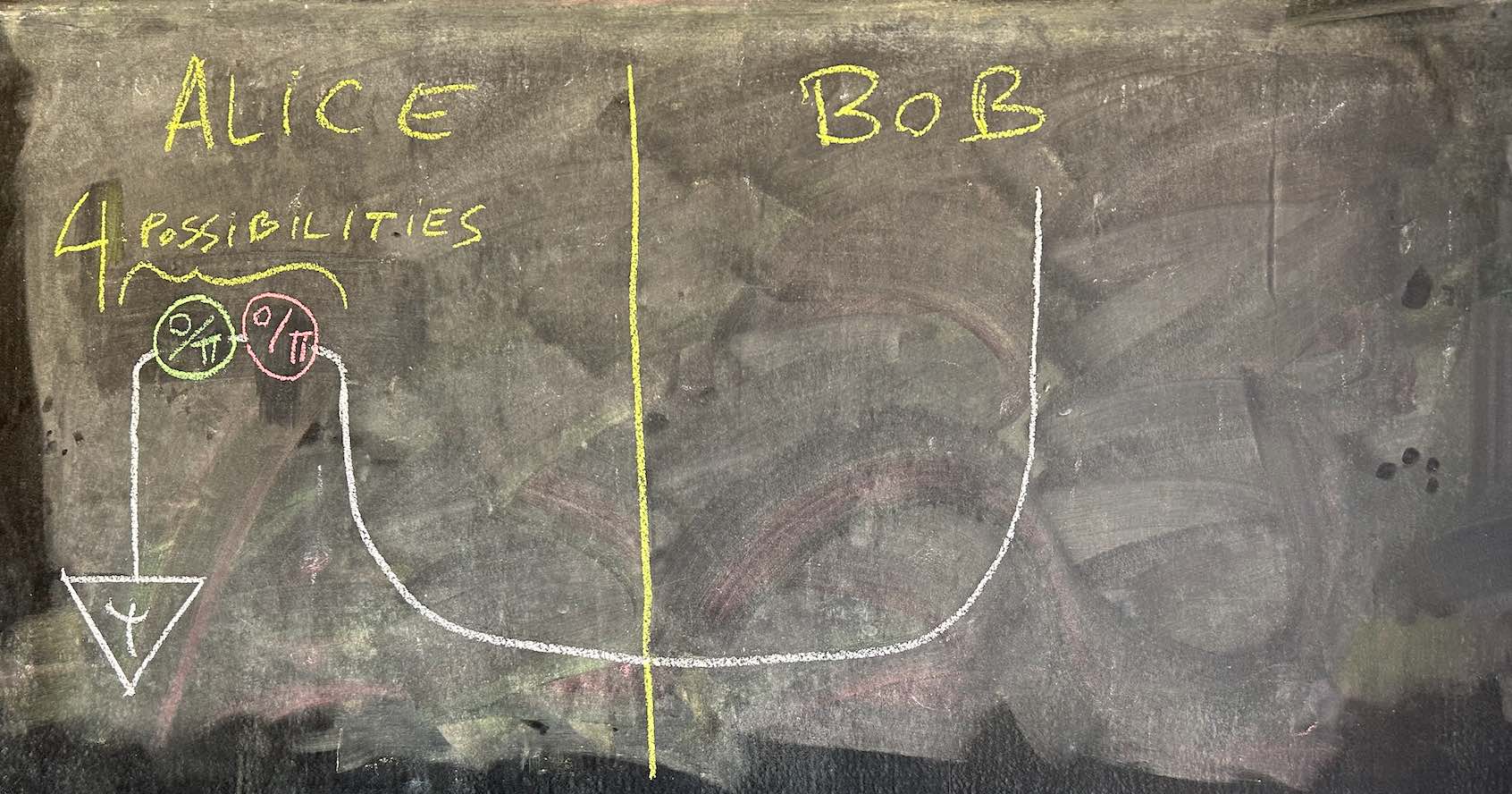,width=320pt} 
\end{center} 
Fortunately, we can undo the corresponding unitaries at Bob's end:
\begin{center}
\epsfig{figure=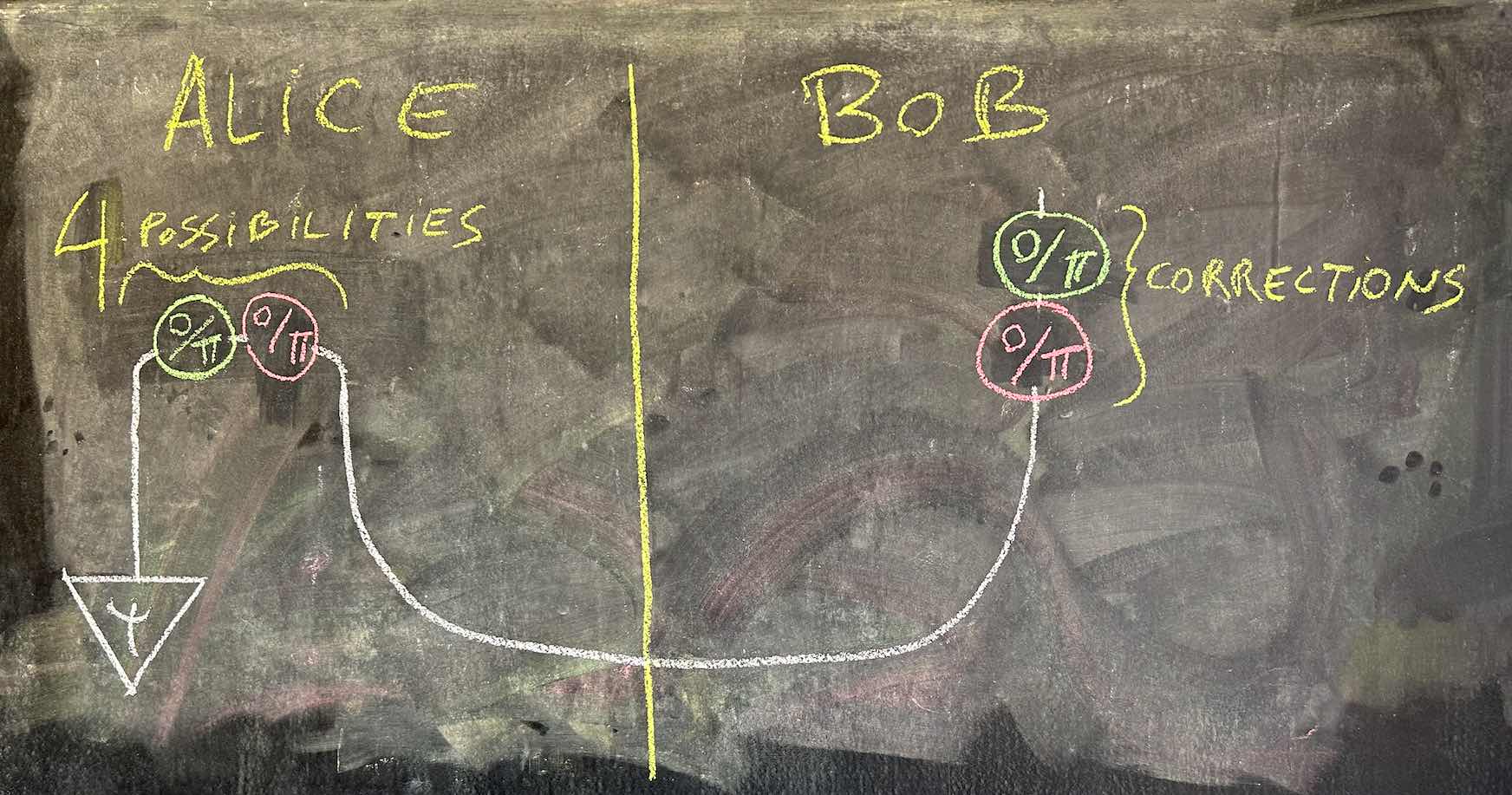,width=320pt} 
\end{center} 
which requires communicating two bits from Alice to Bob:
\begin{center}
\epsfig{figure=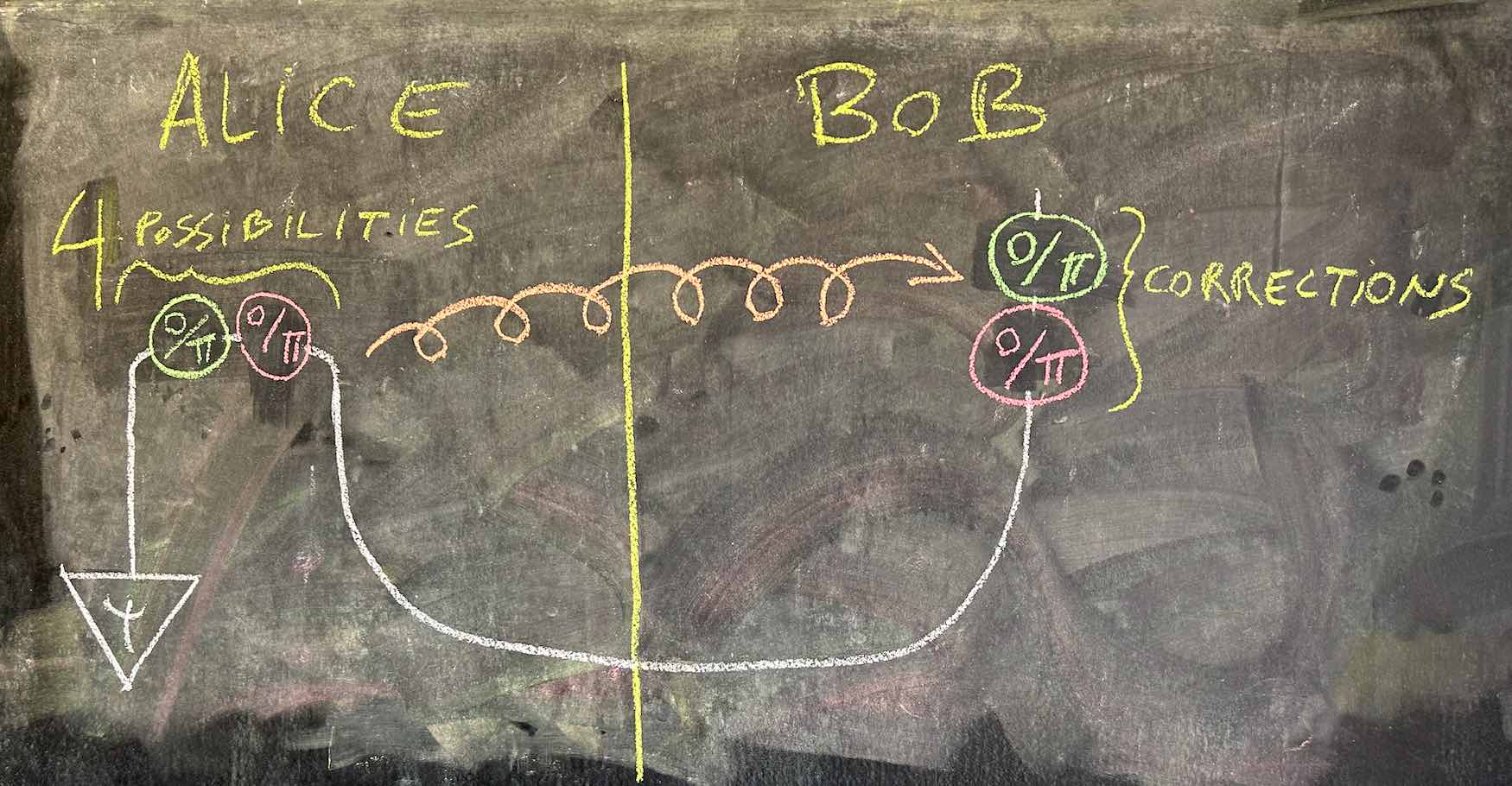,width=320pt} 
\end{center} 
The original paper on quantum teleportation is \cite{Tele}.     

\subsection{Measurement-based quantum computing}
 
We can also pass a state as follows, which now only involves two qubits:
\begin{center}
\epsfig{figure=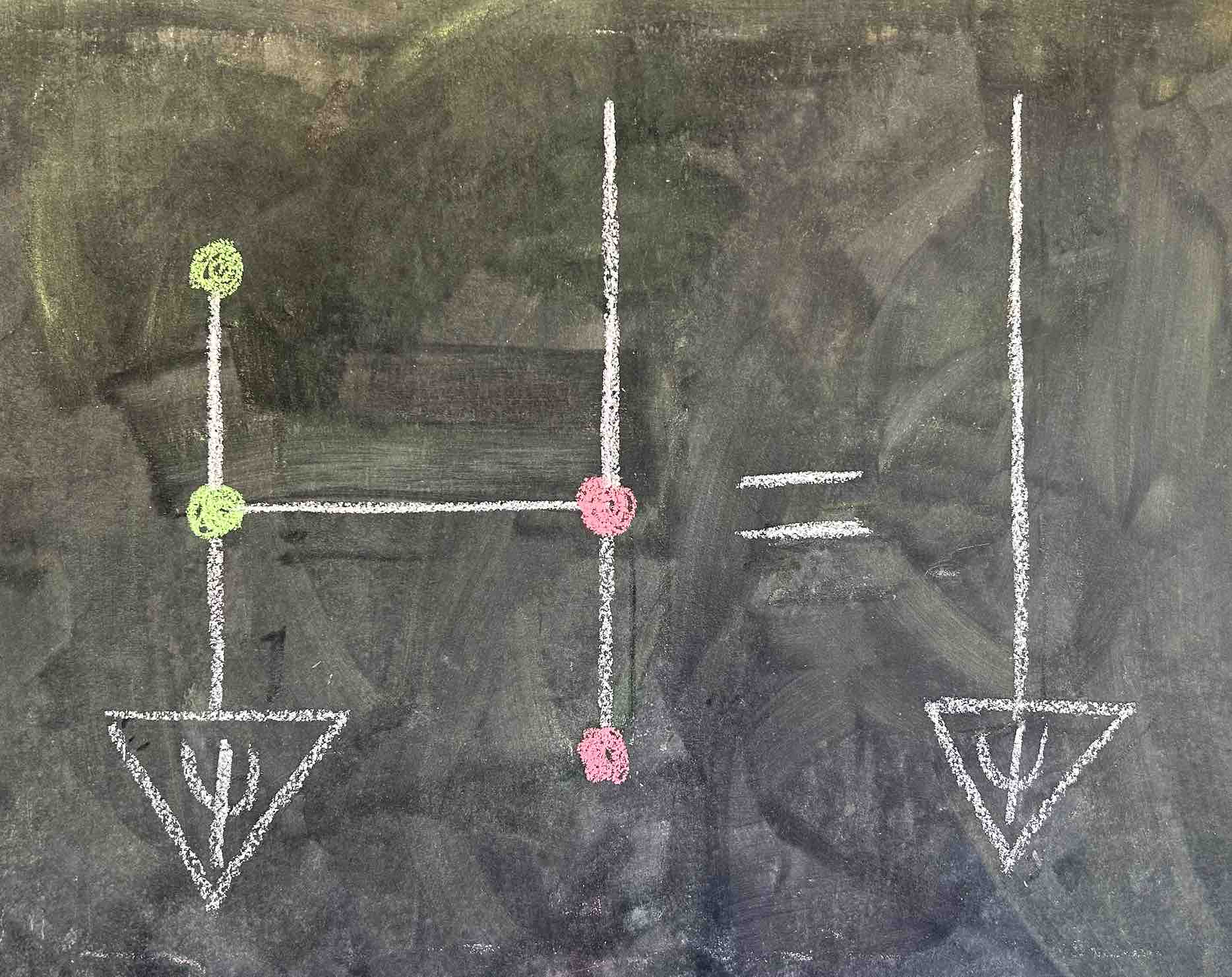,width=180pt} 
\end{center} 
If instead of an X-measurement, we use a measurement corresponding to a basis with a phase, then that phase is applied to the state as a phase gate:
\begin{center}
\epsfig{figure=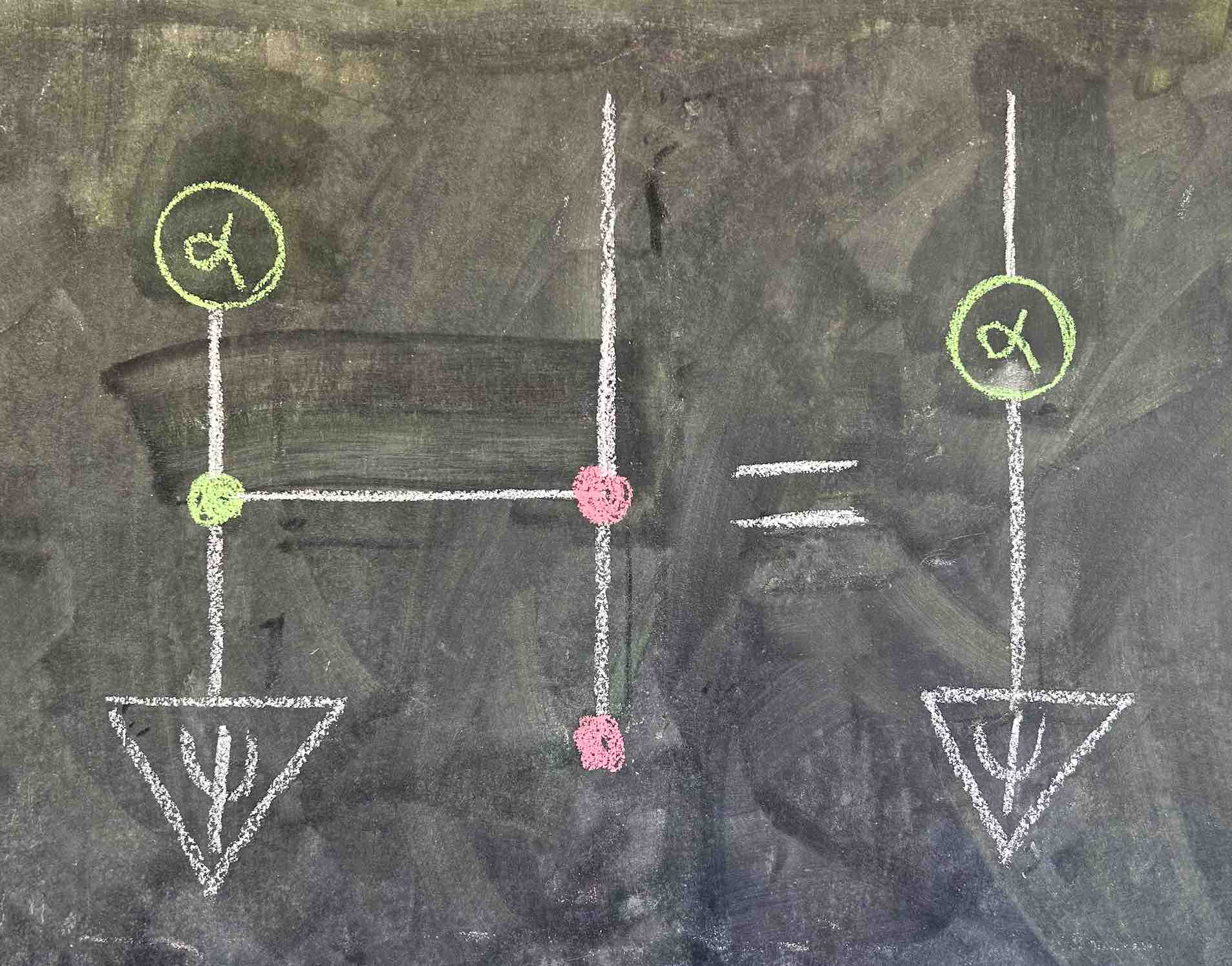,width=180pt} 
\end{center} 
Again accounting for two possible measurement outcomes, we can do:
\begin{center}
\epsfig{figure=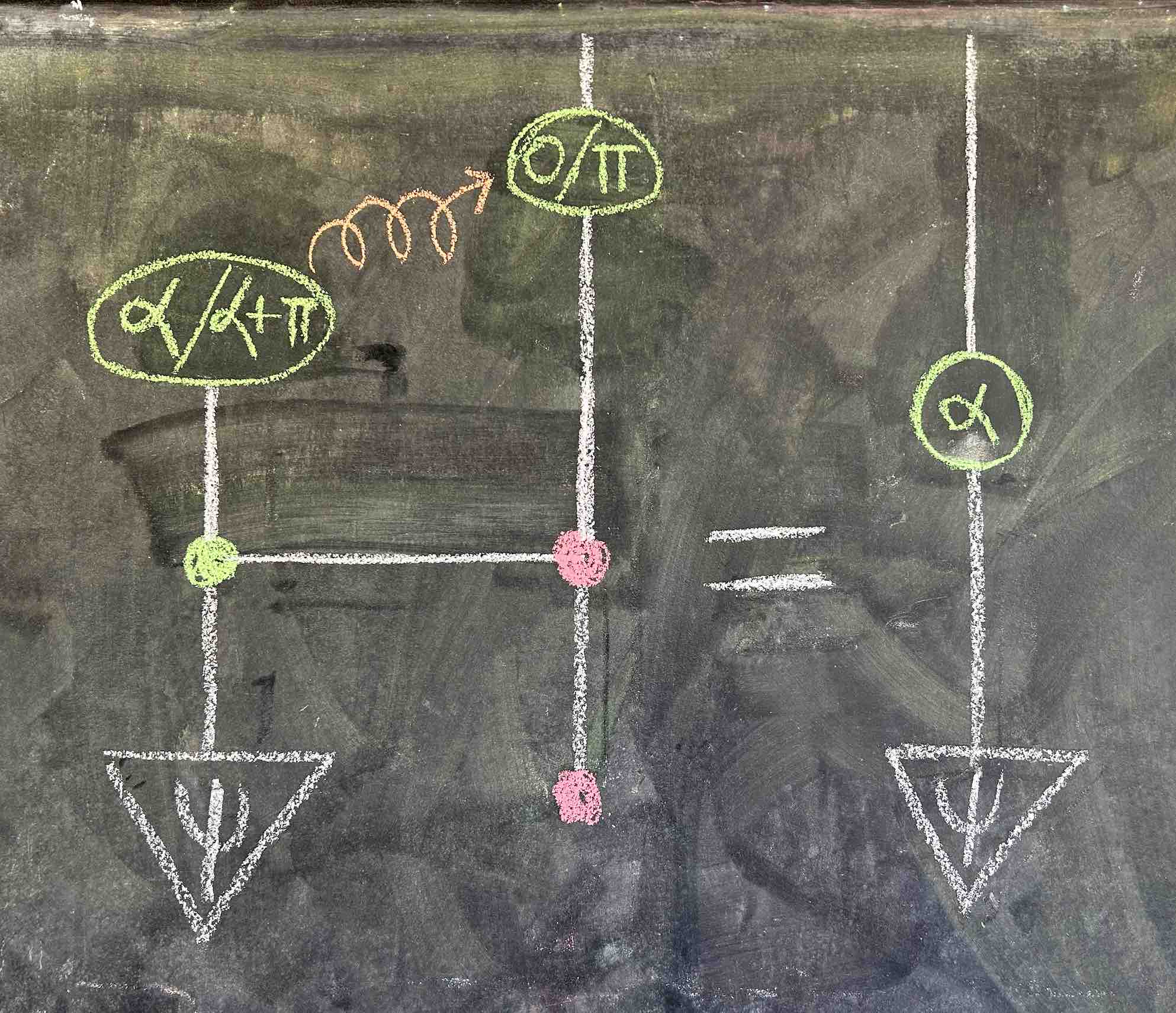,width=180pt}  
\end{center} 
and the gate is now applied for every measurement outcome.  This manner of applying   a phase gate is called \bM measurement-based quantum computing\e.  Using the following gate:
\begin{center}
\mbox{\bM CZ-gate\e} = \raisebox{-35pt}{\epsfig{figure=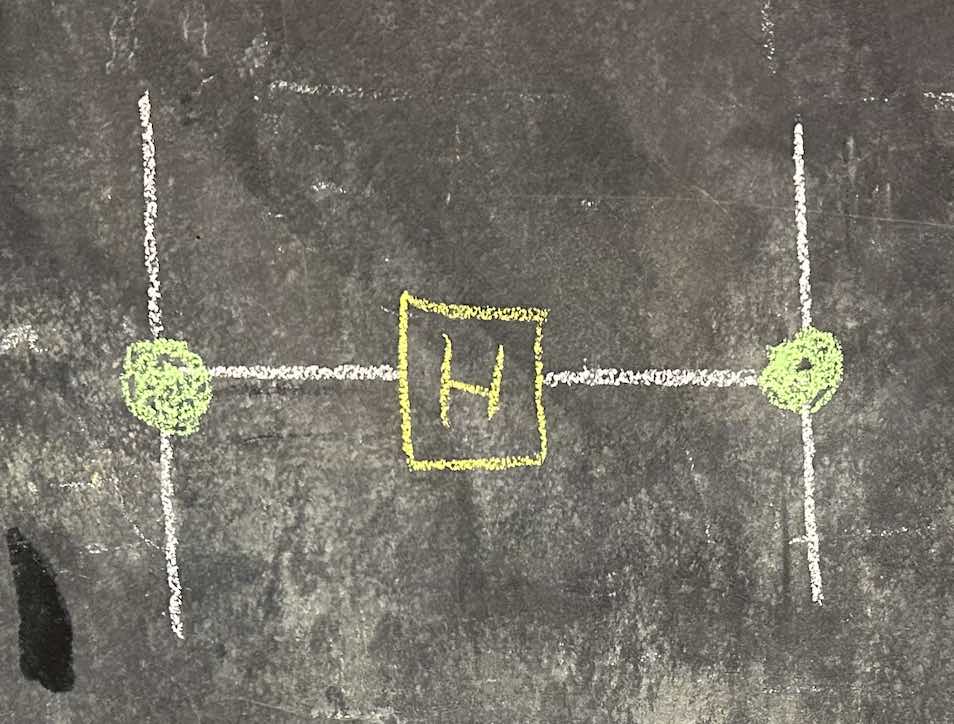,width=100pt}} 
\end{center} 
we can make the following state:
\begin{center}
\epsfig{figure=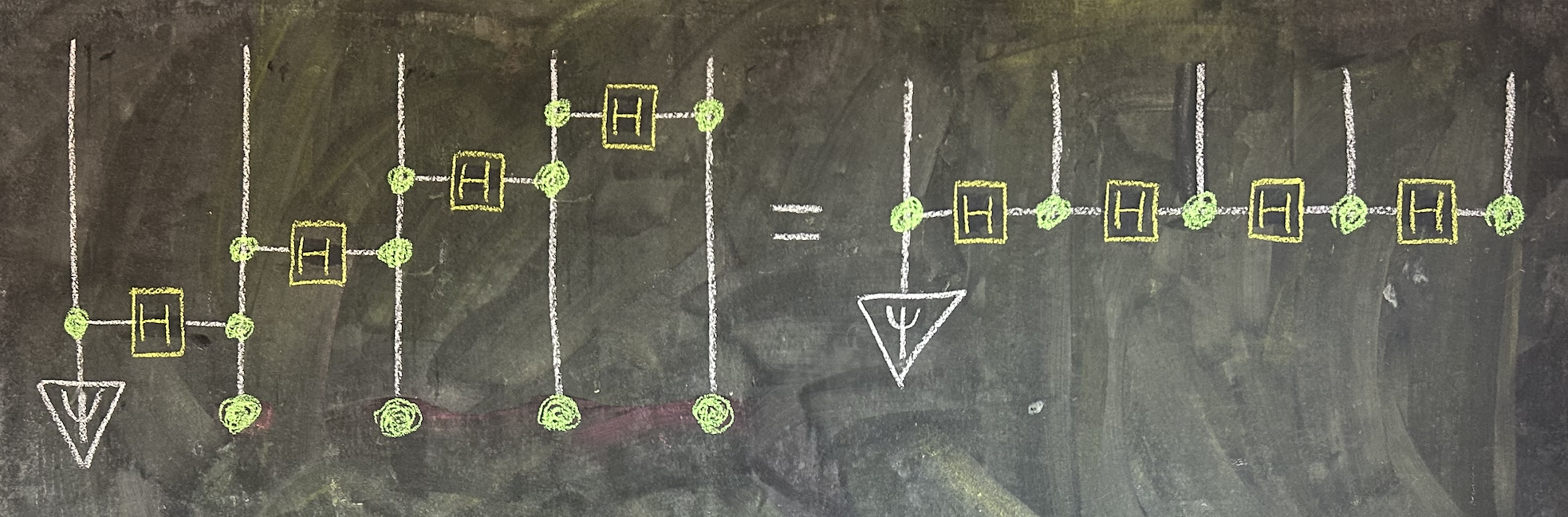,width=400pt} 
\end{center} 
Now, using measurement-based quantum computing we obtain: 
\begin{center}
\epsfig{figure=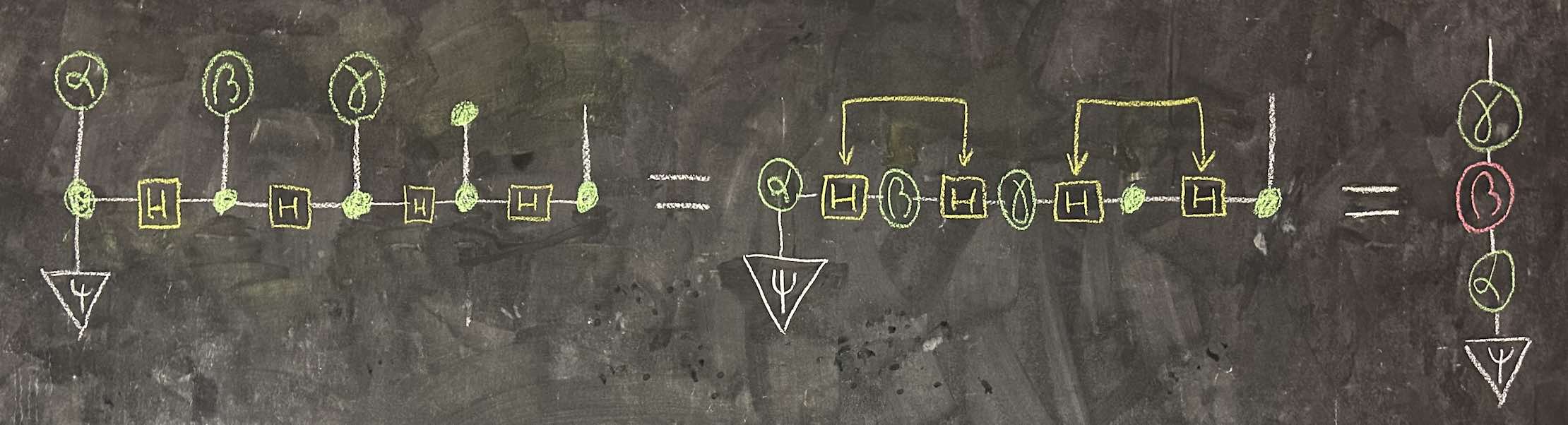,width=450pt} 
\end{center} 
that is, we have applied an arbitrary unitary in terms of its Euler angles.  The two MBQC protocols presented here were first introduced in \cite{perdrix2005} and \cite{MBQC2} respectively.

\section{Classical versus quantum in diagrams}

In order to distinguish between classical and quantum diagrams, we draw classical diagrams as single diagrams, and \bM quantum diagrams as doubled diagrams\e:
\begin{center}
\epsfig{figure=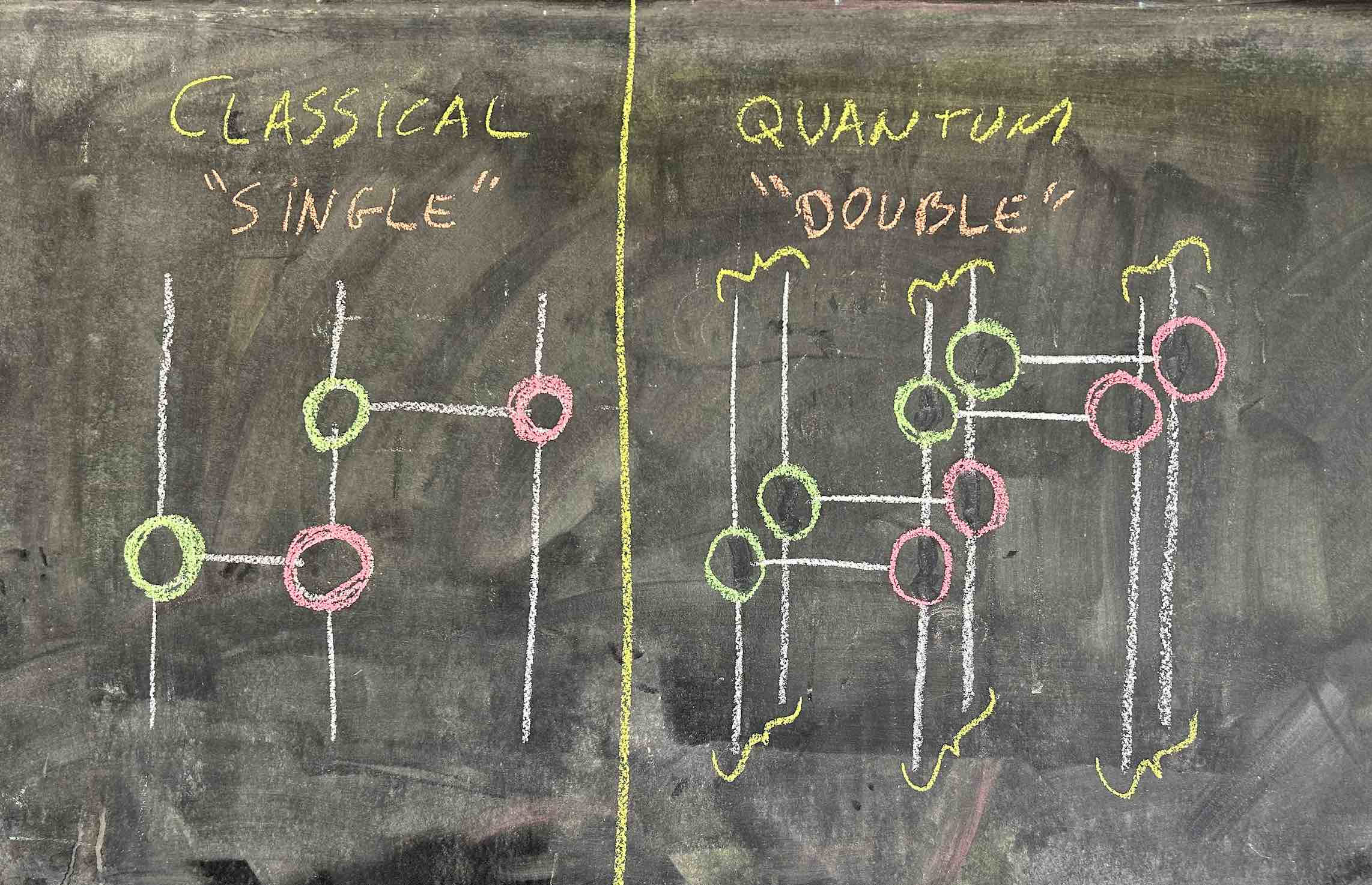,width=300pt} 
\end{center} 
If there are phases, then the double needs a minus:
\begin{center}
\epsfig{figure=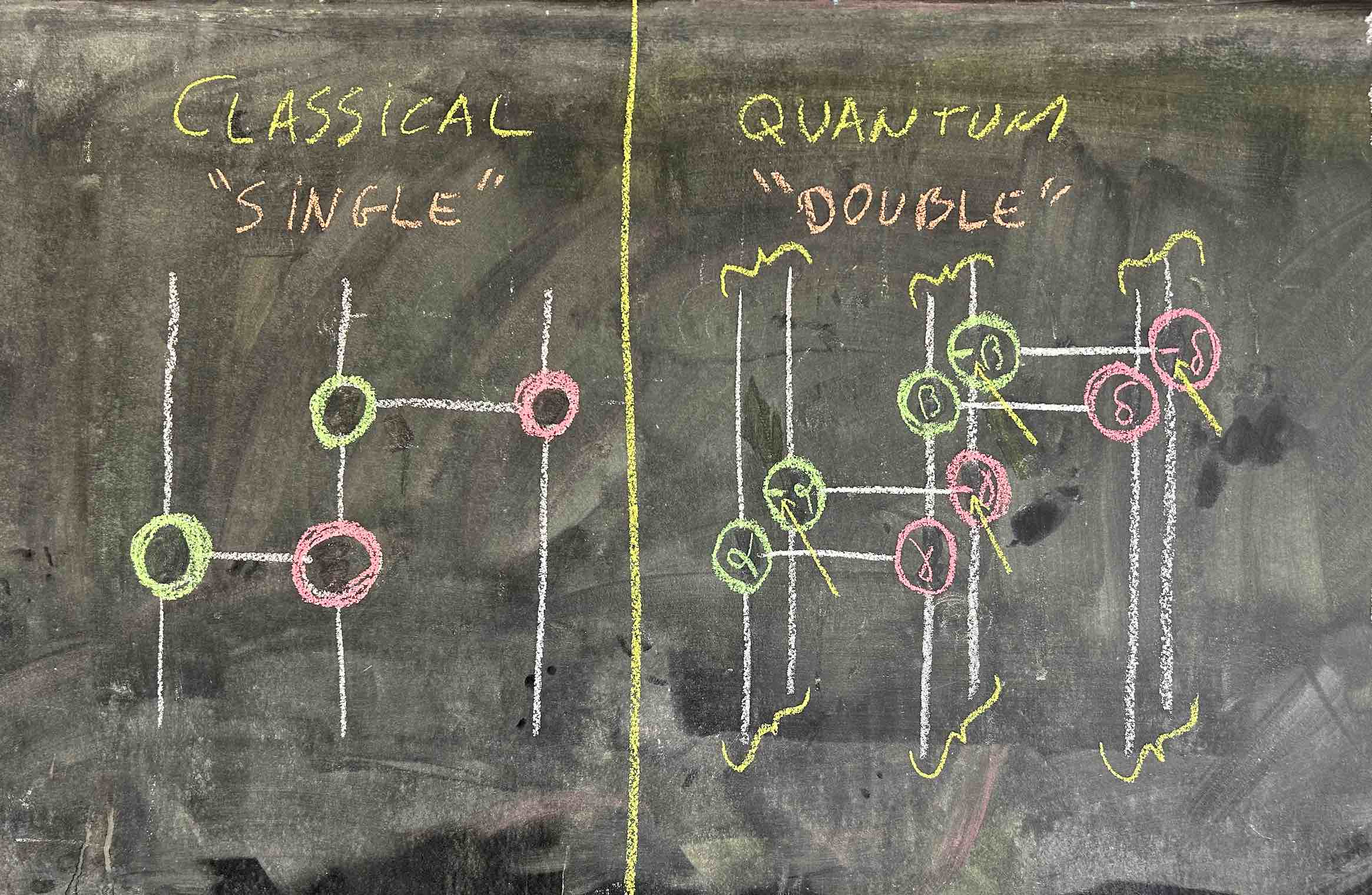,width=300pt} 
\end{center} 
To make drawing easy we just double wires and spider-nodes:
\begin{center}
\epsfig{figure=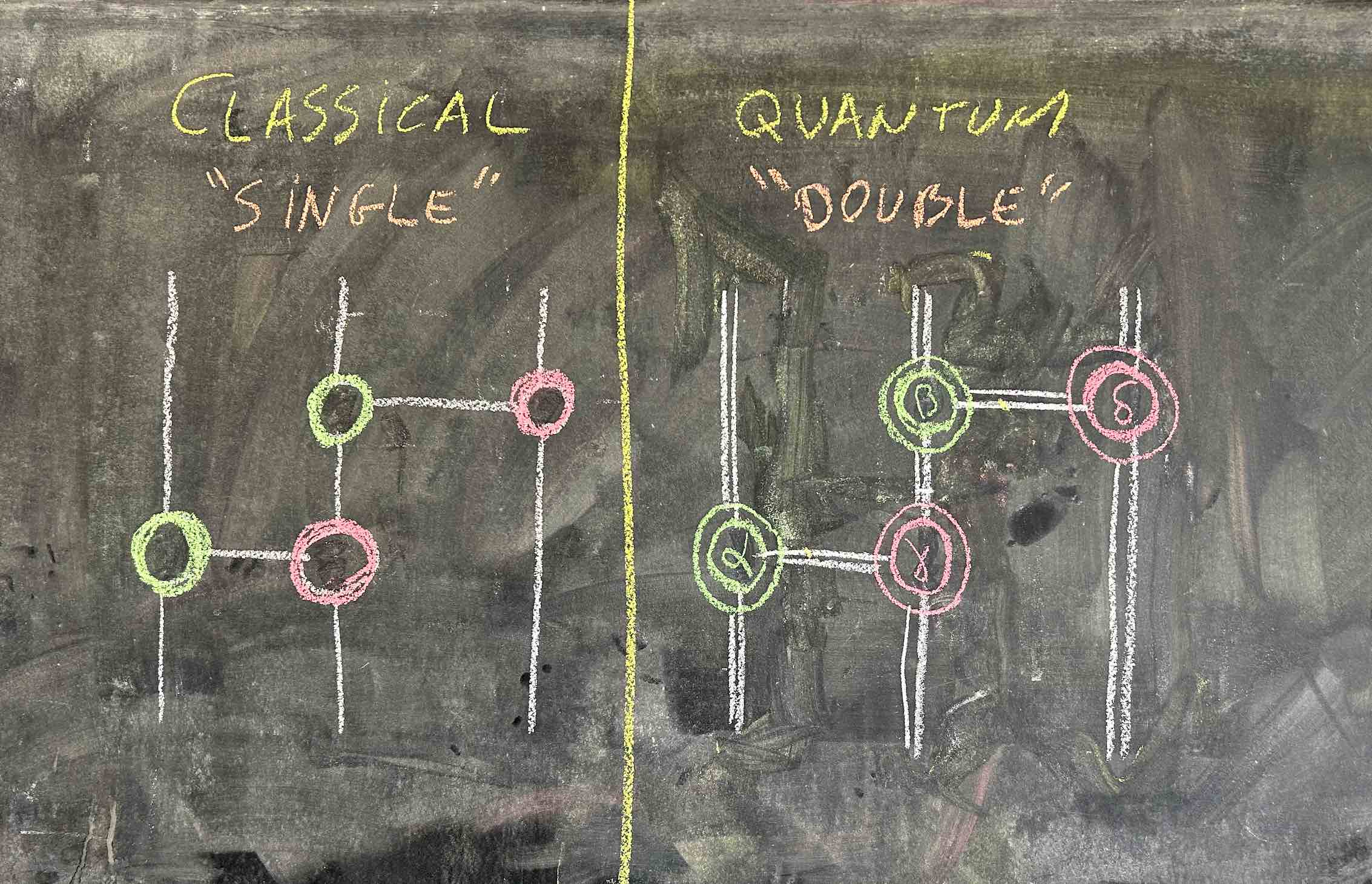,width=300pt} 
\end{center} 
A \bM measurement\e goes from quantum to classical and we can also \bM encode\e classical data as a quantum state as follows:
\begin{center}
\epsfig{figure=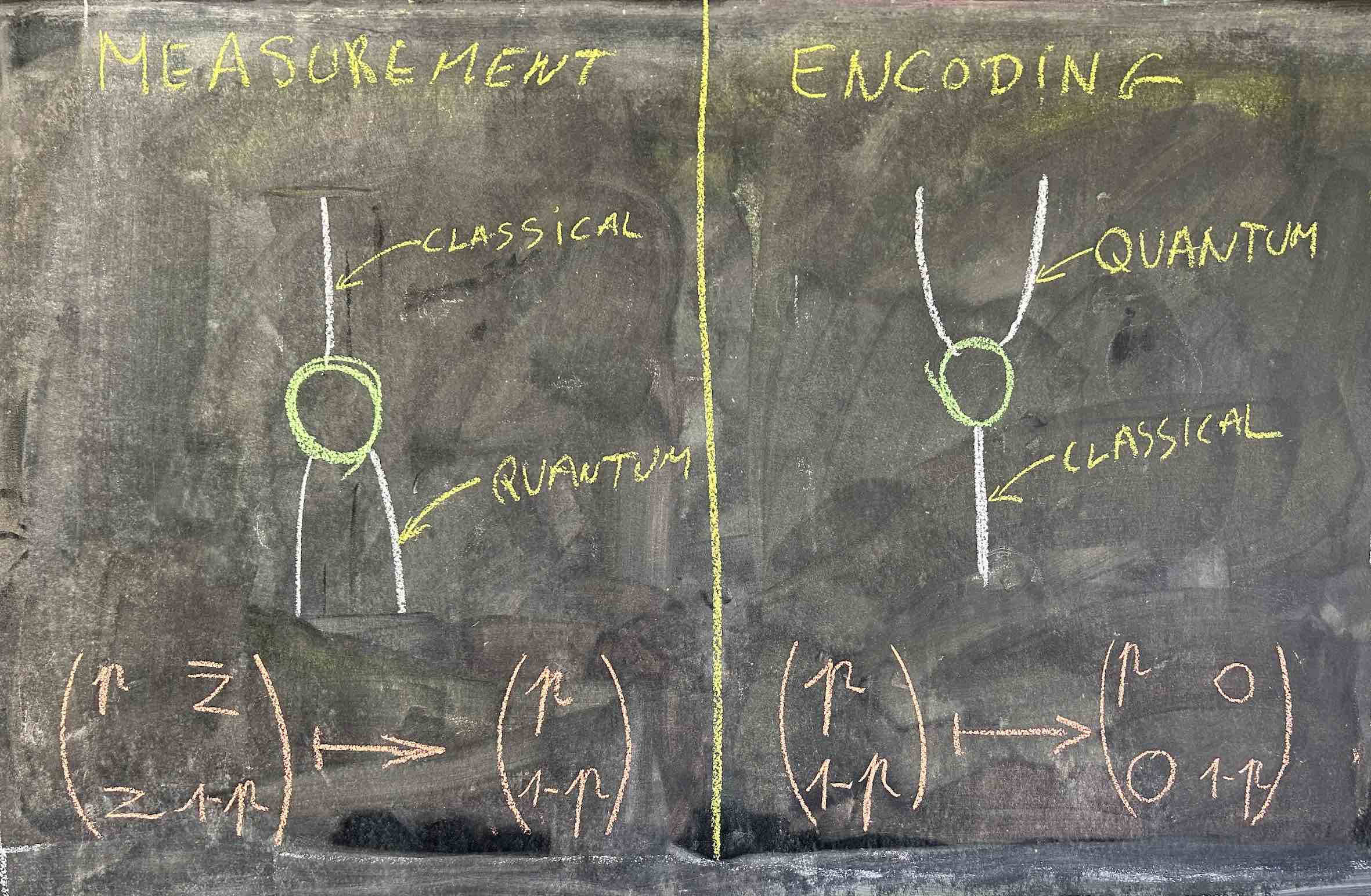,width=340pt} 
\end{center} 
As indicated, a measurement takes in a density matrix and returns a classical probability distribution on the possible outcomes, and encoding takes in a classical probability distributions and returns a density matrix with that probability distribution on the diagonal.

\bigskip\noindent
{\bf Exercise.}  Show that this indeed the case.  To do so, one should think of the density matrix as a two qubit state via the isomorphism between qubit to qubit linear maps and two qubit states.

\bigskip\noindent
Non-demolition measurements are depicted as follows:
\begin{center}
\epsfig{figure=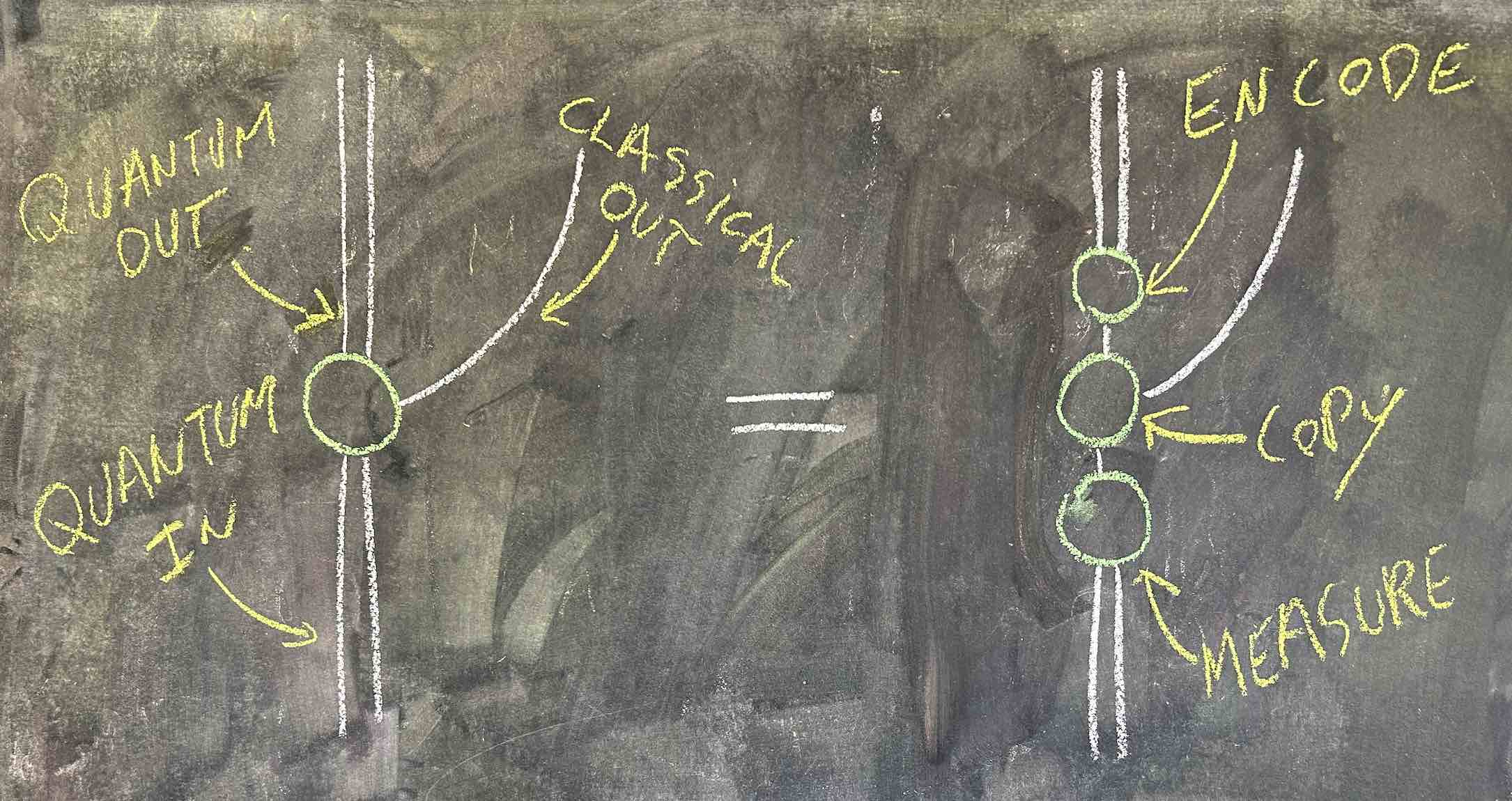,width=340pt} 
\end{center} 
where the decomposition on the right clearly shows how collapse is implicit here.

\section{Doing stuff with ZX-calculus part II}  

We will now see some more ZX-calculus in action.

\subsection{Composing quantum measurements}

Measurement after encoding in the same colour by spider-fusion results in doing nothing:
\begin{center}
\epsfig{figure=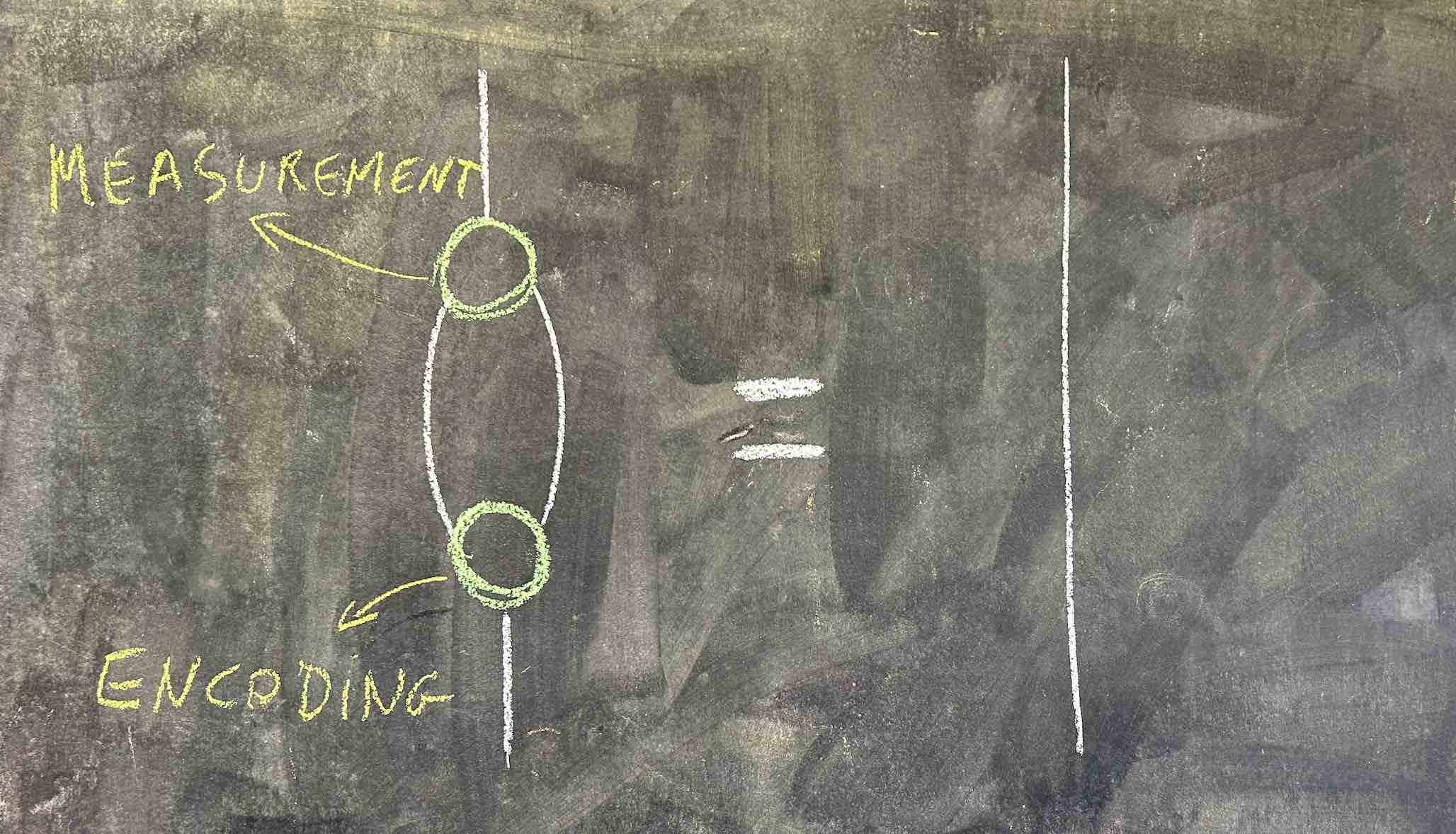,width=260pt} 
\end{center} 
while by the Hopf-rule for different colours we get:
\begin{center}
\epsfig{figure=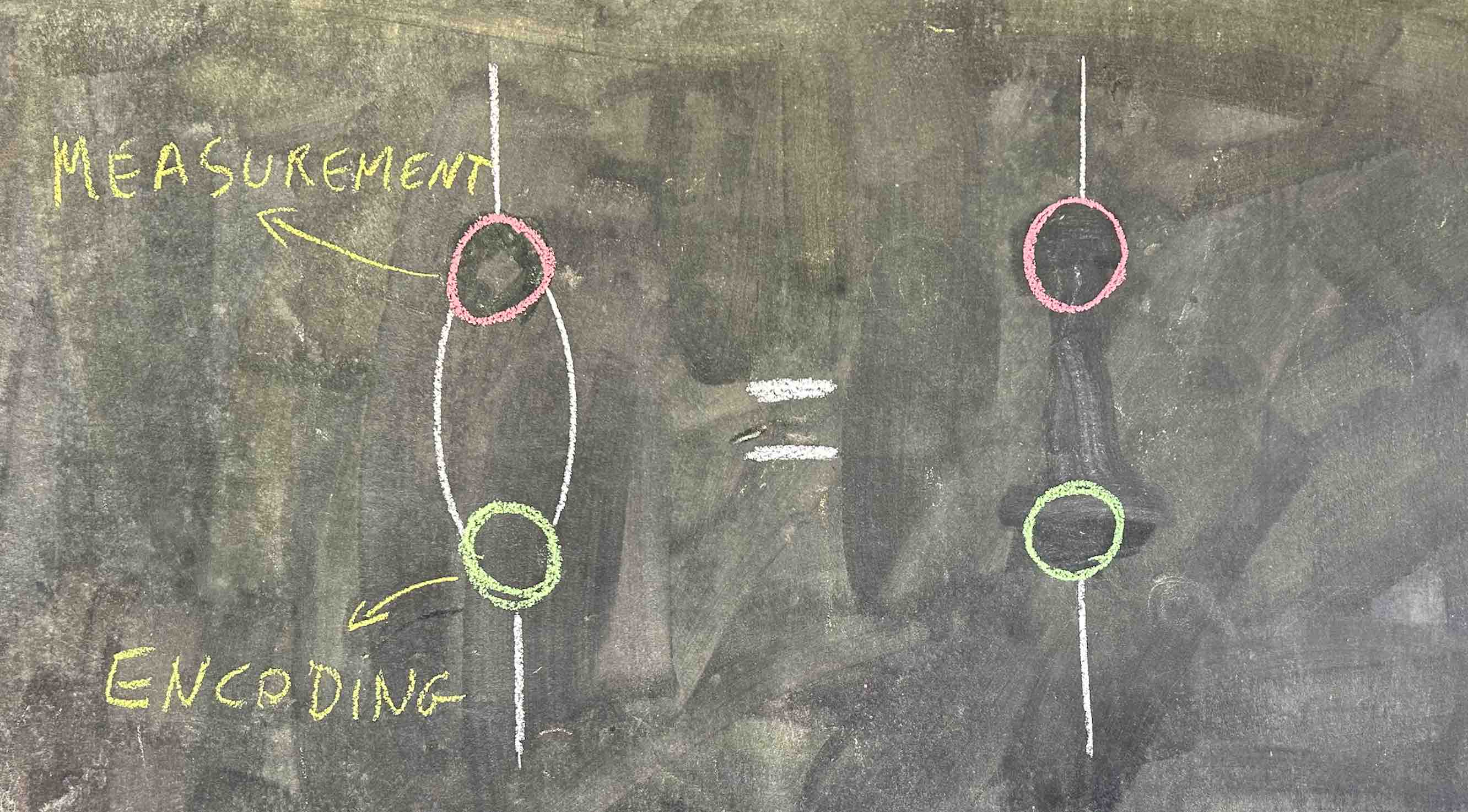,width=260pt} 
\end{center} 
that is, nothing gets through. In fact, the classical states:
\begin{center}
\epsfig{figure=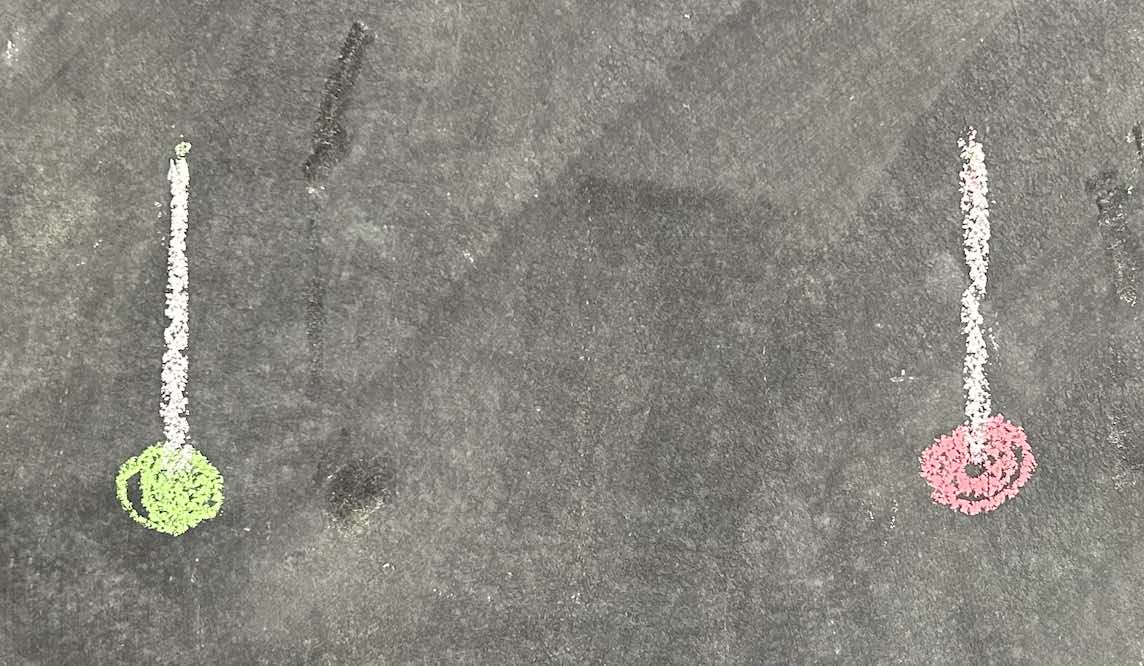,width=105pt}
\end{center}
stand for the uniform probability distribution, for classical data obtained  in a Z-measurement and an X-measurement respectively.  In our ZX-representation it is indeed the case that classical data is encoded in the basis that it is measured in \cite[\S 9.2.4]{CKbook}, recalling that:
 \begin{center}
\epsfig{figure=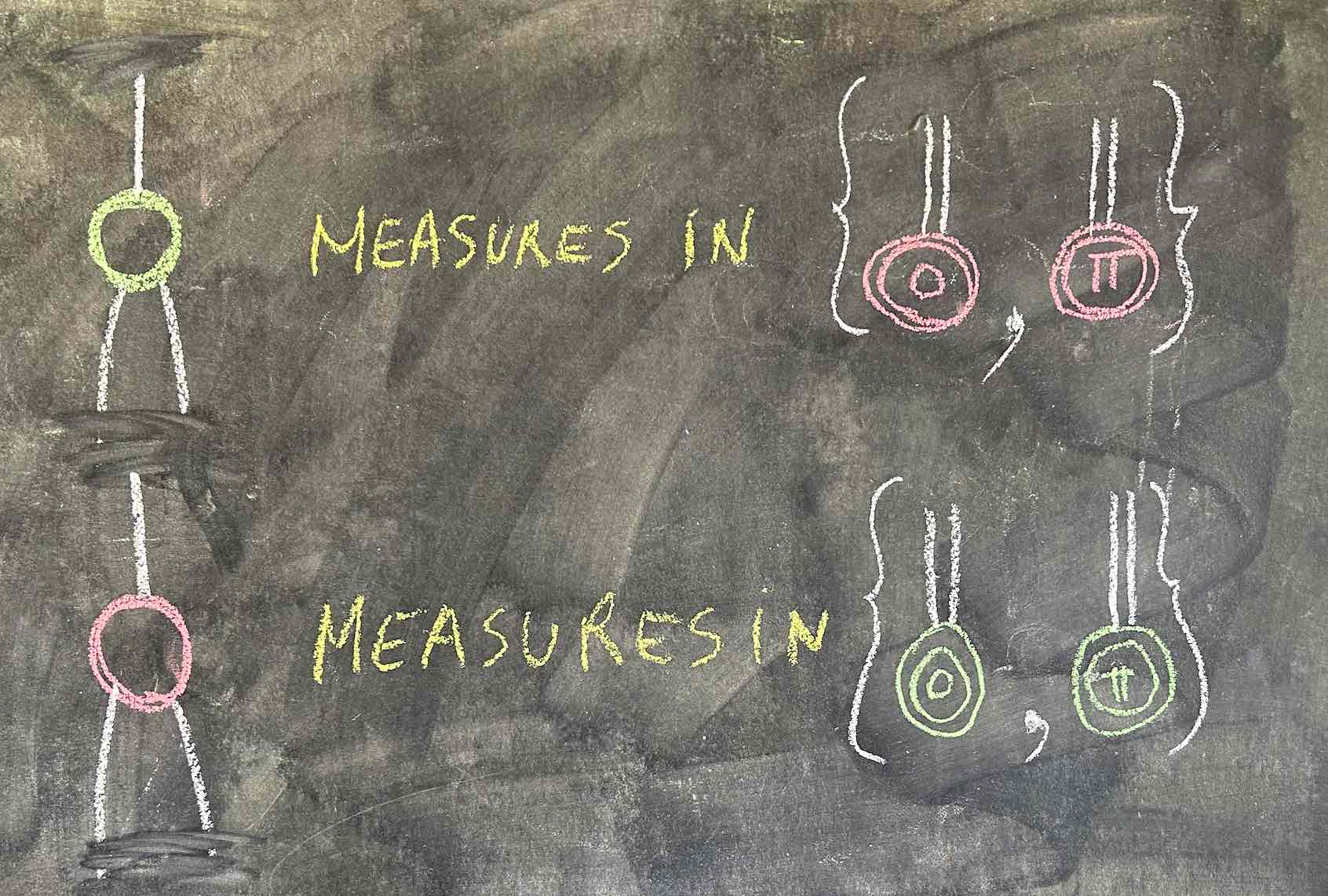,width=260pt}  
\end{center} 
When we measure a phase: 
\begin{center}
\epsfig{figure=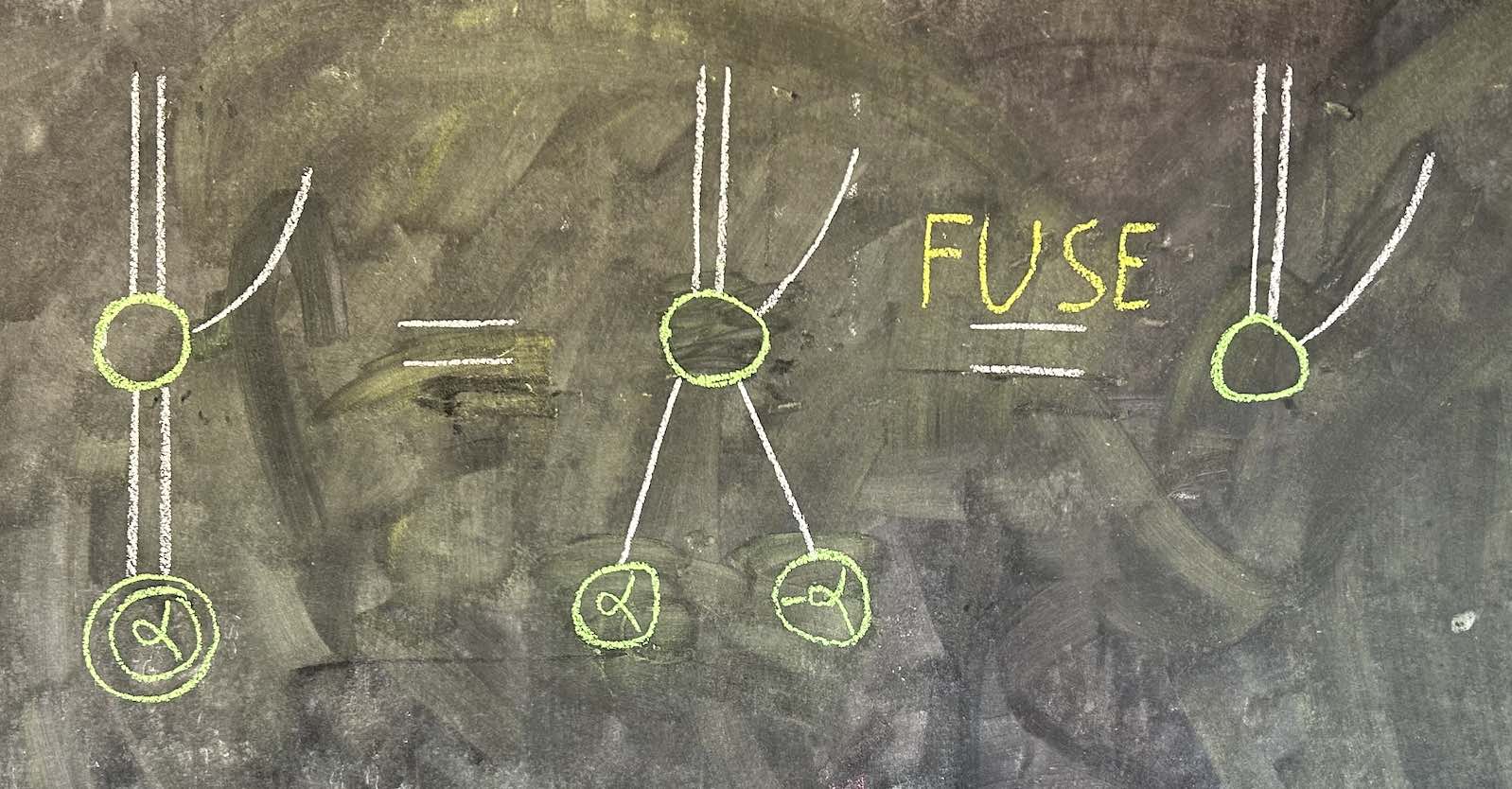,width=280pt} 
\end{center} 
it vanishes, and if we measure Z, then X, and then Z again:
\begin{center}
\epsfig{figure=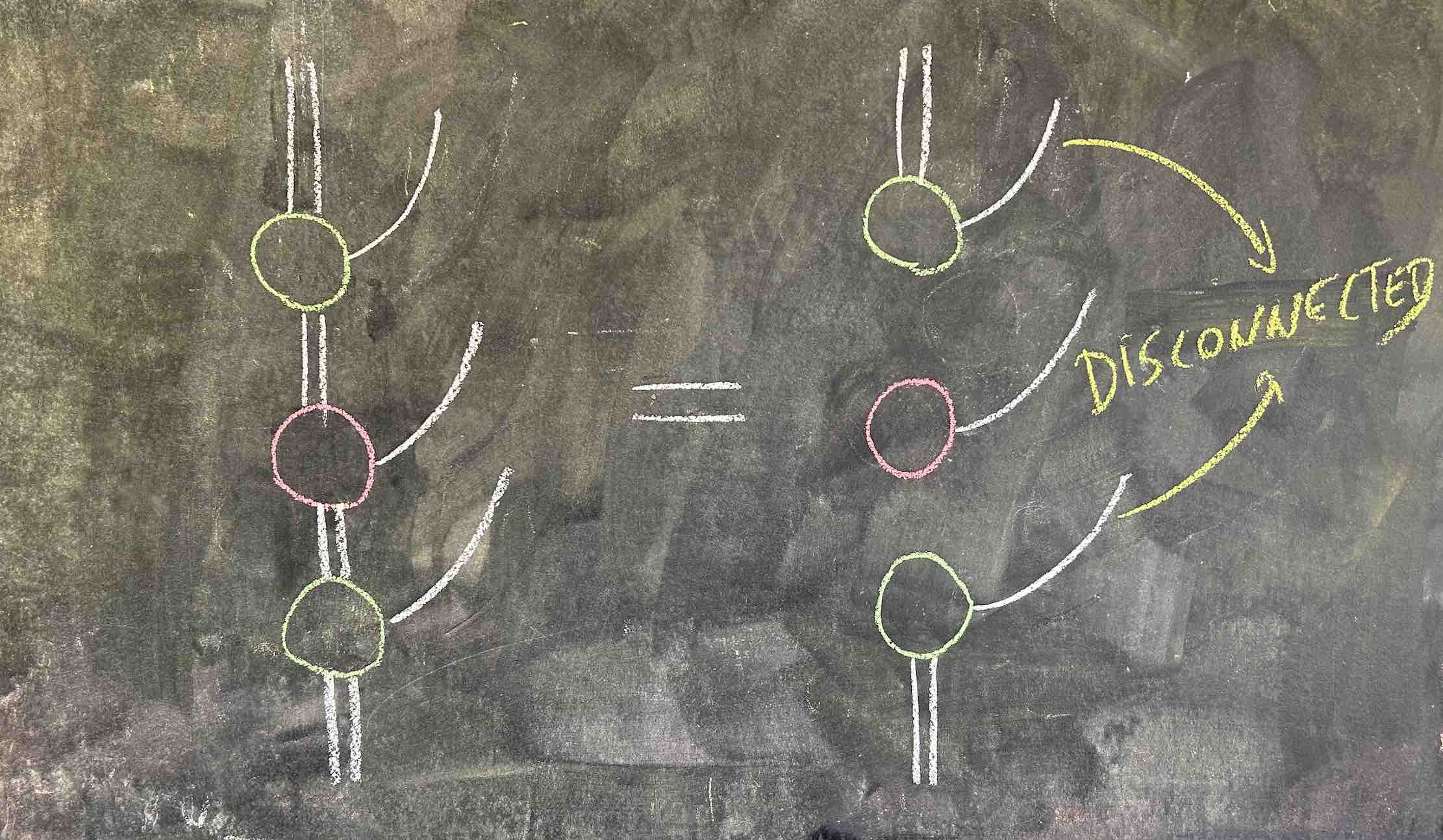,width=280pt} 
\end{center} 
so the two outcomes of the Z-measurements are unrelated.  As shown in \cite{CPer}, this simple results is directly underpinning the security of quantum key distribution in the BB84 form \cite{BB84}, and by using cups we obtain the Ekert 91 form \cite{Ekert91}.

\subsection{Classical communication}

We now revisit quantum teleportation again in order to make the communication between Alice and Bob part of the diagram.  Flipping what we saw earlier upside down, we can represent the Bell-measurement as a CNOT, a Z-measurement and an X-measurement:
\begin{center}
\epsfig{figure=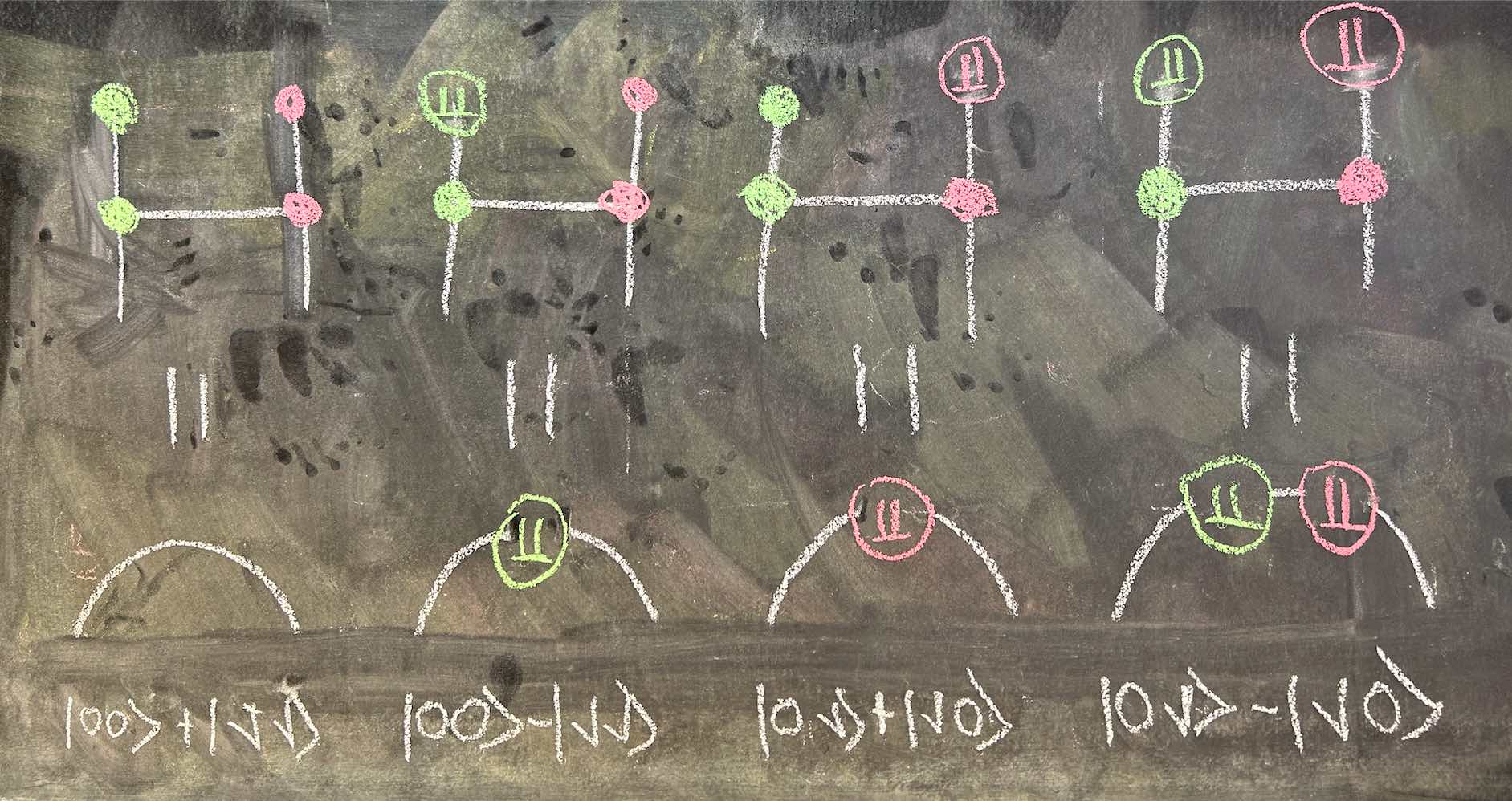,width=260pt} 
\end{center}  
so we obtain:
\begin{center}
\epsfig{figure=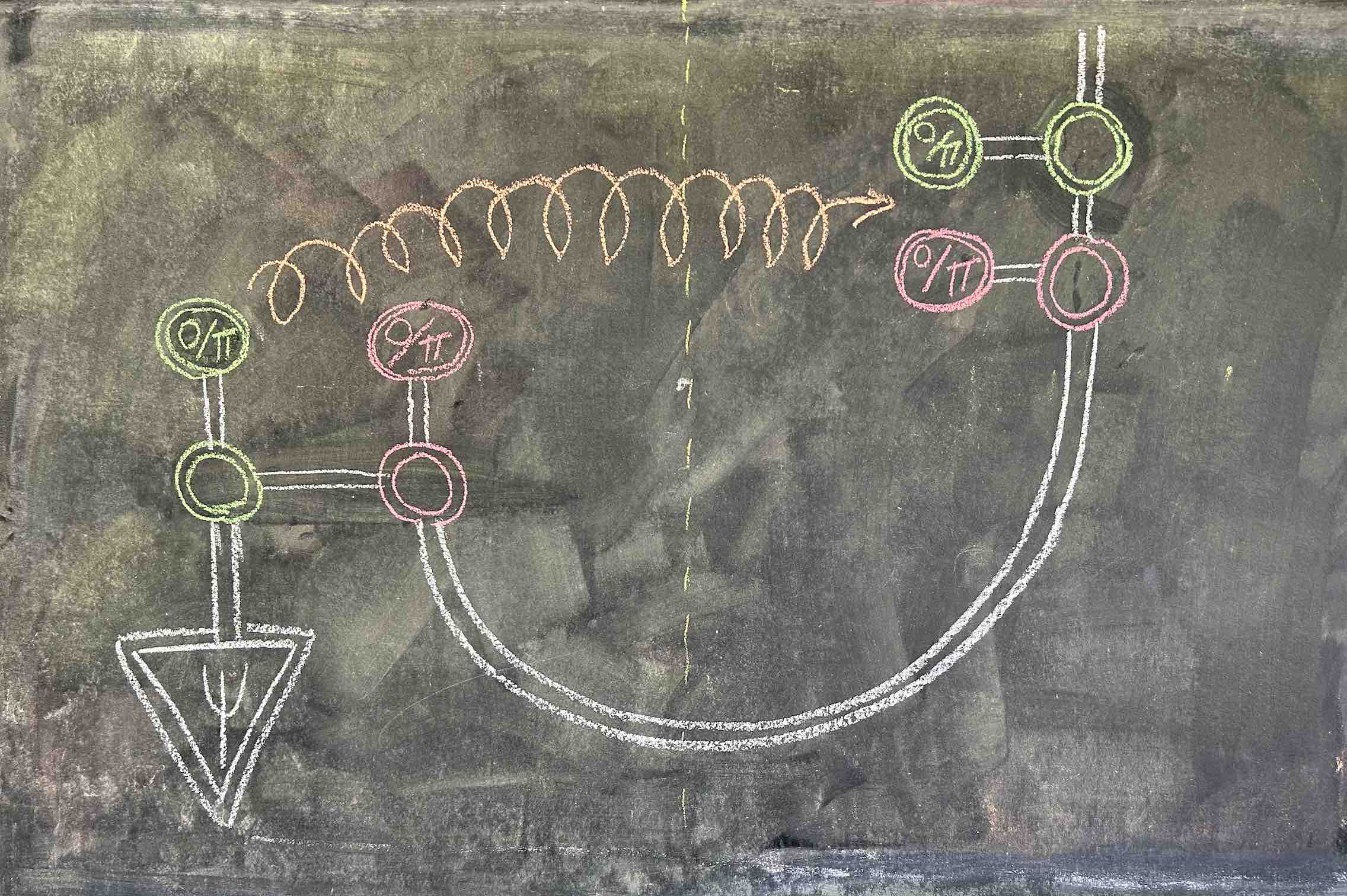,width=260pt} 
\end{center} 
and now using spiders as measurements and encoding we get:
\begin{center}
\epsfig{figure=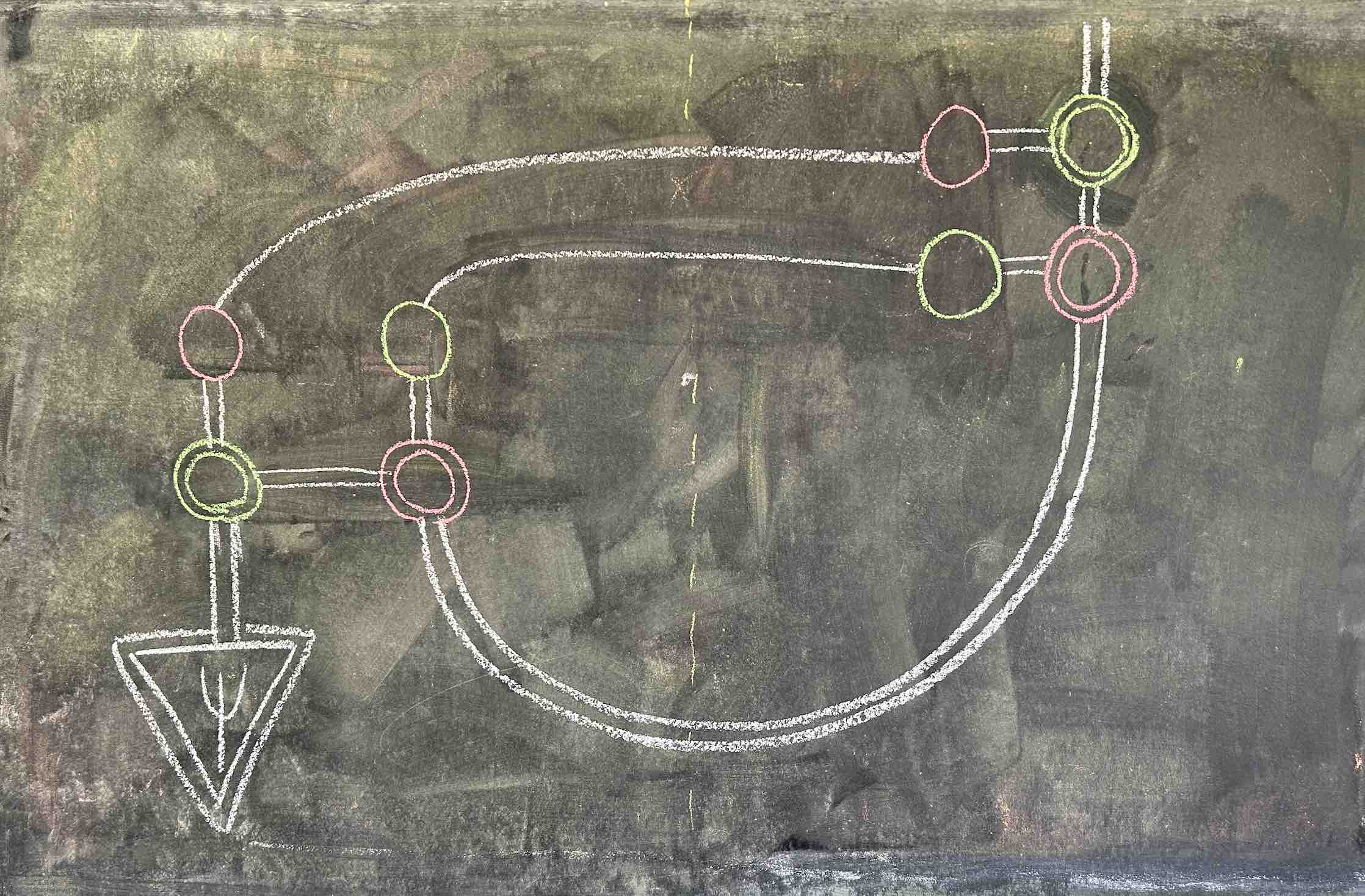,width=260pt} 
\end{center} 
where the colour change is due to the fact that green/red measures in the red/green basis. We now know what the dodo on the cover of \cite{CKbook} was all the time thinking of:
\begin{center}
\epsfig{figure=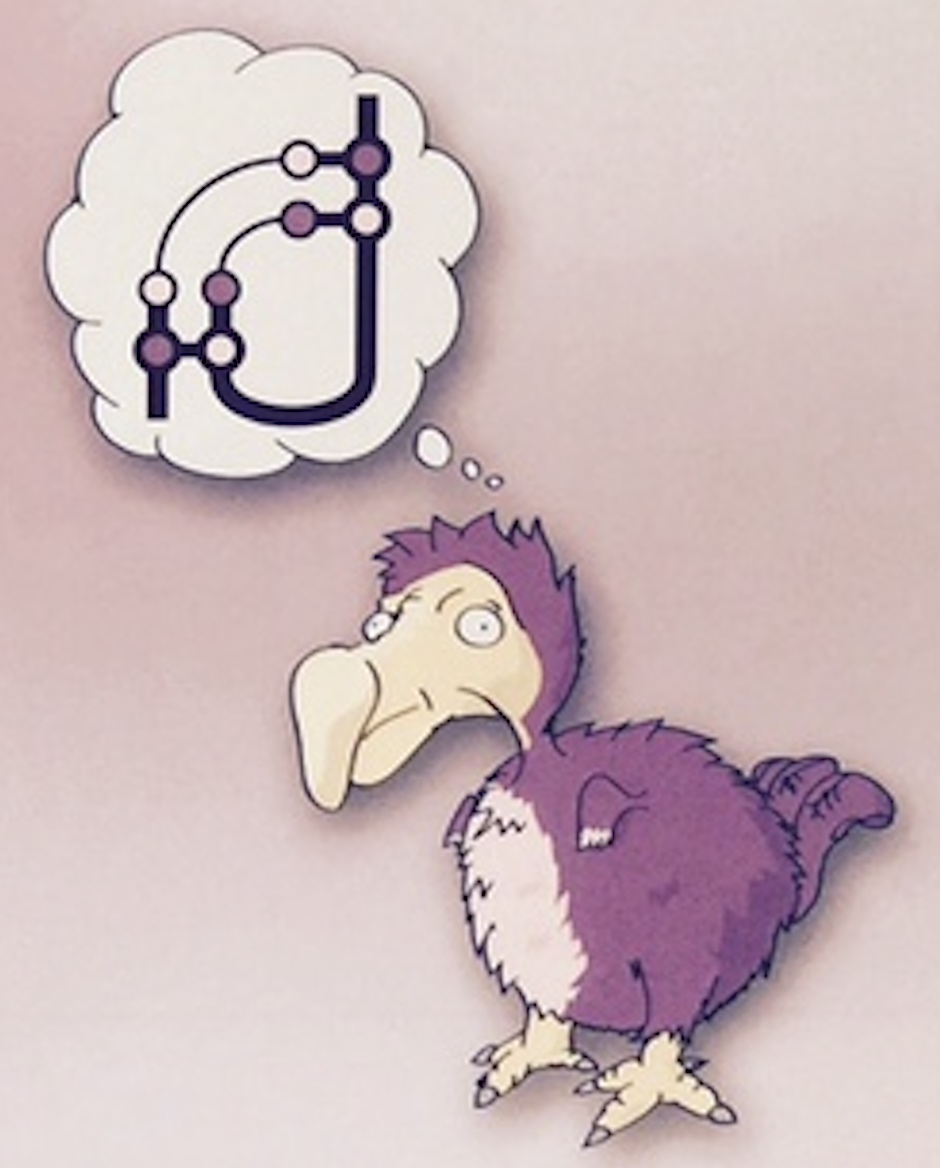,width=180pt} 
\end{center} 
If you can't get enough of ZX-calculus, then this `dodo-book' is also a  place to continue reading.  In the case that its 900 pages scare you -- do note that we teach it in only 20 lectures at Oxford University -- you may want to start with Quantum in Pictures \cite{QiP}:  
\begin{center}
\epsfig{figure=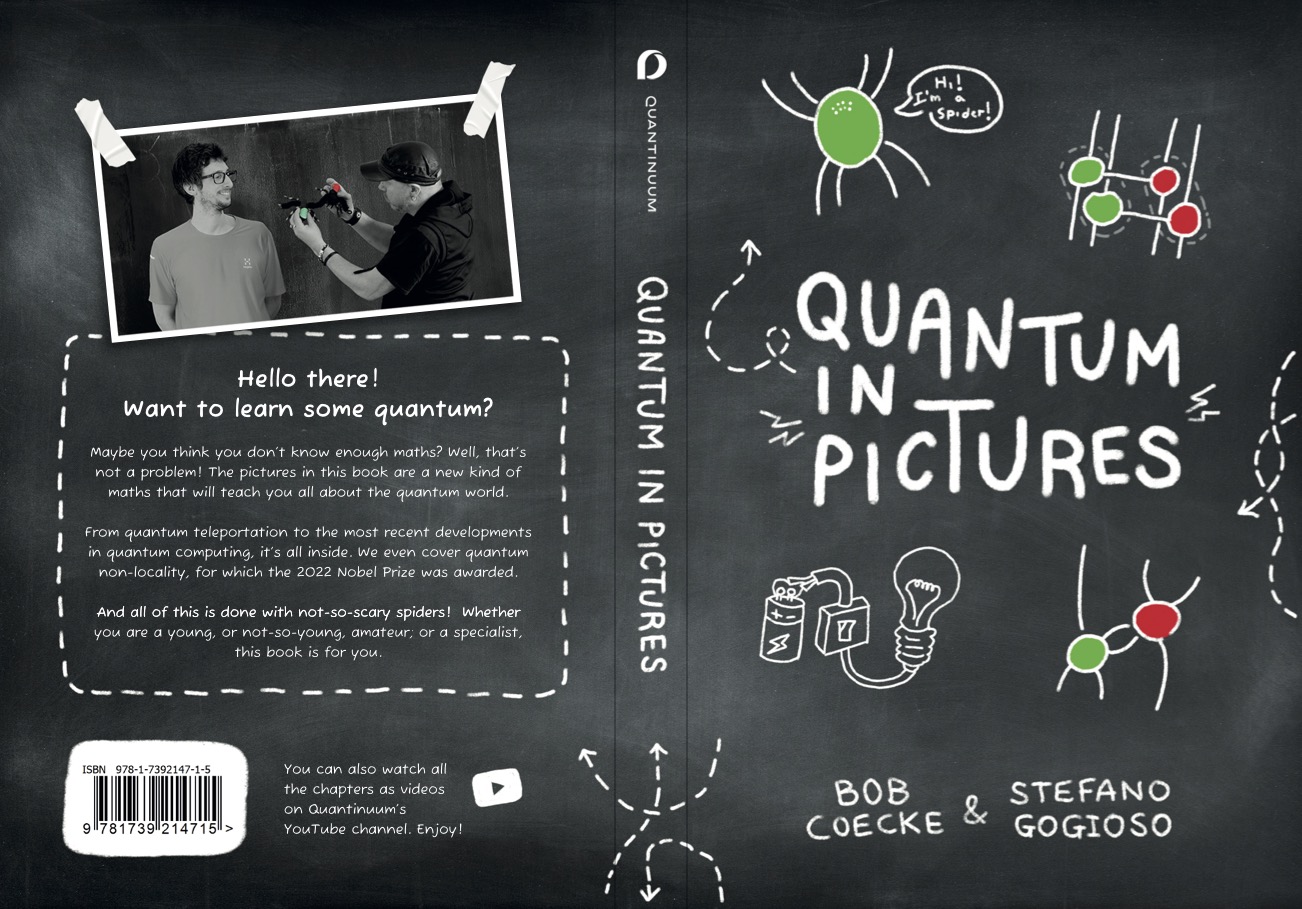,width=360pt}   
\end{center} 
Both \cite{CKbook} and \cite{QiP} contain a purely diagrammatic proof of quantum non-locality, and lot's of other stuff not presented here.   In fact, QiP  tells the pictorial ZX-story without any reference to Hilbert space nor any other mathematics or physics prerequisites.  The very fact that this is even possible is an important statement in its own right, both from a scientific and educational perspective.  A somewhat similar to this one but older tutorial  is \cite{coecke2012tutorial}.  An overview of some more recent applications of ZX-calculus and also some more philosophy can be found in \cite{coecke2021kindergarden}. Professionals may also want to go for the more advanced \cite{JohnSurvey}.  All the ZX-publications can be found at:
\begin{center}
https://zxcalculus.com/publications.html
\end{center}
as well as a link for the ZX-calculus Discord page.  

\section{Acknowledgements} 

We thank Marcello Benedetti and Lia Yeh for carefully proofreading these notes, and Matty Hoban for additional comments.

%%%%%%%%%%%%%%%%%%%%%%%%%%%%%%%%%%%
\bibliographystyle{plain} 
\bibliography{mainNOW}

\begin{thebibliography}{10}

\bibitem{Backens}
M.~Backens.
\newblock The {ZX}-calculus is complete for stabilizer quantum mechanics.
\newblock {\em New Journal of Physics}, 16:093021, 2014.
\newblock {arXiv:1307.7025}.

\bibitem{Backens2}
M.~Backens.
\newblock The {ZX}-calculus is complete for the single-qubit {C}lifford+{T}
  group.
\newblock In B.~Coecke, I.~Hasuo, and P.~Panangaden, editors, {\em Proceedings
  of the 11th workshop on Quantum Physics and Logic}, volume 172 of {\em
  Electronic Proceedings in Theoretical Computer Science}, pages 293--303. Open
  Publishing Association, 2014.

\bibitem{BB84}
C.~H. Bennett and G.~Brassard.
\newblock Quantum cryptography: Public key distribution and coin tossing.
\newblock In {\em Proceedings of IEEE International Conference on Computers,
  Systems and Signal Processing}, pages 175--179. IEEE, 1984.

\bibitem{Tele}
C.~H. Bennett, G.~Brassard, C.~Crepeau, R.~Jozsa, A.~Peres, and W.~K. Wootters.
\newblock {Teleporting an unknown quantum state via dual classical and
  Einstein-Podolsky-Rosen channels}.
\newblock {\em Physical Review Letters}, 70(13):1895--1899, 1993.

\bibitem{ContPhys}
B.~Coecke.
\newblock Quantum picturalism.
\newblock {\em Contemporary Physics}, 51:59--83, 2009.
\newblock {a}rXiv:0908.1787.

\bibitem{QNLP-foundations}
B.~Coecke, G.~de~Felice, K.~Meichanetzidis, and A.~Toumi.
\newblock Foundations for near-term quantum natural language processing, 2020.
\newblock arXiv preprint arXiv:2012.03755.

\bibitem{CD0}
B.~Coecke and R.~Duncan.
\newblock A graphical calculus for quantum observables.
\newblock {\em zxcalculus.com/publications.html}, 2007.

\bibitem{CD1}
B.~Coecke and R.~Duncan.
\newblock Interacting quantum observables.
\newblock In {\em Proceedings of the 37th International Colloquium on Automata,
  Languages and Programming (ICALP)}, Lecture Notes in Computer Science, 2008.

\bibitem{CD2}
B.~Coecke and R.~Duncan.
\newblock Interacting quantum observables: categorical algebra and
  diagrammatics.
\newblock {\em New Journal of Physics}, 13:043016, 2011.
\newblock {arXiv:quant-ph/09064725}.

\bibitem{coecke2012tutorial}
B.~Coecke and R.~Duncan.
\newblock Tutorial: Graphical calculus for quantum circuits.
\newblock In {\em International Workshop on Reversible Computation}, pages
  1--13. Springer, 2012.

\bibitem{QiP}
B.~Coecke and S.~Gogioso.
\newblock {\em Quantum in Pictures}.
\newblock Quantinuum, 2022.

\bibitem{coecke2021kindergarden}
B.~Coecke, D.~Horsman, A.~Kissinger, and Q.~Wang.
\newblock Kindergarden quantum mechanics graduates (... or how {I} learned to
  stop gluing lego together and love the {ZX}-calculus).
\newblock {\em arXiv preprint arXiv:2102.10984}, 2021.

\bibitem{CKpaperI}
B.~Coecke and A.~Kissinger.
\newblock Categorical quantum mechanics {I}: causal quantum processes.
\newblock In E.~Landry, editor, {\em Categories for the Working Philosopher}.
  Oxford University Press, 2016.
\newblock ar{X}iv:1510.05468.

\bibitem{CKbook}
B.~Coecke and A.~Kissinger.
\newblock {\em Picturing Quantum Processes. A First Course in Quantum Theory
  and Diagrammatic Reasoning}.
\newblock Cambridge University Press, 2017.

\bibitem{CPer}
B.~Coecke and S.~Perdrix.
\newblock Environment and classical channels in categorical quantum mechanics.
\newblock In {\em Proceedings of the 19th EACSL Annual Conference on Computer
  Science Logic (CSL)}, volume 6247 of {\em Lecture Notes in Computer Science},
  pages 230--244, 2010.
\newblock Extended version: {a}rXiv:1004.1598.

\bibitem{DBLP:conf/rc/CoeckeW18}
B.~Coecke and Q.~Wang.
\newblock {ZX}-rules for 2-qubit {C}lifford+{T} quantum circuits.
\newblock In Jarkko Kari and Irek Ulidowski, editors, {\em Reversible
  Computation - 10th International Conference, {RC} 2018, Leicester, UK,
  September 12-14, 2018, Proceedings}, volume 11106 of {\em Lecture Notes in
  Computer Science}, pages 144--161. Springer, 2018.

\bibitem{cowtan2019phase}
A.~Cowtan, S.~Dilkes, R.~Duncan, W.~Simmons, and S.~Sivarajah.
\newblock Phase gadget synthesis for shallow circuits.
\newblock {\em arXiv preprint arXiv:1906.01734}, 2019.

\bibitem{de2017zx}
N.~de~Beaudrap and D.~Horsman.
\newblock The {ZX} calculus is a language for surface code lattice surgery.
\newblock {\em arXiv preprint arXiv:1704.08670}, 2017.

\bibitem{defeliceQuantumLinearOptics2022}
G.~de~Felice and B.~Coecke.
\newblock Quantum {{Linear Optics}} via {{String Diagrams}}.
\newblock arXiv:2204.12985.

\bibitem{clifford-simp}
R.~Duncan, A.~Kissinger, S.~Perdrix, and J.~Van De~Wetering.
\newblock Graph-theoretic simplification of quantum circuits with the
  {ZX}-calculus.
\newblock {\em Quantum}, 4:279, 2020.

\bibitem{Ekert91}
A.~K. Ekert.
\newblock {Quantum cryptography based on {B}ell's theorem}.
\newblock {\em Physical Review Letters}, 67(6):661--663, 1991.

\bibitem{Gidney2019}
C.~Gidney and A.~G. Fowler.
\newblock Efficient magic state factories with a catalyzed {$|CCZ\rangle$} to
  {$2|T\rangle$} transformation.
\newblock {\em {Quantum}}, 3:135, April 2019.

\bibitem{Amar}
A.~Hadzihasanovic.
\newblock A diagrammatic axiomatisation for qubit entanglement.
\newblock In {\em Proceedings of the 30th Annual IEEE Symposium on Logic in
  Computer Science (LICS)}, 2015.
\newblock {arXiv:1501.07082}.

\bibitem{hadzihasanovic2018two}
A.~Hadzihasanovic, K.~F. Ng, and Q.~Wang.
\newblock Two complete axiomatisations of pure-state qubit quantum computing.
\newblock In {\em Proceedings of the 33rd Annual ACM/IEEE Symposium on Logic in
  Computer Science}, pages 502--511. ACM, 2018.

\bibitem{jeandel2018complete}
E.~Jeandel, S.~Perdrix, and R.~Vilmart.
\newblock A complete axiomatisation of the {ZX}-calculus for {C}lifford+{T}
  quantum mechanics.
\newblock In {\em Proceedings of the 33rd Annual ACM/IEEE Symposium on Logic in
  Computer Science}, pages 559--568, 2018.
\newblock arXiv preprint arXiv:1705.11151.

\bibitem{JS}
A.~Joyal and R.~Street.
\newblock The geometry of tensor calculus {I}.
\newblock {\em Advances in Mathematics}, 88:55--112, 1991.

\bibitem{khesinGraphicalQuantumCliffordencoder2023}
A.~B. Khesin, J.~Z. Lu, and P.~W. Shor.
\newblock Graphical quantum {{Clifford-encoder}} compilers from the {{ZX}}
  calculus.
\newblock arXiv:2301.02356.

\bibitem{KissingerTcount}
A.~Kissinger and J.~van~de Wetering.
\newblock Reducing the number of non-{C}lifford gates in quantum circuits.
\newblock {\em Phys. Rev. A}, 102:022406, Aug 2020.

\bibitem{litinski2022active}
D.~Litinski and N.~Nickerson.
\newblock Active volume: An architecture for efficient fault-tolerant quantum
  computers with limited non-local connections.
\newblock {\em arXiv preprint arXiv:2211.15465}, 2022.

\bibitem{perdrix2005}
S.~Perdrix.
\newblock State transfer instead of teleportation in measurement-based quantum
  computation.
\newblock {\em International Journal of Quantum Information}, 3(01):219--223,
  2005.

\bibitem{MBQC2}
R.~Raussendorf, D.E. Browne, and H.J. Briegel.
\newblock Measurement-based quantum computation on cluster states.
\newblock {\em Physical Review A}, 68(2):22312, 2003.

\bibitem{JohnSurvey}
J.~van~de Wetering.
\newblock {ZX}-calculus for the working quantum computer scientist.
\newblock {\em arXiv:2012.13966}, 2020.

\bibitem{vilmart2018near}
R.~Vilmart.
\newblock A near-optimal axiomatisation of {ZX}-calculus for pure qubit quantum
  mechanics.
\newblock {\em arXiv preprint arXiv:1812.09114}, 2018.

\bibitem{wangDifferentiatingIntegratingZX2022}
Q.~Wang, R.~Yeung, and M.~Koch.
\newblock Differentiating and {{Integrating ZX Diagrams}} with {{Applications}}
  to {{Quantum Machine Learning}}.
\newblock arXiv:2201.13250.

\bibitem{zhao2021analyzing}
Chen Zhao and Xiao-Shan Gao.
\newblock Analyzing the barren plateau phenomenon in training quantum neural
  networks with the zx-calculus.
\newblock {\em Quantum}, 5:466, 2021.

\end{thebibliography}
%%%%%%%%%%%%%%%%%%%%%%%%%%%%%%%%%%%

\end{document}